\documentclass[12pt,english]{article}
\usepackage[T1]{fontenc}
\usepackage[latin1]{inputenc}
\usepackage{a4wide}
\usepackage{amsmath}
\usepackage{color}
\usepackage{graphicx}
\usepackage{amssymb}

\usepackage{babel}

\newcommand {\beq}{\begin{equation}}
\newcommand {\eeq}{\end{equation}}
\newcommand {\bea}{\begin{eqnarray}}
\newcommand {\eea}{\end{eqnarray}}
\newcommand {\nn}{\nonumber \\}
\newcommand {\Tr}{{\rm Tr\,}}

\newcommand {\e}{{\rm e}}

\newcommand {\m}{\mu}
\newcommand {\n}{\nu}
\newcommand {\pl}{\partial}

\newcommand {\al}{\alpha}
\newcommand {\be}{\beta}

\newcommand {\la}{\lambda}
\newcommand {\La}{\Lambda}

\newcommand {\om}{\omega}

\newcommand {\ep}{\epsilon}

\newcommand {\na}{\nabla}
\newcommand {\del}  {\delta}
\newcommand {\Del}  {\Delta}
\newcommand {\mn}{{\mu\nu}}

\newcommand {\half}{ {\frac{1}{2}} }

\newcommand {\fourth} {\frac{1}{4} }

\newcommand {\Ecal}{{\cal E}}
\newcommand {\Fcal}{{\cal F}}
\newcommand {\Lcal}{{\cal L}}

\newcommand {\Pcal}{{\cal P}}
\newcommand {\Dcal}{{\cal D}}

\newcommand {\avec}{{\vec a}}

\newcommand {\rvec}{{\vec r}}

\newcommand {\omvec}{{\vec \omega}}

\newcommand {\ctil}{{\tilde c}}

\newcommand {\Ltil}  {{\tilde L}}

\newcommand {\ptil} {{\tilde p}}
\newcommand {\ktil} {{\tilde k}}

\newcommand {\omtil}{{\tilde \omega}}

\newcommand {\Lhat}{{\hat L}}

\newcommand {\What}{{\hat W}}

\newcommand {\delh} {{\hat \delta}}

\newcommand {\bdot}{\dot{b}}
\newcommand {\rdot}{\dot{r}}
\newcommand {\rddot}{\ddot{r}}
\newcommand {\udot}{\dot{u}}
\newcommand {\bfZ} {{\bf Z}}
\newcommand {\K}{{\bf K}}
\newcommand {\I}{{\bf I}}

\newcommand {\intfx} {{\int d^4x}}

\newcommand {\intxz} {{\int d^4xdz}}

\newcommand {\intp} {{\int \frac{d^4p}{(2\pi)^4}}}
\newcommand {\intpE} {{\int \frac{d^4p_E}{(2\pi)^4}}}

\newcommand {\intt} {{\int_{0}^{\infty}\frac{dt}{t}}}
\newcommand {\change} {\leftrightarrow}
\newcommand {\ra} {\rightarrow}

\newcommand {\pr}   {{\quad .}}
\newcommand {\com}  {{\quad ,}}
\newcommand {\q}    {\quad}

\newcommand {\qqq}   {\quad\quad\quad}

\newcommand {\nl}    {\newline}

\newcommand {\NP}   {Nucl.Phys.}
\newcommand {\PL}   {Phys.Lett.}
\newcommand {\PR}   {Phys.Rev.}
\newcommand {\PRt}   {Phys.Rept.}
\newcommand {\PRL}   {Phys.Rev.Lett.}

\newcommand {\JHEP}  {J. High Energy Phys.}

\newcommand {\PTP}  {Prog.Theor.Phys.}

\newcommand {\CQG}  {Class.Quantum.Grav.}

\newcommand {\Pla} {\frac{{\tilde p}}{\omega}}

\newcommand {\Tev} {\frac{{\tilde p}}{T}}

\newcommand {\tnpb}  {{\frac{2n\pi}{\be}}}

\newcommand {\Hora} {Ho\u{r}ava}      

\begin{document}

\title{Casimir Energy of 5D Warped System and 
Sphere Lattice Regularization}

\author{S. Ichinose}

\maketitle
\begin{center}\emph{
Laboratory of Physics, School of Food and Nutritional Sciences, 
University of Shizuoka\\
Yada 52-1, Shizuoka 422-8526, Japan
}\end{center}

\begin{abstract}
Casimir energy is calculated for the 5D electromagnetism and 5D scalar theory
in the {\it warped} geometry. 
It is compared with the flat case(arXiv:0801.3064). 
A new regularization, 
called {\it sphere lattice regularization}, is taken. 
In the integration over the 5D space, we introduce two boundary 
curves (IR-surface and UV-surface) based on the {\it minimal area principle}. 
It is a {\it direct} realization of the geometrical approach 
to the {\it renormalization group}.  
The regularized configuration is {\it closed-string like}. 
We do {\it not} take the KK-expansion approach. Instead, 
the position/momentum propagator is exploited, 
combined with the {\it heat-kernel method}. All expressions
are closed-form (not KK-expanded form). 
The {\it generalized} P/M propagators are introduced. 
Rigorous quantities 
are only treated (non-perturbative treatment).  
The properly regularized form of Casimir energy, is expressed in a closed form. 
We numerically evaluate $\La$(4D UV-cutoff), $\om$(5D bulk curvature, 
warp parameter)
and $T$(extra space IR parameter) dependence of the Casimir energy. We present 
two {\it new ideas} in order to define the 5D QFT:\  
1) the summation (integral) region over the 5D space is {\it restricted} by two minimal surfaces 
(IR-surface, UV-surface) ; or 
2) we introduce a {\it weight function} and require the dominant contribution, in the summation, 
is given by the {\it minimal surface}. 
Based on these, 
5D Casimir energy is {\it finitely} obtained after the {\it proper renormalization
procedure.} 
The {\it warp parameter} $\om$ suffers from the {\it renormalization effect}. 
The IR parameter $T$ does not. 
In relation to characterizing the dominant path, 
we {\it classify} all paths 
(minimal surface curves) in AdS$_5$ space. 
We examine the meaning of the weight function and finally 
reach a {\it new definition} of the Casimir energy where {\it the 4D momenta( or coordinates) 
are quantized} with the extra coordinate as the Euclidean time (inverse temperature). 
We comment on the cosmological constant term and present an answer to the problem at the end. 
Dirac's large number naturally appears. 
\end{abstract}

PACS: 
PACS NO:
04.50.+h,\ 
11.10.Kk,\ 
11.25.Mj,\ 
12.10.-g 
11.30.Er,\ 

Keywords: Warped geometry, position/momentum propagator, 
heat-kernel, Casimir energy, sphere lattice, 
Renormalization of boundary parameters, Cosmological constant
\section{Introduction\label{intro}}

\begin{figure}
\caption{
The configuration of the Casimir energy measurement. 
The radiation cavity bounded by two parallel perfectly-conducting plates
separated by $2l$. The plate size is $2L\times 2L$. 
        }
\begin{center}
\includegraphics[height=8cm]{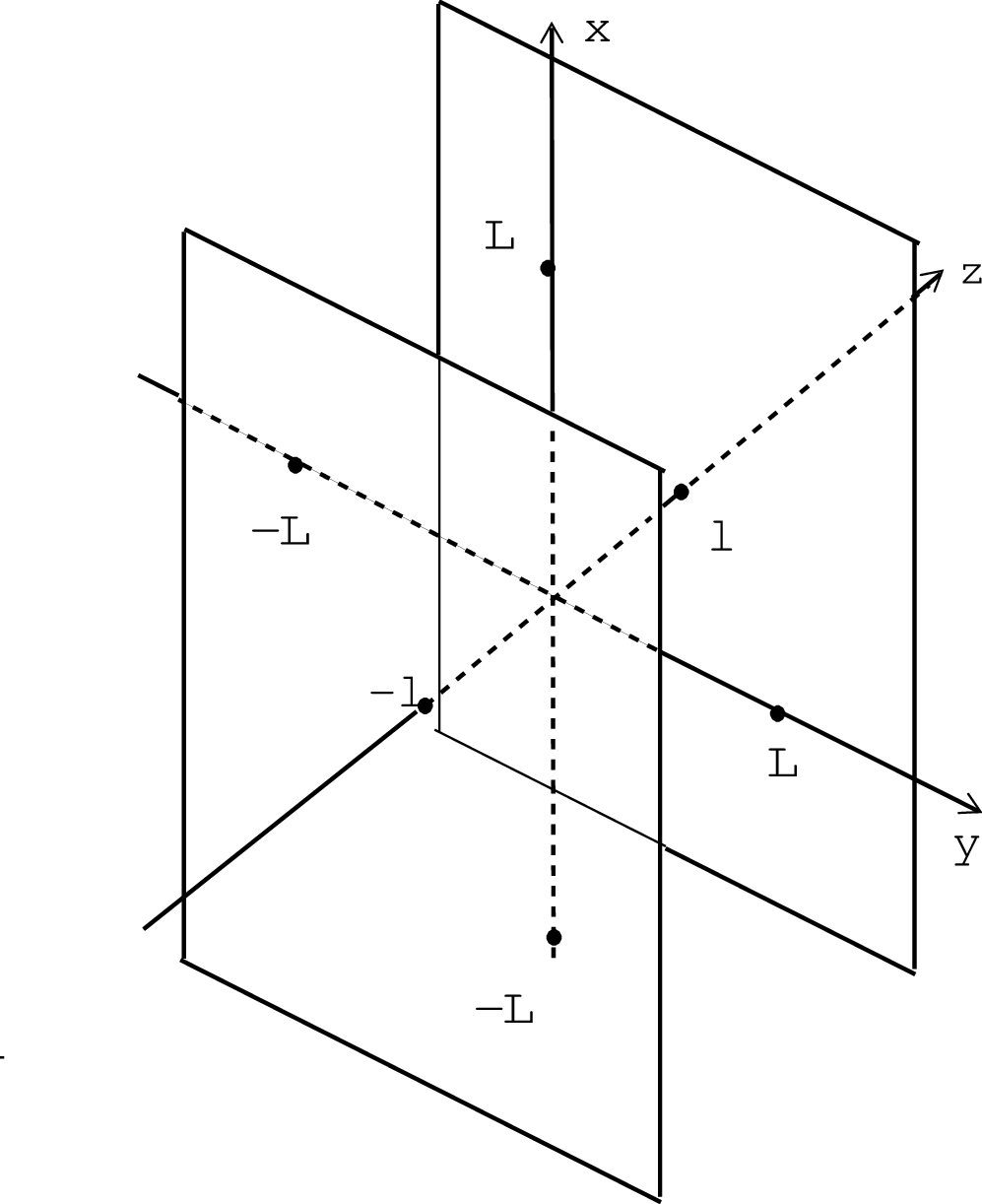}
\end{center}
\label{2Plates}
\end{figure}
In the dawn of the quantum theory, the {\it divergence} problem of
the specific heat of the radiation cavity was the biggest one 
(the problem of the blackbody radiation). It is historically so famous 
that the difficulty was solved by Planck's idea that the energy is quantized. 
In other words, the phase space of the photon field dynamics is not continuous but 
has the "cell" or "lattice" structure with the unit area ($\Del x\cdot\Del p$) of the size $2\pi\hbar$ (Planck constant). 
The radiation energy is composed of two parts, $E_{Cas}$ and $E_\be$:
\bea
E_{4dEM}=E_{Cas}+E_\be\com\nn
E_{Cas}=\sum_{m_x,m_y,n\in\bfZ}\omtil_{m_xm_yn}\com\q
E_{\be}=2\sum_{m_x,m_y,n\in\bfZ}\frac{\omtil_{m_xm_yn}}{\e^{\be\omtil_{m_xm_yn}}-1}\com\nn
{\omtil_{m_xm_yn}}^2=(m_x\frac{\pi}{L})^2+(m_y\frac{\pi}{L})^2+(n\frac{\pi}{l})^2
\com\q\q l\ll L\com
\label{4dEM18X}
\eea 
where the parameter $\be$ is the inverse temperature, $l$ is the separation length 
between two perfectly-conducting plates, and $L$ is the IR regularization parameter 
of the plate-size. See Fig.\ref{2Plates}. 
The second part $E_\be$ 
is, essentially, Planck's radiation formula. The first one $E_{Cas}$ is 
the vaccuum energy of the radiation field, that is, the Casimir energy. 
It is a very delicate quantity. The quantity is formally {\it divergent}, hence it 
must be defined with careful {\it regularization}. (See App.B). $E_{Cas}/(2L)^2$ does 
depend only on the {\it boudary} parameter $l$. The quantity is a quantum effect and 
, at the same time, depends on the {\it global} (macro) parameter $l$. 
\footnote{
See the recent reviews on Casimir energy: Ref.\cite{Cas01t06}.
}
\bea
\frac{E_{Cas}}{(2L)^2}=\frac{\pi^2}{(2l)^3}\frac{B_4}{4!}=-\frac{\pi^2}{720}\frac{1}{(2l)^3}\com\q
B_4\mbox{(the fourth Bernoulli number)}=-\frac{1}{30}
\com
\label{4dEM24x}
\eea 

\begin{figure}
\caption{
Graph of Planck's radiation formula.  
$ \Pcal (\be,k)=\frac{1}{(c\hbar)^3}\frac{1}{\pi^2}k^3/(\e^{\be k}-1)\ \ 
(1\leq\be\leq 2,\ 0.01\leq k\leq 10),\ \  
k\mbox{(photon energy)}=\hbar \om=h\n=\hbar c\frac{2\pi}{\la}, \ \be=1/k_BT$ where $c\hbar=2000~\mbox{eV~\AA}, \ k_B\mbox{(Boltzmann's constant)}=8.6\times 
10^{-5}~\mbox{eV/K}$. $k$=(1,10)~eV correspond to $\frac{\la}{2\pi}$=(2000,200)\AA, 
$\be$=(1,2)~eV$^{-1}$ correspond to T=($10^5/8.6,~10^5/17.2$) K. 
$\Pcal=0.1$ corresponds to $\Pcal\Del k=(2000)^{-3}\times 0.1\times \Del k$[eV/\AA $^3$]. 
        }
\begin{center}
\includegraphics[height=8cm]{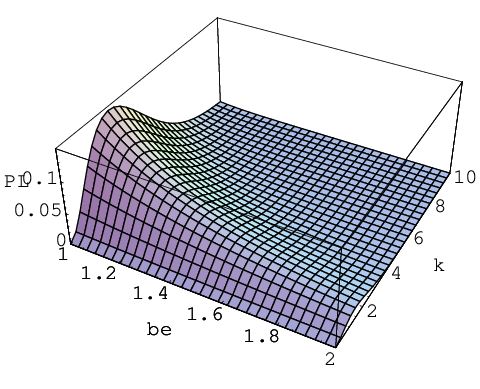}
\end{center}
\label{PlanckDistB}
\end{figure}

In Fig.\ref{PlanckDistB}, Planck's radiation spectrum distribution is shown. 
Introducing the axis of the inverse temperature($\be$), besides the photon energy or 
frequency ($k$), 
it is shown stereographically. Although we will examine the 5D version of 
the zero-point part (the Casimir energy), the calculated quantities in this paper 
are much more related to this Planck's formula. 
\footnote{
Planck's formula depends only on the temperature $1/\be$, not on $l$. 
(Comparatively the Casimir energy part does not depend on $\be$, but on the separation $l$. )
It is known that $\be$ can be regarded as the periodicity for the axis of 
the inverse temperature (Euclidean time). The axis corresponds to the {\it extra axis} in the 
following text.  
} 
We see, near the $\be$-axis, a sharply-rising surface, which is the Rayleigh-Jeans 
region (the energy density is proportional to the {\it square} of the photon frequency). 
The {\it damping} region in high $k$ is the Wien's region.
\footnote{
We recall that the old problem of the {\it divergent} specific heat was 
solved by the Wien's formula. This fact strongly supports 
the present idea of introducing the {\it weight function} 
(see Sec.\ref{uncert}). 
} 
The ridge (the line of peaks at each $\be$) 
forms the {\it hyperbolic} curve (Wien's displacement law). When we will, in this paper, 
deal with the 
energy distribution over the 4D momentum and the extra-coordinate, we will see the similar 
behavior (although top and bottom appear in the opposite way). In order to compare this 4D case with 
the 5D case of the present paper, we do some preparation for the Casimir energy in App.B. 

In the quest for the unified theory, the higher dimensional (HD) approach 
is a fascinating one from the geometrical point. Historically the initial successful 
one is the Kaluza-Klein model\cite{Kal21,Klein26}, which unifies the photon, graviton and dilaton
from the 5D space-time approach. 
The HD theories
\footnote{
The HD theories we consider here, are the HD generalization of the familiar 
renormalizable (in 4D) theories such as 5D free scalar theory, 5D QED and 5D Yang-Mills theory.
}
, however, generally have the serious defect as the quantum field
theory(QFT) : un-renormalizability. 
\footnote{
Note that the ordinary power counting criterion is about the divergence degree for 
the coupling expansion. In this sense, there are some 
renormalizable HD field theories such as 6D $\Phi^3$-theory. 
In the present paper, however, we have focused on the coupling-independent 
part, the Casimir part. This part is generally divergent in the higher dimensions. 
%
%
}
The HD quantum field theories, at present, 
are not defined within the QFT. One can take the standpoint 
that the more fundamental formulation, such as the string theory and D-brane 
theory, can solve the problem. In the present paper, we have the {\it new standpoint} 
that the HD theories should be defined by themselves within the QFT.  In order to 
escape the dimension requirement D=10 or 26 
from the quantum consistency (anomaly cancellation)\cite{StringText}, 
we treat the gravitational (metric) field only as the background one. 
This does {\it not} mean the space-time
is not quantized. See later discussions (Sec.\ref{weight}). We present a way to define 5D quantum field theory 
through the analysis of the Casimir energy of 5D electromagnetism. 

In 1983, the Casimir energy in the Kaluza-Klein theory was calculated 
by Appelquist and Chodos\cite{AC83}. They took the cut-off ($\La$) regularization and 
found the quintic ($\La^5$) divergence and the finite term. The divergent 
term shows the {\it unrenormalizability} of the 5D theory, but the finite term looks 
meaningful
\footnote{
The gauge independence was confirmed in Ref.\cite{SI85PLB}. 
}
 and, in fact, is widely regarded as the right vacuum energy 
which shows {\it contraction} of the extra axis. In this decade, triggered by 
the development of the string and D-brane theories, new treatments or new ideas 
were introduced to calculate the vacuum energy or the effective potential in HD. 
(The motivation is to settle the stability problem of the moduli 
parameters\cite{GoldWise9907,GoldWise9911}. )
One is to regard the system as the bulk and boundary, and do renormalization
in both parts\cite{Toms00,GoldWise01}. Various regularization methods were carefully re-examined 
for the bulk-boundary theory\cite{GoldRoth00}. They are 
applied to various theories including realistic models\cite{FlachiToms01,GPT01,PonPop01,FGPT03,FlaPujo03}. 
From the regularization viewpoint, the zeta-function (or dimensional) regularization 
combined with some summation formula is most commonly taken. The renormalization 
procedure, however, does {\it not} seem satisfactory. They succeed in 
calculating the properly regularized quantity and in separating  
the divergent terms. They found the finite part, but its physical meaning is 
obscure because the treatment of the divergent part is {\it not established}. 
They simply say, based on the analogy to the case of the ordinary (4D) renormalizable thoeries, 
the {\it local counterterms} can cancel divergences.
\footnote{
See the first footnote of Sec.8 about the present treatment of the local counterterms.
}
 They try to absorb 
divergences by the renormalization of 
parameters such as the brane tension (cosmological constant) and 
the gravitational constant. But 
it is fair to say that the $\La^5$-divergence problem, 
posed by Appelquist and Chodos, is {\it not} yet solved.  
All these come from the unsatisfactory situation 
of the quantum treatment of the brane dynamics and the HD quantum field theories. 

In the development of the string and D-brane theories, a new approach 
to the renormalization group was found. It is called {\it holographic renormalization} 
\cite{SusWit9805,HenSken9806,SkenTown9909,DFGK9909,FGPW9904,dBVV9912}. 
We regard the renormalization flow as a curve 
in the bulk (HD space). The flow goes along the extra axis. 
The curve is derived as a dynamical equation 
such as Hamilton-Jacobi equation. 
It originated from the AdS/CFT correspondence\cite{Malda9711,GKP9802,Witten9802}. 
Spiritually the present basic idea overlaps with this approach. 
The characteristic points of this paper are:\ 
a) We do {\it not} rely on the 5D supergravity;\ 
b) We do {\it not} quantize the gravitational(metric) field;\ 
c) The divergence problem is solved by reducing the degree of freedom of the system, 
where we require, not higher symmetries, but some restriction based on the 
{\it minimal area principle};\ 
d) No local counterterms are necessary. 

\begin{figure}
\caption{
IR-regularized geometry of 5D flat space (\ref{5dEM1}). The 4D world 
(3-brane) is Euclideanized and is shown as 4D ball (shaded disk region) surrounded
by $S^3$ sphere with radius $\mu^{-1}$. $\mu$ is the 4D IR regularization parameter, 
and is taken to be $\mu=1/l$. UV-regularization is introduced, in Sec.\ref{UIreg}, 
by replacing the 4D ball
with the "sphere lattice" composed of many small (size: $1/\La$) 4D balls. 
See Sec.\ref{surf} for detail.
        }
\begin{center}
\includegraphics[height=8cm]{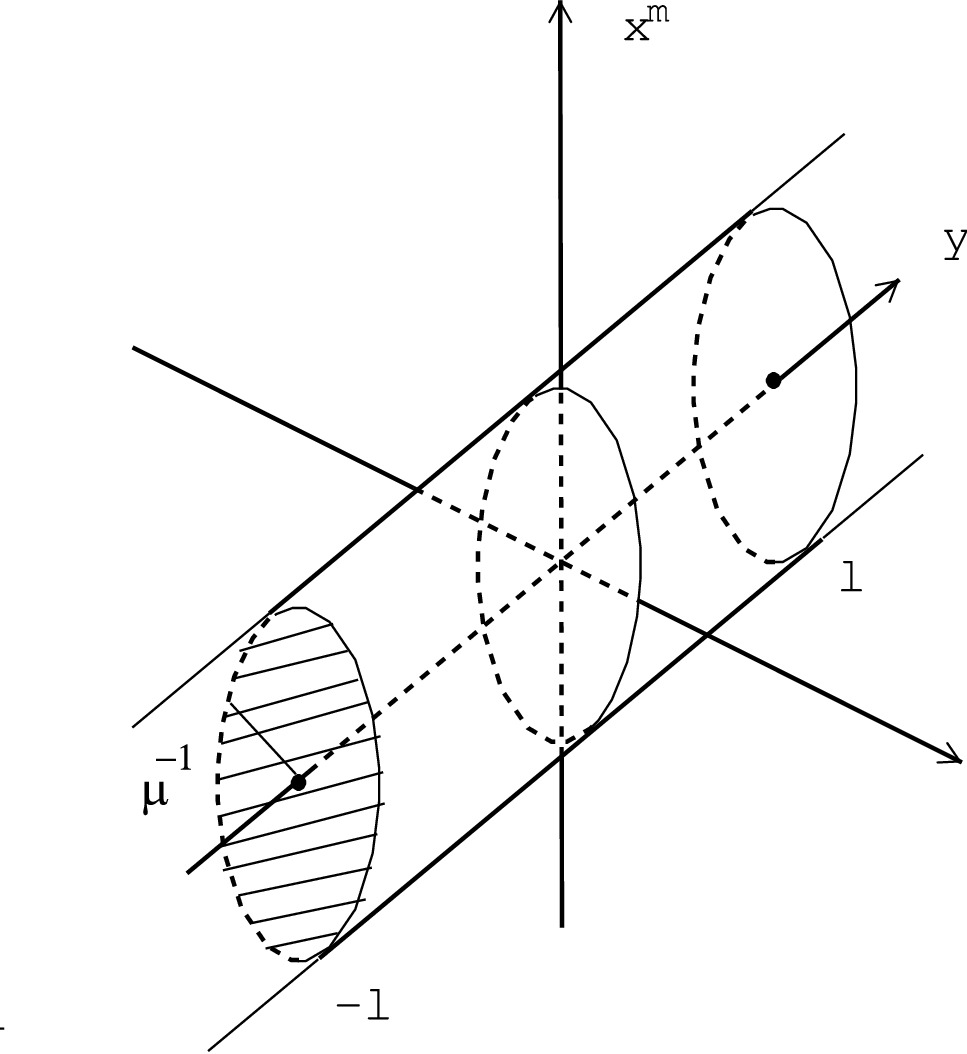}
\end{center}
\label{RegGeomFlat}
\end{figure}

In the previous paper\cite{SI0801}, we investigated the 5D electromagnetism 
in the {\it flat} geometry. For the later use of comparison (with the present {\it warped} case), 
we list the main results here.
\footnote{
We do not require the reader to read ref.\cite{SI0801}. The necessary key procedures are
explained in the text. 
}
 The extra space is {\it periodic} (periodicity $2l$) and
$Z_2$-parity is taken into account: 
\bea
ds^2=\eta_\mn dx^\m dx^\n+dy^2\com\q
-\infty < x^\m, y < \infty\com\q
y\ra y+2l,\ y\change -y\com\nn
(\eta_\mn)=\mbox{diag}(-1,1,1,1)\ ,
(X^M)=(X^\m=x^\m,X^5=y)\equiv (x,y)\ ,\nn 
M,N=0,1,2,3,5;\ \m,\n=0,1,2,3.
\label{5dEM1}
\eea
The IR-regularized geometry of this 5D flat space(-time) is depicted 
in Fig.\ref{RegGeomFlat}.  
The Casimir energy is {\it rigorously} (all KK-modes are taken into account) expressed as
\bea
E_{Cas}(\La,l)=\frac{2\pi^2}{(2\pi)^4}\int_{1/l}^{\La}d\ptil\int_{1/\La}^ldy~\ptil^3 
W(\ptil,y)F(\ptil,y)\com\nn
F(\ptil,y)\equiv F^-(\ptil,y)+4F^+(\ptil,y)
=\int_\ptil^\La d\ktil\frac{-3\cosh\ktil(2y-l)-5\cosh\ktil l}{2\sinh(\ktil l)}
\pr
\label{UIreg2}
\eea
where $\La$ is the 4D-momentum {\it cutoff}, and $W(\ptil,y)$ is the {\it weight function} 
to suppress the IR and UV divergences.
\footnote{
Z$_2$-odd part $F^-$ comes from the quantum fluctuation of the extra component $A_5(X)$, 
while Z$_2$-even part $F^+$ from the 4D components $A_\m(X)$. 
} 
\footnote{
The expression (\ref{UIreg2}) of $E_{Cas}$ is negative definite. The same thing can be 
said about the warped case (\ref{HKA8}) in the later description. 
} 
We obtained the following 
$\La$ and $l$ dependence by the numerical analysis. 

1) Un-weighted case: $W=1$
\bea
\mbox{Un-restricted integral region}:\q\qqq\qqq\nn
E_{Cas}(\La,l)=\frac{1}{8\pi^2}\left[ -0.1249 l\La^5
-(1.41, 0.706, 0.353)\times 10^{-5}~l\La^5\ln (l\La)\right]
\com\nn
\mbox{Randall-Schwartz integral region}:\q
E^{RS}_{Cas}=\frac{1}{8\pi^2}[-0.0894~\La^4]
\pr
\label{UIreg5X}
\eea
The {\it quintic} divergence of the upper one of (\ref{UIreg5X}) shows the 
{\it unrenormalizability} of the 5D theory in the ordinary treatment. 
The triplet data show the unstable situation of numerical results.
\footnote{
The results of (\ref{UIreg5X}) are based on the numerical integral of (\ref{UIreg2}) for 
$l=(10,20,40), \La=10\sim 10^3$. The triplet coefficients correspond to the three values of $l$. 
This unstable situation does {\it not} appear in the present case of warped geometry. 
See (\ref{UIreg5}) and (\ref{AppD3}). The same thing can be said about the weighted case 2) in the 
following. 
}
As for the lower case, the $(\ptil,y)$-integral region is restricted 
to below the {\it hyperbolic curve} $\ptil y =1$.\footnote{
This restriction was taken in Ref.\cite{RS01} to suppress the UV-divergence. 
} 

2) Weighted case

\bea
E^W_{Cas}/\La l=\mbox{\hspace{10cm}}\nn
\nn
\left\{
\begin{array}{cc}
-2.50\frac{1}{l^4}+(-0.142,1.09,1.13)\cdot 10^{-4}\frac{\ln l\La}{l^4} & 
 \mbox                                          {for}\q W=\frac{1}{N_1}\e^{-(1/2) l^2\ptil^2-(1/2) y^2/l^2}\equiv W_1\\
-6.04~10^{-2}\frac{1}{l^4}-(24.7,2.79,1.60)\cdot 10^{-8}\frac{\ln l\La}{l^4} & 
                                                 \mbox{for}\q W=\frac{1}{N_2}\e^{-\ptil y}\equiv W_2\\
-2.51\frac{1}{l^4}+(19.5,11.6,6.68)\cdot 10^{-4}\frac{\ln l\La}{l^4}  & \mbox{for}\q W=\frac{1}{N_8}\e^{-(l^2/2) (\ptil^2+1/y^2)}
\equiv W_8\end{array}
           \right.
\label{uncert1b}
\eea
where some representative cases ($W_1$:\ elliptic, $N_1=1.557/8\pi^2$ ; $W_2$:\ hyperbolic, $N_2=2(l\La)^3/8\pi^2$; 
$W_8$:\ reciprocal, $N_8=0.3800/8\pi^2$) are shown. 
(See ref.\cite{SI0801} for other cases. ) The quantity $\La l$ is the normalization factor 
in the numerical analysis. 
The Casimir energy behavior of the case $W_2$ is consistent with the Randall-Schwartz's one of 1).  
These results imply the {\it renormalization of the compactification size} $l$. 
\bea
E^W_{Cas}/\La l =-\frac{\al}{l^4}\left( 1-4c\ln (l\La) \right) =-\frac{\al}{{l'}^4}\com\q
\be=\frac{\pl}{\pl (\ln \La)}\ln\frac{l'}{l}=c\com
\label{uncert1cc}
\eea
where $\al$ and $c$ should be uniquely fixed by clarifying the meaning of the weight function $W$
and the unstable situation of the triplet data.

\begin{figure}
\caption{
IR-regularized geometry of 5D warped space (\ref{KKexp2}). The 4D world 
(3-brane) is Euclideanized and is shown as a 4D ball (shaded disk region) surrounded
by $S^3$ sphere with radius $\mu^{-1}$. $\mu$ is the 4D IR regularization parameter. 
UV-regularization is introduced  by replacing the 4D ball
with the "sphere lattice" composed of many small (size: $1/\La$) 4D balls. 
See Sec.\ref{surf} for detail.
        }
\begin{center}
\includegraphics[height=8cm]{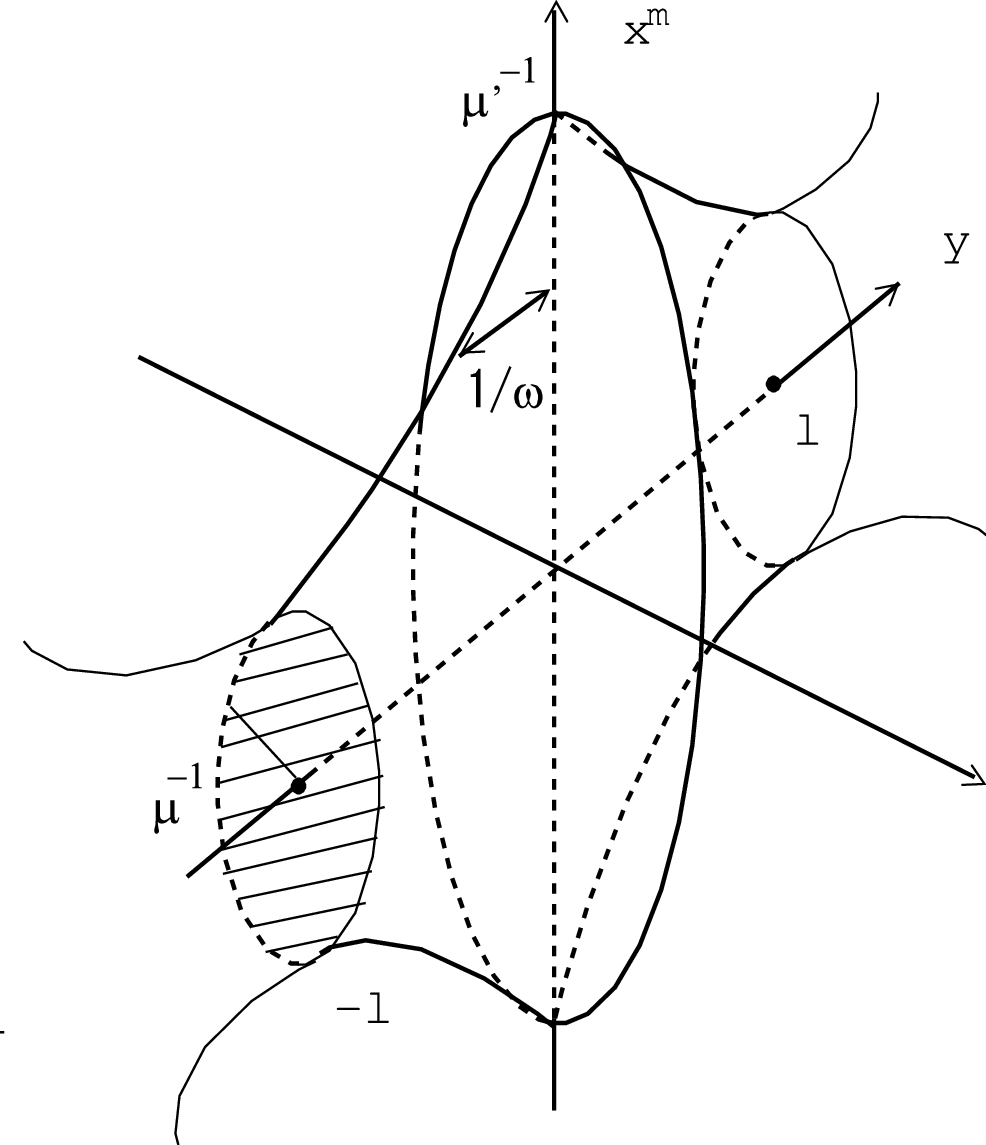}
\end{center}
\label{RegGeomWarp}
\end{figure}

The aim of this paper is to examine how the above results change 
for the 5D warped geometry case. 
The {\it IR-regularized} geometry of the 5D warped space(-time) is depicted 
in Fig.\ref{RegGeomWarp}. One additional massive parameter, that is, 
the warp (bulk curvature) parameter $\om$ appears. The limit $\om\ra 0$ 
leads to the flat case.  This introduction of the "thickness" $1/\om$ 
comes from the expectation that it softens the UV-singularity, which 
is the same situation as in the string theory. See Ref.\cite{SI0909} and \cite{SI0912} besides 
this work.

The content is organized as follows. We start with
the familiar approach to the 5D warped system: the Kaluza-Klein expansion, in 
Sec.\ref{KKexp}. 
In Sec.\ref{HKA}, the same content of Sec.\ref{KKexp} is dealt in the {\it heat-kernel} method and 
the Casimir energy is expressed in a {\it closed} form in terms of the P/M 
propagator. The closed expression of Casimir energy enables us {\it numerically} evaluate 
the quantity in Sec.\ref{UIreg}. Here we 
introduce UV and IR regularization parameters in (4D momentum, extra coordinate)-space. 
A new idea about the UV and IR regularization is presented in Sec.\ref{surf}. The {\it minimal area 
principle} is introduced. The {\it sphere lattice} and the {\it renormalization flow} are explained. In Sec.\ref{uncert}
an improved regularization procedure is presented where a {\it weight function} is 
introduced. Here again the {\it minimal surface principle} is taken. The meaning of the 
weight function is given in Sec.\ref{weight}. In Sec.\ref{conc} we make the concluding remarks. 
{\it Renormalization of the warp parameter $\om$} is explicitly shown. We argue the (4D) coordinates 
or momenta look quantized in the present treatment. The cosmological constant is addressed. 
We prepare 
five appendices to supplement the text. App.A  deals with the {\it classification} of all minimal surface curves in the 5D warped space. 
App.B reviews 4D Casimir energy (the ordinary radiation cavity problem) 
where the features of the cut-off and zeta-function regularizations
are examined. 
App.C explains the numerical confirmation of the (approximate) 
equality of the minimal surface curve and the dominant path in the Casimir energy calculation. 
The results, appearing in this paper, heavily relies on some {\it numerical calculations}. 
We explain them in App.D. 
Normalization constants of various weight functions are explained in App.E.

\section{Kaluza-Klein expansion approach\label{KKexp}}

In order to analyze the 5D EM-theory, we start with 5D {\it massive} vector
theory. 
\bea
S_{5dV}=\intxz\sqrt{-G}(-\fourth F_{MN}F^{MN}-\half m^2A^M A_M)\ ,\ 
F_{MN}=\pl_M A_N-\pl_NA_M\ ,\nn
ds^2=\frac{1}{\om^2z^2}(\eta_\mn dx^\m dx^\n+{dz}^2)=G_{MN}dX^M dX^N\com\q G\equiv \det G_{AB}\com\nn 
(X^M)=(x^\mu,z)\com\q M,N=0,1,2,3,5(\mbox{or }z);\ \mu,\nu=0,1,2,3\pr
\label{KKexp1}
\eea 
The 5D vector mass, $m$, is regarded as a IR-regularization parameter. 
In the limit, $m=0$, the above one has the 5D local-gauge symmetry. 
Casimir energy is given by some integral where the (modified) Bessel functions, 
with the index $\nu=\sqrt{1+\frac{m^2}{\om^2}}$, appear. 
(See, for example, ref.\cite{RS01}.) Hence the 5D EM 
limit is given by $\n=1\ (m=0)$. We consider, however, the {\it imaginary} 
mass case $m=i\om\ (m^2=-\om^2,\ \n=0)$ for the following reasons:\  
1)\ the UV-behavior does not depend on the bulk mass parameter m which is regarded as a IR regularization one;\  
2)\ we can compare the result with the 5D flat case where the 5D scalars
   (4 even-parity modes + 1 odd-parity mode) are considered\cite{SI0801}, 
3)\ $\n=0$ Bessel functions are meaningfully simple in the analysis.
\footnote{
$\n=1/2$ ($m^2=-3\om^2/4$) is another simple case where 
Bessel functions reduce to trigonometric functions. 
} 
We can simplify the model furthermore. 
Instead of analyzing the $m^2=-\om^2$ of the massive vector (\ref{KKexp1}), 
we take the 5D massive {\it scalar} theory
on AdS$_5$ with $m^2=-4\om^2,\ \n=\sqrt{4+m^2/\om^2}=0$. 
\bea
\Lcal=\sqrt{-G}(-\half \na^M\Phi\na_M\Phi-\half m^2\Phi^2)\com
\q G\equiv \det G_{MN}\com\nn
ds^2=G_{MN}dX^MdX^N\com\q\na^M\na_M \Phi-m^2\Phi+J=0\com
\label{KKexp1b}
\eea 
where $\Phi(X)=\Phi(x^\m,z)$ is the 5D scalar field. 
The background geometry is AdS$_5$ which takes the following form, 
in terms of $z$,  
\bea
(G_{MN})=
\left(
\begin{array}{ll}
\frac{1}{\om^2z^2}\eta_{\mn} & 0 \\
0               & \frac{1}{\om^2z^2}
\end{array}
\right)
\com\q
\sqrt{-G}=\frac{1}{(\om |z|)^5}\com\nn
-\frac{1}{T}\leq z \leq -\frac{1}{\om}\q\mbox{or}\q
\frac{1}{\om}\leq z \leq \frac{1}{T}\q 
(-l\leq y \leq l\ ,\ |z|=\frac{1}{\om}\e^{\om |y|})\com\nn
\frac{1}{T}\equiv \frac{1}{\om}\e^{\om l}
\com
\label{KKexp2}
\eea 
where we take into account $Z_2$ symmetry: $z\change -z$.  
$\om$ is the bulk curvature (AdS$_5$ parameter) and $T^{-1}$ is the
size of the extra space (Infrared parameter). 

In this section, we do the standard analysis of the warped system, that is, 
the Kaluza-Klein expansion approach. 

The Casimir energy $E_{Cas}$ is given by
\bea
\e^{-T^{-4}E_{Cas}}=\int\Dcal\Phi\exp\{i\int d^5X\Lcal\}\nn
=\int\Dcal\Phi(X)\exp\left[
i\intfx dz\frac{1}{(\om|z|)^5}\half\Phi\{
\om^2z^2\pl_a\pl^a\Phi+(\om |z|)^5\frac{\pl}{\pl z}\frac{1}{(\om z)^3}\pl_z\Phi
                           -m^2\Phi
                                       \} 
                     \right]
\pr
\label{KKexp3}
\eea 
Here we introduce, instead of $\Phi(X)$, the partially (4D world only)
Fourier transformed field $\Phi_p(z)$. 
\bea
\Phi(X)=\intp \e^{ipx}\Phi_p(z)\com
\label{KKexp4}
\eea 
Eq.(\ref{KKexp3}) can be rewritten as
\bea
\e^{-T^{-4}E_{Cas}}
=\int\Dcal\Phi_p(z)\times\nn
\exp\left[
i\intp 2\int_{1/\om}^{1/T}dz\left\{
\half\Phi_p(z)\{
-\frac{1}{(\om z)^3}p^2+\frac{d}{dz}\frac{1}{(\om z)^3}\frac{d}{dz}
                           -\frac{m^2}{(\om z)^5}
                \}\Phi_p(z)
                             \right\} 
                          \right]
\com
\label{KKexp5}
\eea 
where we have used the $Z_2$-property, defined in (\ref{KKexp7}), of $\Phi_p(z)$. 
From the above expression, we can read the measure function 
$s(z)$ and the extra-space kinetic operator $\Lhat_z$. 
\bea
s(z)=\frac{1}{(\om z)^3}\com\q
\Lhat_z\equiv\frac{d}{dz}\frac{1}{(\om z)^3}\frac{d}{dz}
                           -\frac{m^2}{(\om z)^5}
\com
\label{KKexp6}
\eea 
and consider the Bessel eigen-value problem.
\bea
\{s(z)^{-1}\Lhat_z+{M_n}^2\}\psi_n(z)=0\com\nn
\psi_n(z)=-\psi_n(-z)\q\mbox{for}\q P=-\q ;\q
\psi_n(z)=\psi_n(-z)\q\mbox{for}\q P=+ 
\com
\label{KKexp7}
\eea
with the appropriate b.c. at fixed points. Because the set $\{\psi_n(z)\}$ 
constitute the orthonormal and complete system, we can express $\Phi_p(z)$ as
\bea
\Phi_p(z)=\sum_{n}c_n(p)\psi_n(z)
\pr
\label{KKexp8}
\eea
Eq.(\ref{KKexp5}) can be further rewritten as 
\bea
\e^{-T^{-4}E_{Cas}}
=\int\Dcal\Phi_p(z)\exp\left[
i\intp 2\int_{1/\om}^{1/T}dz\left\{
\half\Phi_p(z)s(z)( {s(z)}^{-1}\Lhat_z-p^2)\Phi_p(z)
                             \right\} 
                          \right]\nn
=\int\prod_n dc_n(p)\exp\left[
\intpE\sum_n\{
-\half c_n(p)^2(p_E^2+M_n^2)
             \}
                         \right]\nn
=\exp\sum_{n,p}\{-\half\ln (p_E^2+M_n^2) \}
\com
\label{KKexp9}
\eea 
where the orthonormal relation
\bea
2\int_{\frac{1}{\om}}^{\frac{1}{T}}\psi_n(z)s(z)\psi_m(z)dz
=\del_{nm}
\com
\label{KKexp10}
\eea 
is used. 
This shows that $s(z)$, defined in (\ref{KKexp6}), plays the role of 
"inner product measure" in the function space 
$\{\psi_n(z), 1/\om \leq z\leq 1/T\}$. 
In (\ref{KKexp9}), Wick's rotation is done for the time-component of $\{ p^\m\}$. 
\bea
ip^0\q\ra\q p^4\com\nn
idp^0dp^1dp^2dp^3\q\ra\q dp^4dp^1dp^2dp^3\equiv d^4p_E\com\nn
p^2=-(p^0)^2+(p^1)^2+(p^2)^2+(p^3)^2\q\ra\q (p^4)^2+(p^1)^2+(p^2)^2+(p^3)^2\equiv p_E^2\com
\label{KKexp11}
\eea 
The expression (\ref{KKexp9}) is the familiar one 
of the Casimir energy.

\section{Heat-Kernel Approach and Position/Momentum Propagator\label{HKA}}

Eq.(\ref{KKexp9}) is the expression of $E_{Cas}$ by the KK-expansion.
In this section, the same quantity is re-expressed in a {\it closed} form
using the heat-kernel method and the P/M propagator. 
\bea
\e^{-T^{-4}E_{Cas}}
=\int\Dcal\Phi_p(z)\exp\left[
i\intp 2\int_{1/\om}^{1/T}dz\left\{
\half\Phi_p(z)s(z)( {s(z)}^{-1}\Lhat_z-p^2)\Phi_p(z)
                             \right\} 
                          \right]\nn
=\exp\left[
T^{-3}\intpE 2\int_{1/\om}^{1/T}dz s(z)\left\{
-\half\ln ( -{s(z)}^{-1}\Lhat_z+p_E^2)
                             \right\} 
      \right]                      \nn
=\exp\left[
T^{-3}\intpE 2\int_{1/\om}^{1/T}dz s(z)\left\{
\half\int_0^\infty\frac{1}{t}\e^{ t( {s(z)}^{-1}\Lhat_z-p_E^2)}dt +\mbox{const}
                            \right\} 
      \right]                      \nn
\com
\label{HKA1}
\eea 
where we have used a formula\cite{Schwinger51}. 
\bea
\int_0^\infty\frac{\e^{-t}-\e^{-tM}}{t}dt=\ln M\com\q \mbox{det}~M>0\com\q
M\ :\ \mbox{a matrix}
\pr
\label{HKA1b}
\eea
(The factors $T^{-4}$ and $T^{-3}$ in (\ref{HKA1}) come from the dimensional analysis. $T$ has 
the meaning of the renormalization point. )  
The above formal result can be {\it precisely} defined using the heat equation. 
\bea
\e^{-T^{-4} E_{Cas}}
=(\mbox{const})\times\exp \left[ 
T^{-4}\intpE 2\int_{0}^{\infty}\half\frac{dt}{t}\mbox{Tr}~H_{p_E}(z,z';t) 
                          \right] \com\nn
\mbox{Tr}~H_p(z,z';t)=\int_{1/\om}^{1/T}s(z)H_p(z,z;t)dz\com\q
\{\frac{\pl}{\pl t}-(s^{-1}\Lhat_z-p^2) \}H_p(z,z';t)=0
\pr
\label{HKA2}
\eea
The heat kernel $H_p(z,z';t)$ is formally solved, using the
Dirac's bra and ket vectors $(z|, |z)$, as
\bea
H_p(z,z';t)=(z|\e^{-(-s^{-1}\Lhat_z+p^2)t}|z')
\pr
\label{HKA3}
\eea
(The bra and ket vectors $(z|, |z)$ are {\it precisely} defined by the orthonormal and complete
set of $\Lhat_z$: $\{\psi_n(z)\}$.
\footnote{
\bea
\mbox{(1) Definition}\q
\psi_n(z)\equiv (n|z)=(z|n)\com\nn
\mbox{(2) }Z_2-\mbox{property}\q
|-z)=P |z)\com\q (-z|=P (z|\com\q P=\mp 1\nn
\mbox{(3) Orthogonality}\nn
\left(\int_{-\frac{1}{T}}^{-\frac{1}{\om}}+\int_{\frac{1}{\om}}^{\frac{1}{T}}\right)
\frac{dz}{(\om |z|)^3}(n|z)(z|k)=
2\int_{\frac{1}{\om}}^{\frac{1}{T}}\frac{dz}{(\om z)^3}(n|z)(z|k)=\del_{n,k},\q
(n|k)=\del_{n,k}\ ,\nn
(z|z')=\left\{
\begin{array}{ll}
(\om |z|)^3\ep(z)\ep(z')\delh (|z|-|z'|) & \mbox{for\ \ P=}-1 \\
(\om |z|)^3\delh (|z|-|z'|) & \mbox{for\ \ P=}1 
\end{array}
        \right.\nn
\mbox{(4) Completeness}\nn
\left(\int_{-\frac{1}{T}}^{-\frac{1}{\om}}+\int_{\frac{1}{\om}}^{\frac{1}{T}}\right)
\frac{dz}{(\om |z|)^3}|z)(z|=
2\int_{\frac{1}{\om}}^{\frac{1}{T}}\frac{dz}{(\om z)^3}|z)(z|={\bf 1}\ ,\nn
\sum_n|n)(n|={\bf 1}
\com
\label{HKA3b}
\eea 
         }
 )
Using the set $\{\psi_n(z)\}$ defined in (\ref{KKexp7}), the explicit solution of (\ref{HKA2}) is
given by
\bea
H_p(z,z';t)=\sum_{n\in \bfZ}\e^{-(M_n^2+p^2)t}
\half\{ \psi_n(z)\psi_n(z')-\psi_n(z)\psi_n(-z') \}\com\q P=-\com\nn
E_p(z,z';t)=\sum_{n\in \bfZ}\e^{-(M_n^2+p^2)t}
\half\{ \psi_n(z)\psi_n(z')+\psi_n(z)\psi_n(-z') \}\com\q P=+
\com
\label{HKA4}
\eea
where we have used the dimensionality of $H_p$ and $E_p$ read from (\ref{HKA2}). 
([$E_p$]=[$H_p$]=$L^{-1}$). 

The above heat-kernels satisfy the following b.c..
\bea
\lim_{t\ra +0}H_p(z,z';t)=\sum_{n\in \bfZ}
\half\{ \psi_n(z)\psi_n(z')-\psi_n(z)\psi_n(-z') \}  
\equiv (\om|z|)^3\ep(z)\ep(z')\delh(|z|-|z'|)\ ,\ P=-\ ,\nn
\lim_{t\ra +0}E_p(z,z';t)=\sum_{n\in \bfZ}
\half\{ \psi_n(z)\psi_n(z')+\psi_n(z)\psi_n(-z') \}  
\equiv (\om|z|)^3\delh(|z|-|z'|)\ ,\ P=+\
\com
\label{HKA5}
\eea
where $\ep(z)$ is the sign function. The above equation defines 
$\delh(|z|-|z'|)$. 
We here introduce the position/momentum propagators $G^{\mp}_p$ as
follows.
\bea
G^-_p(z,z')\equiv
\int_0^\infty dt~ H_p(z,z';t)=
\sum_{n\in \bfZ}\frac{1}{M_n^2+p^2}
\half\{ \psi_n(z)\psi_n(z')- \psi_n(z)\psi_n(-z') \}\com\nn
G^+_p(z,z')\equiv
\int_0^\infty dt~ E_p(z,z';t)=
\sum_{n\in \bfZ}\frac{1}{M_n^2+p^2}
\half\{ \psi_n(z)\psi_n(z')+ \psi_n(z)\psi_n(-z') \}
\pr
\label{HKA6}
\eea
They satisfy the following differential equations of {\it propagators}.
\bea
(\Lhat_z-p^2s(z))G^{\mp}_p(z,z')=
\sum_{n\in \bfZ}
\half\{ \psi_n(z)\psi_n(z')\mp \psi_n(z)\psi_n(-z') \}\nn
=\left\{
\begin{array}{ll}
\ep(z)\ep(z')\delh (|z|-|z'|) & \mbox{for\ \ P=}-1 \\
\delh (|z|-|z'|) & \mbox{for\ \ P=}1 
\end{array}
        \right.
\label{HKA7}
\eea

Therefore the Casimir energy $E_{Cas}$ is, from (\ref{HKA2}) and (\ref{HKA4}), given by
\bea
-E^{-}_{Cas}(\om,T)=
\intpE 2\intt 2\int_{1/\om}^{1/T} dz~s(z)H_{p_E}(z,z;t)\nn
=\intpE 2\intt 2\int_{1/\om}^{1/T} dz~s(z)\left\{
\sum_{n\in \bfZ}\e^{-(M_n^2+p_E^2)t}\psi_n(z)^2
                                       \right\}
\com
\label{HKA8}
\eea
where $s(z)=1/(\om z)^3$ (\ref{KKexp6}). 
This expression leads to the same treatment as the previous section. 
Note that the above expression shows the {\it negative definiteness} of 
$E^{-}_{Cas}$. 
\footnote{
We notice the subtraction of positive infinity (M-independent term) 
in the formula (\ref{HKA1b}) is essential for this negative definiteness. 
This should be compared with the expansion-expression $E_{Cas}$ of (\ref{KKexp9}). 
}

Here we introduce the {\it generalized} P/M propagators, $I_\al$(P=$-$) and $J_\al$(P=+) as
\bea
I_\al (p^2;z,z')\equiv
\int_0^\infty\frac{dt}{t^\al} H_p(z,z';t)                     \nn
=\int_0^\infty\frac{dt}{t^\al}\sum_{n\in \bfZ}
\e^{-(M_n^2+p^2)t}\half\{\psi_n(z)\psi_n(z')-\psi_n(z)\psi_n(-z')\}\com\q \mbox{P=}-   \com\nn
J_\al (p^2;z,z')\equiv
\int_0^\infty\frac{dt}{t^\al} E_p(z,z';t)                     \nn
=\int_0^\infty\frac{dt}{t^\al}\sum_{n\in \bfZ}
\e^{-(M_n^2+p^2)t}\half\{\psi_n(z)\psi_n(z')+\psi_n(z)\psi_n(-z')\}\com\q \mbox{P=}+ 
\com
\label{HKA9}
\eea
where $\al$ is the arbitrary real number. 
Then we have the following relations. 
\bea
I_0(p^2;z,z')=G_p^-(z,z')\com\q J_0(p^2;z,z')=G_p^+(z,z')\com\nn
\frac{\pl I_\al(p^2;z,z')}{\pl p^2}=-I_{\al-1}(p^2;z,z')\com\q
\int_{p^2}^\infty dk^2 I_\al (k^2;z,z')=I_{\al+1}(p^2;z,z')\com\nn
\frac{\pl J_\al(p^2;z,z')}{\pl p^2}=-J_{\al-1}(p^2;z,z')\com\q
\int_{p^2}^\infty dk^2 J_\al (k^2;z,z')=J_{\al+1}(p^2;z,z')\com\nn
(p^2-r(z)^{-1}\Lhat_z)I_\be(p^2;z,z')=-\be I_{\be+1}(p^2;z,z')\ ,\ 
(p^2-r(z)^{-1}\Lhat_z)J_\be(p^2;z,z')=-\be J_{\be+1}(p^2;z,z')\ ,\nn
 \be\neq 0\com\nn
\frac{\pl I_\al(p^2;z,z')}{\pl \al}=-\int_0^\infty\frac{\ln t}{t^\al}dt H_p(z,z';t)
\pr
\label{HKA10}
\eea
Finally we obtain the following useful expression of the Casimir energy for $P=\mp$.
\bea
-E^-_{Cas}(\om,T)
=\intpE\{
\mbox{Tr}~I_1(p_E^2;z,z')
       \} 
=\intpE\int_{p_E^2}^\infty\{\Tr I_0(k^2;z,z')\}dk^2\nn
=\intpE\int_{p_E^2}^\infty\{\Tr G_k^-(z,z') \}dk^2
=\intpE\int_{1/\om}^{1/T}dz~ s(z)\int_{p_E^2}^\infty\{G_k^-(z,z) \}dk^2\com\nn
-E^+_{Cas}(\om,T)
=\intpE\{
\mbox{Tr}~J_1(p_E^2;z,z')
       \} 
=\intpE\int_{p_E^2}^\infty\{\Tr J_0(k^2;z,z')\}dk^2\nn
=\intpE\int_{p_E^2}^\infty\{\Tr G_k^+(z,z') \}dk^2
=\intpE\int_{1/\om}^{1/T}dz~ s(z)\int_{p_E^2}^\infty\{G_k^+(z,z) \}dk^2
\pr
\label{HKA11}
\eea

Here we list the dimensions of various quantities appeared above.
\[
  \begin{array}{c|c|c|c|c|c|c|c|c|c}
  L^{1-2\al}& L^{-4} & L^{-3/2} & L^{-1} & L^{-1/2} & L^0& L & L^2 & L^{5/2} & L^3\\
\hline
            & E_{Cas}&          & \La,p,\om,T  &          && z,l& t   &&        \\
I_\al,J_\al &        &          & H_p,E_p&          && G^\mp_p&  &&        \\
           &         & \Phi &         &         & s(z) &    &    & \Phi_p & \\
            &        &          &\delh(z-z')&|z),(z|,\psi_n(z)&  |n),(n|    &  &   & & c_n(p)
   \end{array}
\]
($\La$ is a regularization parameter defined below.)

\begin{figure}
\caption{
Behavior of $\ln |\half\Fcal^-(\ktil,z)|=\ln |\ktil~ G^-_k(z,z)/(\om z)^3|$. 
$\om=10^4, T=1, \La=2\times 10^4$. $1.0001/\om \leq z \leq 0.9999/T$. $\La T/\om \leq \ktil \leq \La$. 
Note $\ln |(1/2)\times (1/2)|\approx -1.39$.  
}
\includegraphics[height=8cm]{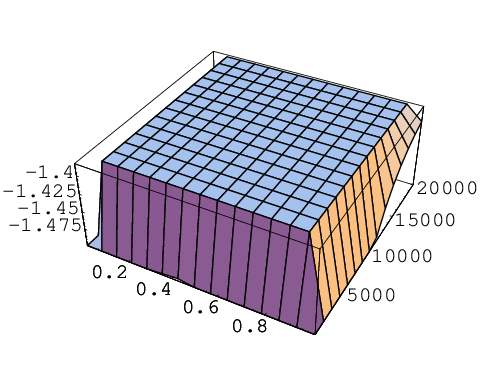}
\label{FcalmHT1k10p4}
\end{figure}

\begin{figure}
\caption{
Behavior of $\ln |\half\Fcal^+(\ktil,z)|=\ln |\ktil~ G^+_k(z,z)/(\om z)^3|$. 
$\om=10^4, T=1, \La=2\times 10^4$. $1.0001/\om \leq z \leq 0.9999/T$. $\La T/\om \leq \ktil \leq \La$ 
}
\includegraphics[height=8cm]{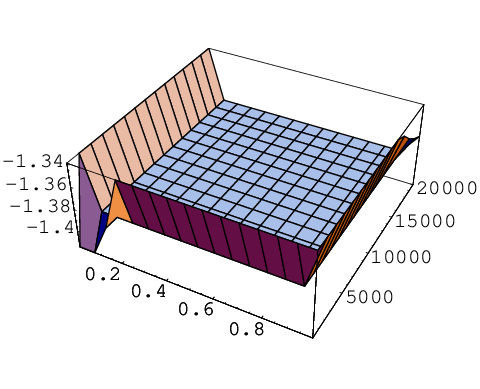}
\label{FcalpHT1k10p4}
\end{figure}

The P/M propagators $G_p^\mp$ in (\ref{HKA6}), (\ref{HKA10}) and (\ref{HKA11}) can be expressed in a {\it closed} form.
(See, for example, \cite{IM0703}.) 
Taking the {\it Dirichlet} condition at all fixed points, the expression
for the fundamental region ($1/\om \leq z\leq z'\leq 1/T$) is given by
\bea
G_p^\mp(z,z')=\mp\frac{\om^3}{2}z^2{z'}^2
\frac{\{\I_0(\Pla)\K_0(\ptil z)\mp\K_0(\Pla)\I_0(\ptil z)\}  
      \{\I_0(\Tev)\K_0(\ptil z')\mp\K_0(\Tev)\I_0(\ptil z')\}
     }{\I_0(\Tev)\K_0(\Pla)-\K_0(\Tev)\I_0(\Pla)}\com\nn
\ptil\equiv\sqrt{p^2}\com\q p^2\geq 0\ (\mbox{space-like})
\pr
\label{HKA12}
\eea 
We can express the $\La$-regularized Casimir energy in terms of the following functions $F^\mp(\ptil,z)$. 
\bea
-E^{\La,\mp}_{Cas}(\om,T)
=\left.\intpE\right|_{\ptil\leq\La}\int_{1/\om}^{1/T}dz~F^\mp(\ptil,z) \com\nn
F^\mp(\ptil,z)\equiv s(z)\int_{p_E^2}^{\La^2}\{G_k^\mp(z,z) \}dk^2\nn
=\frac{2}{(\om z)^3}\int_\ptil^\La\ktil~ G^\mp_k(z,z)d\ktil
\equiv \int_\ptil^\La\Fcal^\mp(\ktil,z)d\ktil
\com
\label{HKA13}
\eea
where $\Fcal^\mp(\ktil,z)$ are the integrands of $F^\mp(\ptil,z)$ and $\ptil=\sqrt{p_E^2}$. 
Here we introduce the UV cut-off parameter $\La$ for the 4D momentum space. 
In Fig.\ref{FcalmHT1k10p4}  and Fig.\ref{FcalpHT1k10p4}, we show the behavior of $\Fcal^\mp(\ktil,z)$. 
The table-shape graphs say the "Rayley-Jeans" dominance.
\footnote{
The energy density in $(p^a,z)$-space is approximately given by, using (\ref{HKA14}), 
$F^\mp(\ptil,z)\approx -\half\ptil+\half\La$. 
For small $\ptil$, $F^\mp(\ptil,z)\approx \half\La$(const.). 
This should be compared with $E_{\be}$ of (\ref{4dEM18}): $
\omtil/(\e^{\be\omtil}-1)\sim 1/\be~\mbox{(const)}$ for small $\omtil$.   
} 
That is, for the wide-range region $(\ptil,z)$ satisfying both 
$\ptil (z-\frac{1}{\om})\gg 1$ and $\ptil (\frac{1}{T}-z)\gg 1$, 
\bea
\Fcal^-(\ptil,z)\approx \half\com\q
\Fcal^+(\ptil,z)\approx \half
\com\nn
(\ptil,z)\in \{(\ptil,z)| \ptil (z-\frac{1}{\om})\gg 1\ \mbox{and}\ \ptil (\frac{1}{T}-z)\gg 1\}
\pr
\label{HKA14}
\eea

\section{UV and IR Regularization Parameters and Evaluation of Casimir Energy
\label{UIreg}}

\begin{figure}
\caption{
Space of (z,$\ptil$) for the integration. The hyperbolic curve 
will be used in Sec.\ref{surf}.  
}
\includegraphics[height=8cm]{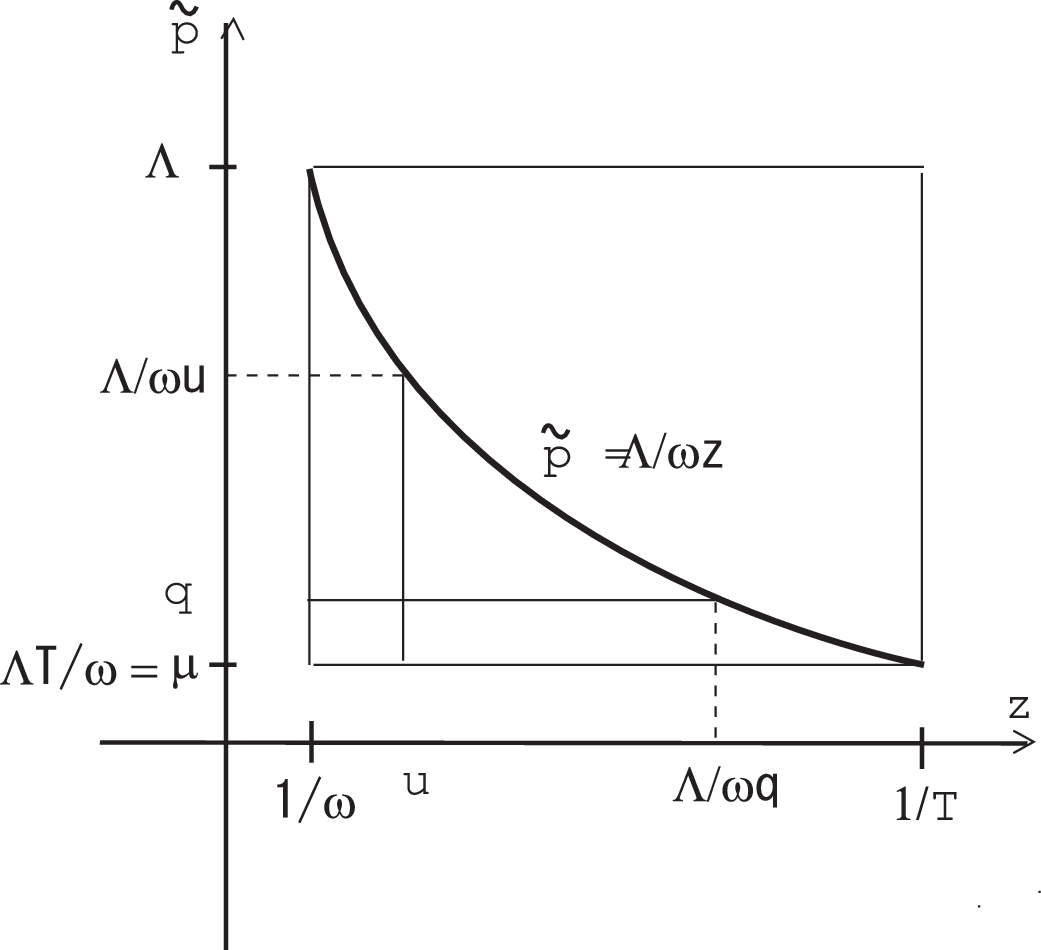}
\label{zpINTregionW}
\end{figure}

The integral region of the above equation (\ref{HKA13}) is displayed in Fig.\ref{zpINTregionW}. 
In the figure, we introduce the regularization cut-offs for the 4D-momentum integral, 
$\m\leq\ptil\leq\La$. 
As for the extra-coordinate integral, it is the finite interval, 
$1/\om\leq z\leq 1/T=\e^{\om l}/\om$, hence we need not introduce further
regularization parameters. 
For simplicity, we take
the following IR cutoff of 4D momentum.
\footnote{
If we take the following relation furthermore
\bea
\La=\om
\com
\label{HKA15B}
\eea
then $\m=T$and we need not any additional regularization parameters. 
We do {\it not} take this relation. The choice of the regularization parameters
affects the counting of the divergence degree. See later discussion of eq.(\ref{surfM1})
}
:
\bea
\m=\La\cdot\frac{T}{\om}=\La \e^{-\om l}
\pr
\label{HKA15}
\eea
Hence the new regularization parameter is $\La$ only. 

Let us evaluate the ($\La,T$)-{\it regularized} value of (\ref{HKA13}). 
\bea
-E_{Cas}^{\La,\mp}(\om,T)=\frac{2\pi^2}{(2\pi)^4}\int_{\m}^{\La}d\ptil\int_{1/\om}^{1/T}dz~\ptil^3 F^\mp (\ptil,z)\com\nn
F^\mp (\ptil,z)
=\frac{2}{(\om z)^3}\int_\ptil^\La\ktil~ G^\mp_k(z,z)d\ktil
\pr
\label{UIreg2b}
\eea
The integral region of ($\ptil,z$) is the {\it rectangle} shown in Fig.\ref{zpINTregionW}
. 
\begin{figure}
\caption{
Behaviour of $(-1/2)\ptil^3F^-(\ptil,z)$ (\ref{UIreg2b}). $T=1, \om=10^4, \La=10^4$.  
$1.0001/\om\leq z<0.9999/T$, $\La T/\om\leq\ptil\leq \La$ . 
}
\includegraphics[height=8cm]{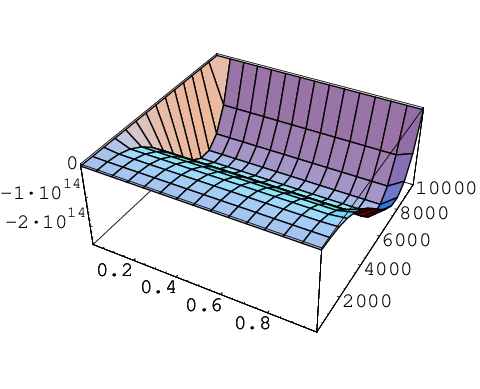}
\label{p3FmL10000}
\end{figure}
\begin{figure}
\caption{
Behavior of $(-1/2)\ptil^3F^-(\ptil,z)$ (\ref{UIreg2b}). $T=1, \om=10^4, \La=2\cdot10^4$.  
$1.0001/\om\leq z<0.9999/T$, $\La T/\om\leq\ptil\leq \La$ . 
}
\includegraphics[height=8cm]{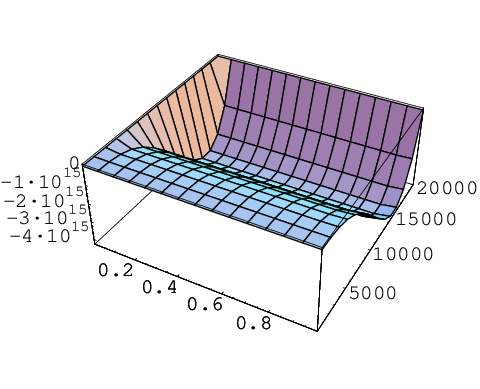}
\label{p3FmL20000}
\end{figure}
\begin{figure}
\caption{
Behavior of $(-1/2)\ptil^3F^-(\ptil,z)$ (\ref{UIreg2b}). $T=1, \om=10^4, \La=4\cdot10^4$.  
$1.0001/\om\leq z<0.9999/T$, $\La T/\om\leq\ptil\leq \La$ . 
}
\includegraphics[height=8cm]{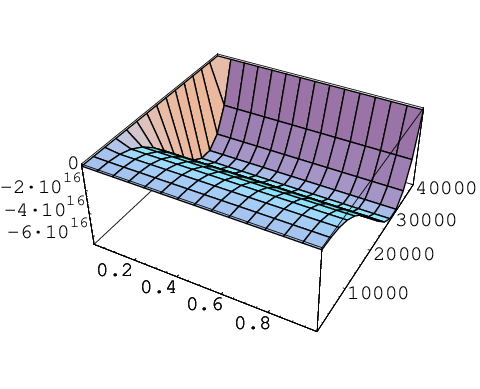}
\label{p3FmL40000}
\end{figure}

Note that eq.(\ref{UIreg2b}) is the {\it rigorous} expression of the $(\La,T)$-regularized Casimir energy. 
We show the behavior of $(-1/2)\ptil^3F^-(\ptil,z)$ taking 
the values $\om=10^4, T=1$ in Fig.\ref{p3FmL10000}($\La=10^4$), Fig.\ref{p3FmL20000}($\La=2\cdot 10^4$) and 
Fig.\ref{p3FmL40000}($\La=4\cdot 10^4$). 
\footnote{
The requirement for the three parameters $\om, T, \La$ is $\La\gg\om\gg T$. 
See ref.\cite{SI01CQG} for the discussion about the hierarchy $\La, \om, T$.  
In the application to the real world, the most interesting choice is $T\sim 1 \mbox{TeV}=10^3 \mbox{GeV}$(TeV physics), 
$\om\sim 10^{15} \mbox{GeV}$(GUT scale), and $\La\sim 10^{19} \mbox{GeV}$(Planck mass,$M_{pl}$). 
In the numerical calculation, however, we must be content with the 
appropriate numbers, shown in the text, due to the purely technical reason. 
Another interesting choice is 
$\om\sim 10^{-3}\mbox{eV}$(neutrino mass, $m_\n\sim \sqrt{M_{pl}/R_{cos}}$,$R_{cos}$: cosmological size), 
$\La\sim 10^{19}\mbox{GeV}=10^{28}\mbox{eV}$
(Planck mass) and 
$T\sim 10^{-20}\mbox{eV}$($\sim R_{cos}^{~-1}(M_{pl}R_{cos})^{1/5}$), 
See the discussion about the cosmological term 
in the concluding section. 
}
All three graphs have a common shape. 
(We confirm the graphs do not depend on the choice of $\om$ and $T$ very much.)
Behavior
along $\ptil$-axis does not so much depend on $z$. A valley runs parallel to the $z$-axis 
with the bottom line 
at the fixed ratio of $\ptil/\La \sim 0.75$. 
\footnote
{
The Valley-bottom line 'path' $\ptil=\ptil(y)\approx 0.75\La$ corresponds to the 
solution of the minimal principle:\  
$\del S_1=0,\ S_1[\ptil(z),z]\equiv (1/8\pi^2)\int dz \ptil(z)^3F(\ptil(z),z), F\approx -(\La-\ptil)/2\ \mbox{(\ref{UIreg6})}$. 
This will be referred in Sec.\ref{uncert} . 
}
The depth of the valley is proportional to $\La^4$. 
Because $E_{Cas}$ is the ($\ptil,z$) 'flat-plane' integral of $\ptil^3F(\ptil,z)$ , the {\it volume}
inside the valley is the quantity $E_{Cas}$ . Hence it is easy to see $E_{Cas}$ is proportional
to $\La^5$. This is the same situation as the flat case (the upper eq. of (\ref{UIreg5X})). 
Importantly, (\ref{UIreg2b}) shows the {\it scaling} behavior for large values of $\La$ and $1/T$. 
From a {\it close} numerical analysis of ($\ptil,z$)-integral (\ref{UIreg2b})
\footnote{
The result (\ref{UIreg5}) is based on the numerical calculation for the following cases:\ 
1) $T=1, \om=10^4, \La=10^4\times (1,2,4,8,16)$;\ 
2) $T=1, \om=10^3\times (1,2,4,8,16), \La=2\times 10^4$;\ 
3) $T=(1,2,4,8,16), \om=10^4, \La=2\times 10^4$.
}
, 
we have confirmed
\bea
E^{\La,-}_{Cas}(\om,T)=\frac{2\pi^2}{(2\pi)^4}\times\left[ -0.0250 \frac{\La^5}{T} \right]
\com
\label{UIreg5}
\eea
which does {\it not} depend on $\om$ and has no $\ln \frac{\La}{T}$-term. 
(Note: $0.025=1/40$. See App.D for the numerical derivation.) 
\footnote{
This numerical result can be checked using the approximate form (\ref{UIreg6}).  
$\int_{\m}^{\La}d\ptil\int_{1/\om}^{1/T}dz~\ptil^3\cdot (-\half)(\La-\ptil)=-0.025\frac{\La^5}{T}(1+O(T/\om)).$ 
}
Compared with the flat case (the upper eq. of (\ref{UIreg5X})), we see the factor $T^{-1}$ 
plays the role of {\it IR parameter} of the
extra space. We note that the behavior of Fig.\ref{p3FmL10000}-\ref{p3FmL40000} 
is similar to the Rayleigh-Jeans's region (small momentum region) of the Planck's radiation 
formula (Fig.\ref{PlanckDistB}) in the sense that 
$\ptil^3F(\ptil,z)\propto \ptil^3$ for $\ptil\ll\La$. 

Finally we notice, from the Fig.\ref{p3FmL10000}-\ref{p3FmL40000}, the approximate form 
of $F(\ptil,z)$ for the large $\La$ and $1/T$ is given by
\bea
F^\mp (\ptil,z)\approx \frac{f}{2} \La (1-\frac{\ptil}{\La})\com\q f=1
\com
\label{UIreg6}
\eea
which does {\it not} depend on $z, \om$ and $T$. $f$ is the degree of freedom. 
The above result is consistent with (\ref{HKA14}).

\section{UV and IR Regularization Surfaces, Principle of 
Minimal Area and Renormalization Flow\label{surf}}

The advantage of the
new approach is that the KK-expansion is replaced by the integral 
of the extra dimensional coordinate $z$ and 
all expressions are written in the {\it closed} ( not expanded )
form. The $\La^5$-divergence, (\ref{UIreg5}), shows the notorious problem
of the higher dimensional theories, as in the flat case (the upper eq. of (\ref{UIreg5X})). 
In spite of all efforts of the past literature, 
we have not succeeded 
in defining the higher-dimensional theories. 
(The divergence causes problems. The famous example is 
the divergent {\it cosmological constant} in the gravity-involving theories.
\cite{AC83} )
Here we notice that the divergence problem can be solved if we find a way to
{\it legitimately restrict the integral region in ($\ptil,z$)-space}. 

One proposal of this was presented by Randall and Schwartz\cite{RS01}. They introduced
the {\it position-dependent cut-off},\ $\mu <\ptil <\La /\om u\ ,\ u\in [1/\om,1/T]$\ , 
for the 4D-momentum integral in the "brane" located at $z=u$. See Fig.\ref{zpINTregionW}.
The total integral region is the lower part of the {\it hyperbolic} curve $\ptil=\La /\om z$. 
They succeeded in obtaining the {\it finite} $\be$-function of the 5D warped vector
model. 
We have confirmed that the value $E_{Cas}$ of (\ref{UIreg2b}), when the Randall-Schwartz 
integral region (Fig.\ref{zpINTregionW}) is taken, is proportional to $\La^5$. 
The close numerical analysis says
\bea
E^{-RS}_{Cas}(\om,T)=
\frac{2\pi^2}{(2\pi)^4}\int_{\m}^{\La}dq\int_{1/\om}^{\La /\om q}dz~q^3 F^- (q,z)
=\frac{2\pi^2}{(2\pi)^4}\int_{1/\om}^{1/T}du\int_{\m}^{\La /\om u}d\ptil~\ptil^3 F^- (\ptil,u)\nn
=\frac{2\pi^2}{(2\pi)^4}\frac{\La^5}{\om}\left\{
-1.58\times 10^{-2}-1.69\times 10^{-4}\ln~\frac{\La}{\om}
                                          \right\}
\com
\label{surfM1}
\eea
which is {\it independent} of $T$ 
.
\footnote{
The approximate form (\ref{UIreg6}) predicts the similar result. 
$\int_{\m}^{\La}dq\int_{1/\om}^{\La/\om q}dz~q^3\cdot (-\half)(\La-q)=-\frac{1}{60}\frac{\La^5}{\om}(1+O((T/\om)^3)).
\ 0.01666\cdots =1/60 .$  
}
\footnote{
The result (\ref{surfM1}) is based on the numerical-integral data for 
$T=(1,2,4,8,16), \om=10^3, \La=2\times 10^4$;  
$\La=10^4\times (1,2,4,8,16), T=1, \om=10^3$;  
$\om=10^2\times(1,2,4,8,16), T=1, \La=2\times 10^4$. 
See App.D for the numerical derivation.  
}
This shows the divergence 
situation does {\it not} improve compared with the non-restricted case of (\ref{UIreg5}). 
{\it $T$ of (\ref{UIreg5}) is replaced by the warp parameter $\om$.} 
This is contrasting with the flat case where $E^{RS}_{Cas}\propto -\La^4$. (the lower eq. of (\ref{UIreg5X})) 
The UV-behavior, however, {\it does improve} if we can choose the parameter $\La$ in the way: $\La\propto\om$.
This fact shows the parameter $\om$ "smoothes" the UV-singularity to some extent. 
\footnote{
In ref.\cite{RS01}, they take $\La=0.5\om, \om, 2\om, \cdots$ and evaluate $\be$-function 
(of the gauge coupling constant) for the
different cases. They regard the parameter $\om$ as the physical cutoff. This choice, however, is {\it not} 
allowed in the present standpoint $\La\gg \om\gg T$. 
The fact that $\om$ appears as (\ref{surfM1}) imply the warp parameter can 
control the UV-behavior to some extent. It matches 
the belief that the theoretical parameter $\om$ physically means the {\it extendedness} of the system configuration 
and smoothes the UV-singularity. 
}

Although they claim the holography is behind the procedure, 
the legitimateness of the restriction looks less obvious. We have proposed 
an alternate approach 
and given a legitimate explanation within the 5D QFT\cite{IM0703,SI07Nara,SI0801,SI0803OCU}. 
Here we closely examine the {\it new regularization}. 
\begin{figure}
\caption{
Space of ($\ptil$,z) for the integration (present proposal). 
}
\begin{center}
\includegraphics[height=8cm]{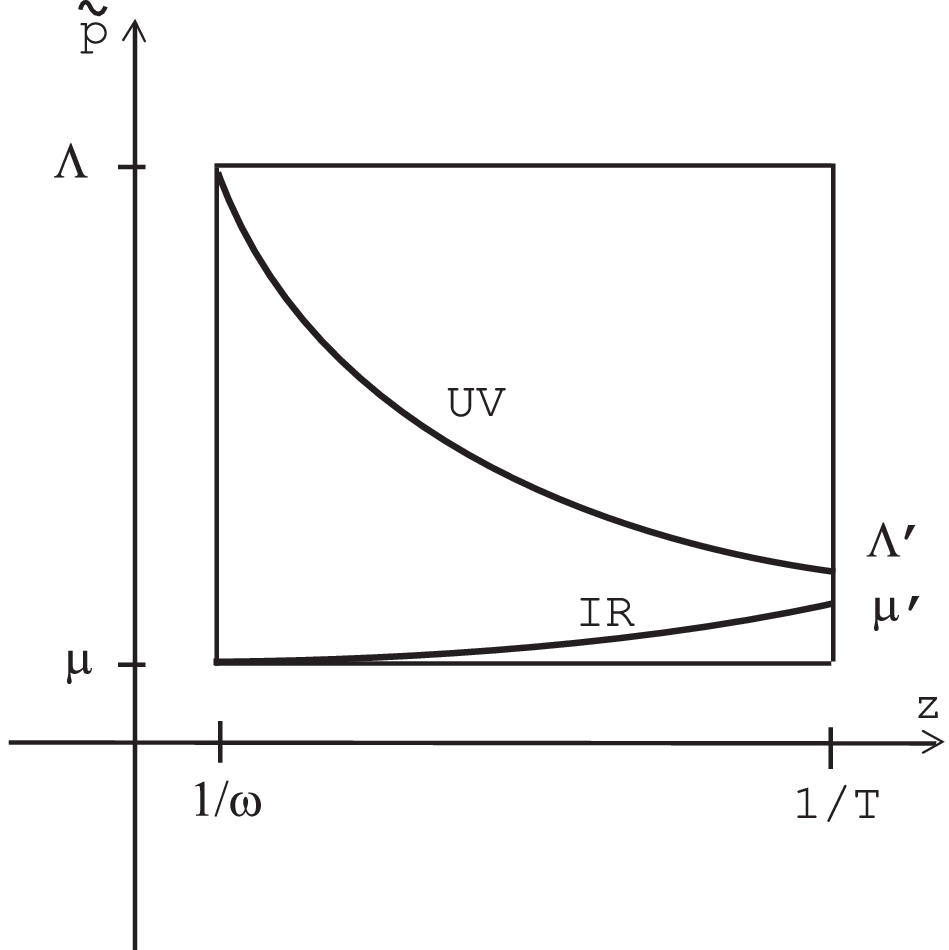}
\end{center}
\label{zpINTregionW2}
\end{figure}

\begin{figure}
\caption{
Regularization Surface $B_{IR}$ and $B_{UV}$ in the 5D coordinate space $(x^\m,z)$. 
The three graphs at the bottom show the  
flow of coarse graining (renormalization) and the sphere lattice regularization 
which will be explained after some paragraphs. 
}
\begin{center}
\includegraphics[height=8cm]{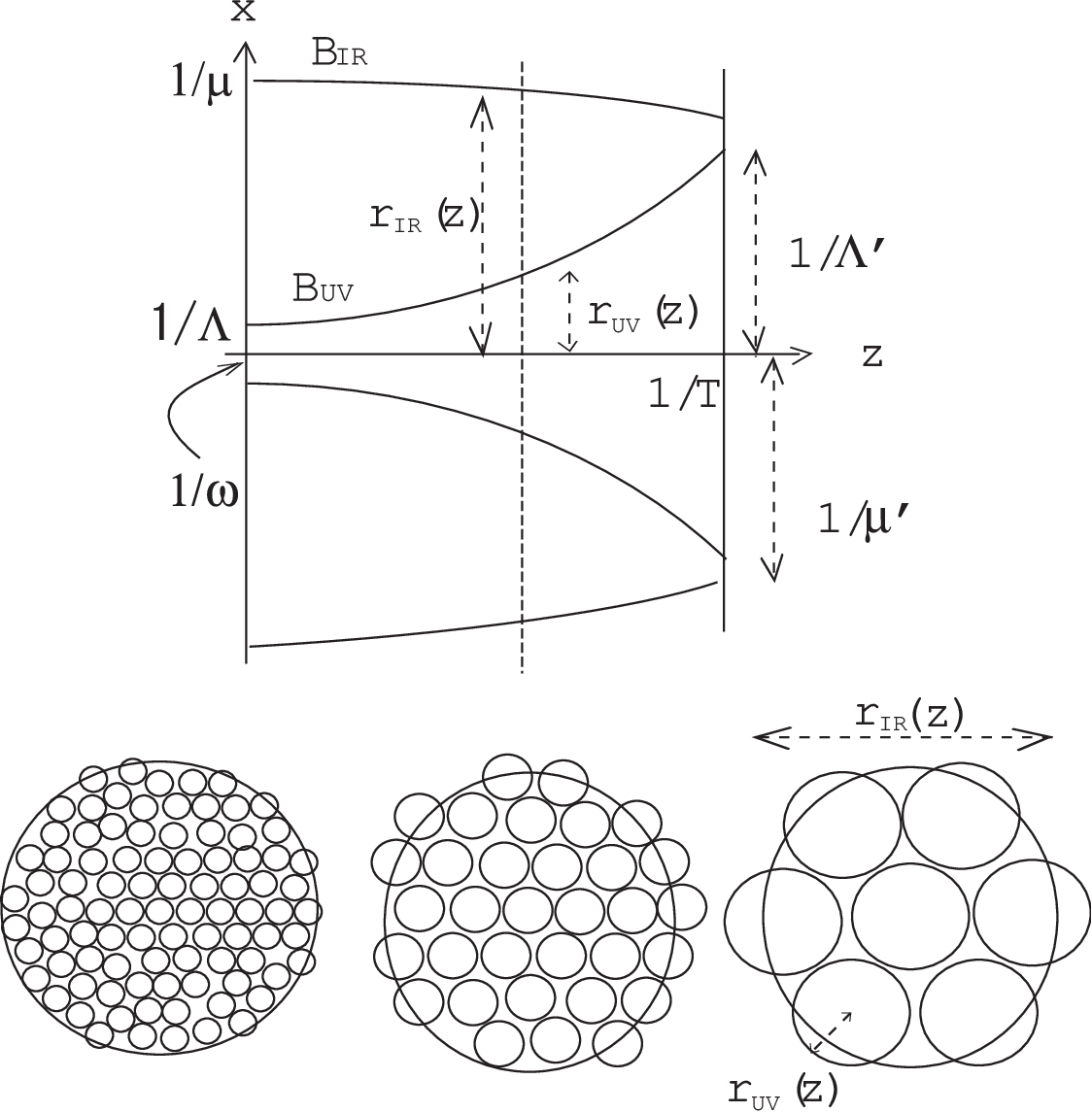}
\end{center}
\label{IRUVRegSurfW}
\end{figure}

\begin{figure}
\caption{
UV regularization surface ($B_{UV}$) in 5D coordinate space. 
}
\includegraphics[height=8cm]{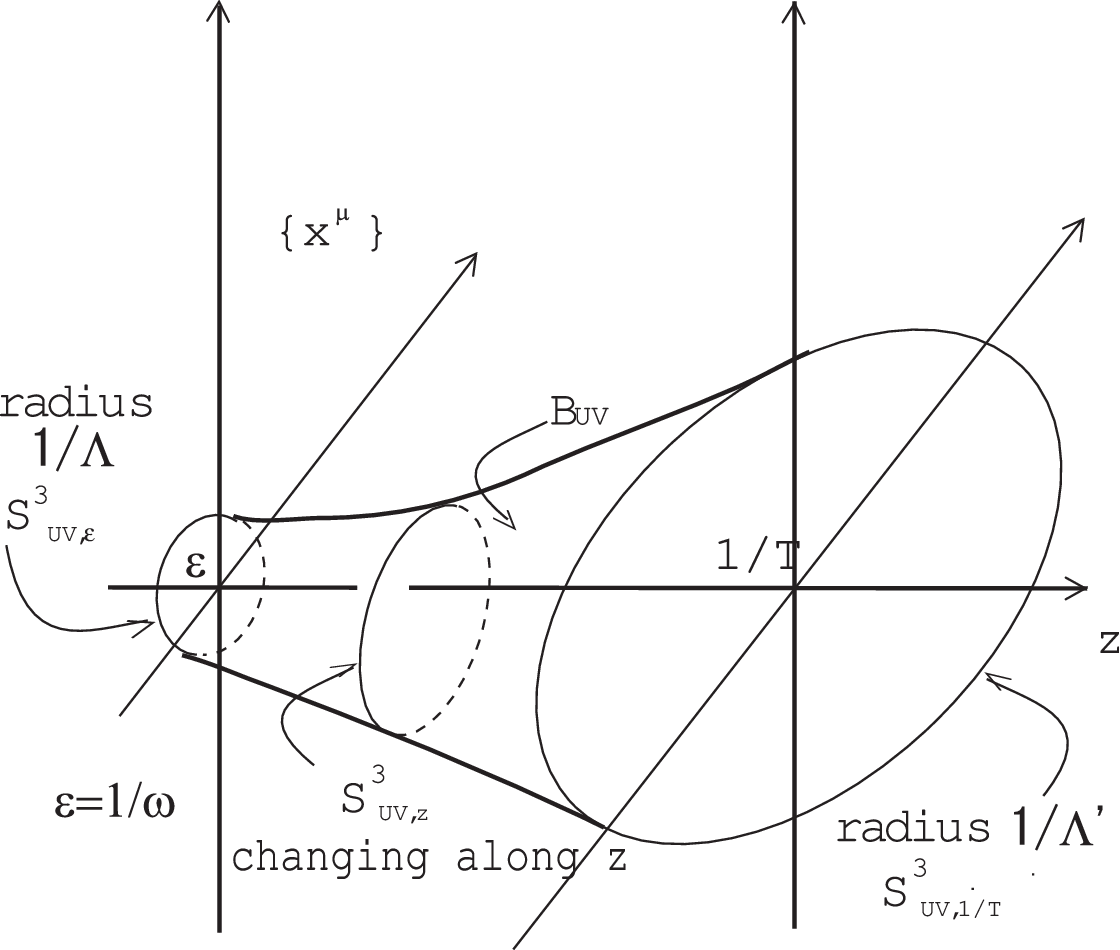}
\label{UVsurfaceW}
\end{figure}

On the "3-brane" at $z=1/\om$, we introduce the IR-cutoff $\mu=\La\cdot\frac{T}{\om}$ and 
the UV-cutoff $\La$\ ($\mu\ll\La$). See Fig.\ref{zpINTregionW2}.  
\bea
\mu\q \ll\q \La\q\q (T\q \ll\q \om)
\pr
\label{surf0a}
\eea
This is legitimate in the sense that we
generally do this procedure in the 4D {\it renormalizable} thoeries. 
(Here we are considering those 5D theories that are {\it renormalizable} in "3-branes". Examples are
5D free theories (present model), 
5D electromagnetism\cite{SI0801}, 5D $\Phi^4$-theory, 5D Yang-Mills theory, e.t.c..)
In the same reason, on the
"3-brane" at $z=1/T$, we may have another set of IR and UV-cutoffs, 
$\mu'$ and $\La'$. 
We consider the case\footnote{
Another interesting case is 
$\mu\leq\La,\ \La\ll \La',\ \mu\sim\mu'$. 
This case gives us the opposite direction flow. 
}: 
\bea
\mu'\leq\La',\ \La'\ll \La,\ \mu\sim\mu'
\pr
\label{surf0b}
\eea
This case will lead us to
introduce the {\it renormalization flow}. (See the later discussion.)
We claim here,  
as for the regularization treatment of the "3-brane" located at other points $z$ ($1/\om<z\leq 1/T$), the regularization 
parameters are determined by the {\it minimal area principle}. 
\footnote{
We do {\it not} quantize the (bulk) geometry, but treat it as the {\it background}. 
The (bulk) geometry fixes the behavior of the {\it regularization} 
parameters in the field quantization. The geometry influences the "boundary" 
of the field-quantization procedure. 
}
To explain it, we move to the 5D coordinate space ($x^\m,z$). See Fig.\ref{IRUVRegSurfW}. 
%
The $\ptil$-expression can be replaced by $\sqrt{x_\m x^\m}$-expression by the 
{\it reciprocal relation}.
\bea
\sqrt{x_\mu(z)x^\mu(z)}\equiv r(z)\q \change \q \frac{1}{\ptil(z)}
\pr
\label{surf0c}
\eea
The UV and IR cutoffs change their values along $z$-axis and their trajectories make
{\it surfaces} in the 5D bulk space $(x^\mu,z)$. 
We {\it require} the two surfaces do {\it not cross} for the purpose 
of the renormalization group interpretation (discussed later).  
We call them UV and IR regularization (or boundary) surfaces($B_{UV},B_{IR}$).
\bea
\mbox{B}_{UV}\q:\q\sqrt{(x^1)^2+(x^2)^2+(x^3)^2+(x^4)^2}=r_{UV}(z)\com
\q \frac{1}{\om}<z<1/T\com\nn
\mbox{B}_{IR}\q:\q\sqrt{(x^1)^2+(x^2)^2+(x^3)^2+(x^4)^2}=r_{IR}(z)
\com\q \frac{1}{\om}<z<1/T\com
\label{surf1}
\eea
where $r_{UV}(z)$ and $r_{IR}(z)$ are some functions of $z$ which are fixed by the minimal area principle. 
The cross sections of the regularization surfaces at $z$ are the spheres $S^3$ with the 
radii $r_{UV}(z)$ and $r_{IR}(z)$. Here we consider the Euclidean space for simplicity.
The UV-surface is stereographically shown in Fig.\ref{UVsurfaceW} and reminds us of the {\it closed string} propagation. 
Note that the boundary surface B$_{UV}$ (and B$_{IR}$) is the 4 dimensional manifold. 

The 5D volume region bounded by $B_{UV}$ and $B_{IR}$ is the integral region 
 of the Casimir energy $E_{Cas}$. 
The forms of $r_{UV}(z)$ and $r_{IR}(z)$ can be
determined by the {\it minimal area principle}.
\bea
\del (\mbox{Surface Area})=0\com\nn
3+\frac{4}{z}r'r-\frac{r''r}{{r'}^2+1}=0
\com\q r'\equiv\frac{dr}{dz}
\com\q r''\equiv\frac{d^2r}{dz^2}
\com\q 1/\om\leq z\leq 1/T
\pr
\label{surf2}
\eea
In App.A, we present the {\it classification} of all solutions (paths). 
It helps to find appropriate minimal surface curves for the renormalization
flow.

\begin{figure}
\caption{
Numerical Solution by Runge-Kutta. (\ref{surf2}), 
Vertical axis: $r$; Horizontal 
axis:  $z$. 
$T=1, \om=10^4, 10^{-4}\leq z\leq 1.0$.  
Upper (B$_{IR}$):\ $r(1)=0.8, r'(1)=1.0, $;\ \  Lower (B$_{UV}$):\ $r(1)=10^{-4}, r'(1)=-1.0$\ . 
Both curves are Graph Type (ia). 
}
\includegraphics[height=8cm]{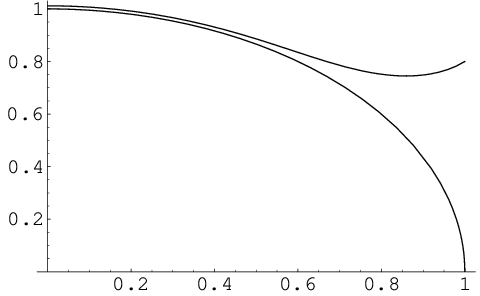}
\label{RegSurfRK}
\end{figure}

In Fig.\ref{RegSurfRK} 
we show two result curves of (\ref{surf2}) taking the following boundary
conditions ($r'\equiv dr/dz$) : 
\bea
\mbox{Fig.11: Coarse Conf. goes to Fine Conf. as $z$ increases}\nn
\begin{array}{cc}
\mbox{IR-curve (upper):} & r[1]=0.8, r'[1]=1.0\q \mbox{type\ (ia)}\\ 
\mbox{UV-curve (lower):} & r[1]=10^{-4}, r'[1]=-1.0\q \mbox{type\ (ia)} 
\end{array}
\label{surf2c}
\eea

They show the flow of renormalization 
\footnote{
The flow direction is opposite 
to the one shown in Fig.\ref{zpINTregionW2}.
          } 
really occurs by the minimal area principle. 
(See the next paragraph for the renormalization flow interpretation.) 
These results imply the {\it boundary conditions} 
determine the property of the renormalization flow. 
\footnote{ 
The minimal area equation (\ref{surf2}) is the 2nd derivative
differential equation. Hence, for given 
two initial conditions (,for example, $r(z=1/\om)$ and $dr/dz|_{z=1/\om}$), 
there exists a unique solution (path). The presented graphs are 
those with these initial conditions. 
Another way of choosing the initial conditions, $r(z=1/\om)$ and $r(z=1/T)$, 
is possible. 
Generally the solution of the second derivative differential equation 
is fixed by two {\it initial conditions}. 
}

The present regularization scheme gives the {\it renormalization group} interpretation
to the change of physical quantities along the extra axis. 
See Fig.\ref{IRUVRegSurfW}. 
\footnote{
This part is contrasting with AdS/CFT approach where 
the renormalization flow comes from the Einstein equation of 5D supergravity. 
} 
In the "3-brane" located at $z$, the UV-cutoff is $r_{UV}(z)$
and the regularization surface is the sphere $S^3$ with the radius $r_{UV}(z)$. 
The IR-cutoff is $r_{IR}(z)$
and the regularization surface is the another sphere $S^3$ with the radius $r_{IR}(z)$. 
We can regard the regularization integral region as the {\it sphere lattice} of
the following properties:
\bea
\mbox{A unit lattice (cell)\ :\ the sphere }S^3\mbox{\ with radius\ }r_{UV}(z) \mbox{and its inside}\com\nn
\mbox{Total lattice\ :\ the sphere }S^3\mbox{with radius\ }r_{IR}(z)\ \mbox{and its inside}\pr\nn
\mbox{It is made of many cells above}\com\nn
\mbox{Total number of cells\ :\ }\mbox{const.}\times\left(\frac{r_{IR}(z)}{r_{UV}(z)} \right)^4
\pr
\label{surf3}
\eea
The total number of cells changes from $(\frac{\La}{\mu})^4$ at $z=1/\om$ to 
$(\frac{\La'}{\mu'})^4$ at $z=1/T$. Along the $z$-axis, the number increases or decreases as
\bea
\left(\frac{r_{IR}(z)}{r_{UV}(z)} \right)^4\equiv N(z)
\pr
\label{surf4}
\eea
For the "scale" change $z\ra z+\Del z$, $N$ changes as 
\bea
\Del (\ln N)=4\frac{\pl}{\pl z}\{\ln (\frac{r_{IR}(z)}{r_{UV}(z)}) \}\cdot \Del z
\pr
\label{surf5}
\eea
When the system has some coupling $g(z)$, 
the renormalization group ${\tilde \be}(g)$-function (along the extra axis) is expressed as
\bea
{\tilde \be}=\frac{\Del(\ln g)}{\Del(\ln N)}=\frac{1}{\Del(\ln N)}\frac{\Del g}{g}
=\frac{1}{4}\frac{1}{\frac{\pl}{\pl z}\ln(\frac{r_{IR}(z)}{r_{UV}(z)})}\frac{1}{g}\frac{\pl g}{\pl z}
\com
\label{surf6}
\eea
where $g(z)$ is a renormalized coupling at $z$. 
\footnote{
Here we consider an interacting theory, such as 5D Yang-Mills theory and 
5D $\Phi^4$ theory, where the coupling $g(z)$ is the renormalized one 
in the '3-brane' at $z$.
}

We have explained, in this section, that the {\it minimal area principle} determines the 
flow of the regularization surfaces. 


\section{Weight Function and Casimir Energy Evaluation\label{uncert}}

In the expression (\ref{HKA11}), the Casimir energy is written by
the integral in the ($\ptil,z$)-space over the range: 
$1/\om\leq z\leq 1/T,\ 0\leq \ptil\leq\infty$. In Sec.\ref{surf}, 
we have seen {\it the integral region should be properly restricted} because the cut-off
region in the 4D world
 {\it changes along the extra-axis} obeying the bulk (warped) geometry 
({\it minimal area principle}). 
We can expect the singular behavior (UV divergences) reduces by the integral-region 
restriction, but the concrete evaluation along the proposed prescription is practically not easy. 
In this section, we consider an alternate approach which respects 
the {\it minimal area principle} and evaluate the Casimir energy. 

\begin{figure}
\caption{
Behavior of $(-N_1/2)\ptil^3W_1(\ptil,z)F^-(\ptil,z)$(elliptic suppression). 
$\La=20000,\ \om=5000,\ T=1$\ . 
$1.0001/\om\leq z\leq 0.9999/T ,\ \m=\La T/\om\leq \ptil\leq \La$. 
}
\begin{center}
\includegraphics[height=8cm]{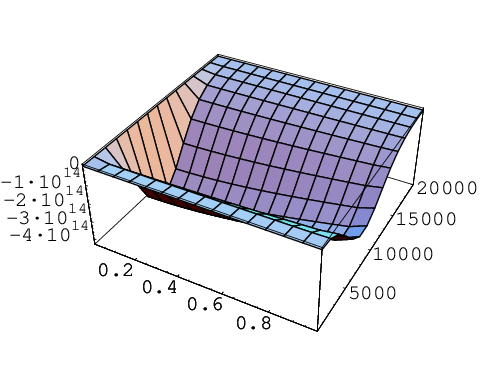}
\end{center}
\label{W1L2mank5senT1}
\end{figure}

\begin{figure}
\caption{
Behavior of $(-N_3/2)\ptil^3W_3(\ptil,z)F^-(\ptil,z)$(hyperbolic suppression2). 
$\La=20000,\ \om=5000,\ T=1$\ . 
$1.0001/\om\leq z\leq 0.9999/T ,\ \m=\La T/\om\leq \ptil\leq \La$. 
}
\begin{center}
\includegraphics[height=8cm]{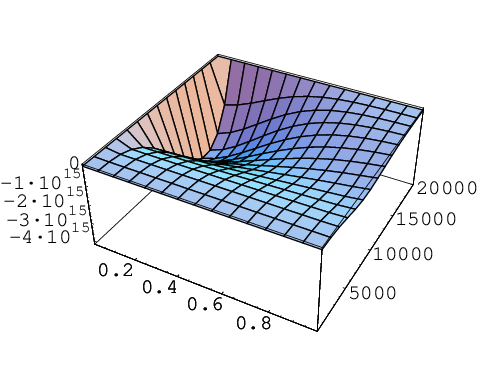}
\end{center}
\label{W3L2mank5senT1}
\end{figure}

\begin{figure}
\caption{
Behavior of $(-N_4/2)\ptil^3W_4(\ptil,z)F^-(\ptil,z)$(linear suppression). 
$\La=100,\ \om=10,\ T=1$\ . 
$1.0001/\om\leq z\leq 0.9999/T ,\ \m=\La T/\om\leq \ptil\leq 25$. 
In order to demonstrate the valley-bottom line is similar to a minimal surface line (See App.C), 
we here take rather small values of $\La$ and $\om$. The contour of this graph will be shown later. 
}
\begin{center}
\includegraphics[height=8cm]{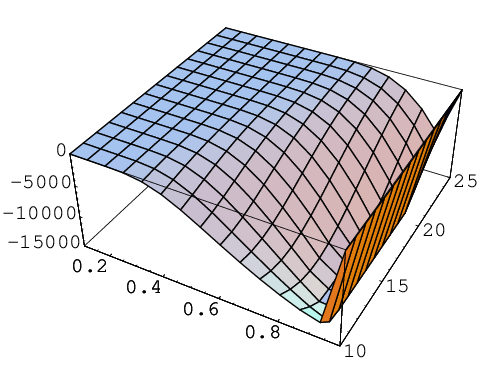}
\end{center}
\label{W4L100k10T1}
\end{figure}

\begin{figure}
\caption{
Behavior of $(-N_6/2)\ptil^3W_6(\ptil,z)F^-(\ptil,z)$(parabolic suppression2). 
$\La=20000,\ \om=5000,\ T=1$\ . 
$1.0001/\om\leq z\leq 0.9999/T ,\ \m=\La T/\om\leq \ptil\leq \La$. 
}
\begin{center}
\includegraphics[height=8cm]{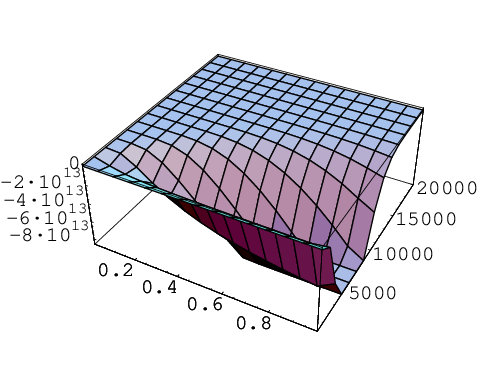}
\end{center}
\label{W6L2mank5senT1}
\end{figure}

We introduce, instead of restricting the integral region, 
a {\it weight function} $W(\ptil,z)$ in the ($\ptil,z$)-space  
for the purpose of suppressing UV and IR divergences of the Casimir Energy. 
\bea
-E^{\mp~W}_{Cas}(\om,T)\equiv\intpE\int_{1/\om}^{1/T}dz~ W(\ptil,z)F^\mp (\ptil,z)\com\q
\ptil=\sqrt{p_4^2+p_1^2+p_2^2+p_3^2}\com\nn
F^\mp(\ptil,z)= s(z)\int_{p^2}^{\infty}\{G_k^\mp(z,z) \}dk^2
=\frac{2}{(\om z)^3}\int_\ptil^\infty\ktil~ G^\mp_k(z,z)d\ktil  \com\nn
\mbox{Examples of}~ W(\ptil,z):\q W(\ptil,z)=\hspace{10cm}\nn
\left\{
\begin{array}{cc}
(N_1)^{-1}\e^{-(1/2) \ptil^2/\om^2-(1/2) z^2 T^2}\equiv W_1(\ptil,z),\ N_1=1.711/8\pi^2 & \mbox{elliptic suppr.}\\
(N_{1b})^{-1}\e^{-(1/2) \ptil^2/\om^2}\equiv W_{1b}(\ptil,z),\ N_{1b}=2/8\pi^2 & \mbox{kinetic-energy suppr.}\\
(N_{2})^{-1}\e^{-\ptil zT/\om}\equiv W_2(\ptil,z),\ N_2=2\frac{\om^3}{T^3}/8\pi^2                   & \mbox{hyperbolic suppr.1}\\
(N_{3})^{-1}\e^{-(1/2) \ptil^2 z^2T^2/\om^2}\equiv W_3(\ptil,z),\ N_3=\frac{2}{3}\frac{\om^3}{T^3}/8\pi^2   &  \mbox{hyperbolic suppr.2}\\
(N_{4})^{-1}\e^{-(1/2) \ptil^2/z^2\om^2 T^2}\equiv W_4(\ptil,z),\ N_4=\frac{2}{5}/8\pi^2               &  \mbox{linear suppr.} \\
(N_{5})^{-1}\e^{-\ptil/z^2\om T^2}\equiv W_5(\ptil,z),\ N_5=\frac{2}{3}/8\pi^2                &  \mbox{parabolic suppr.1}  \\ 
(N_{6})^{-1}\e^{-\ptil^2/2z\om^2T}\equiv W_6(\ptil,z),\ N_6=\frac{2}{3}/8\pi^2                &  \mbox{parabolic suppr.2} \\  
(N_{7})^{-1}\e^{-(1/2) \ptil^4/\om^4}\equiv W_7(\ptil,z),\ N_7=\frac{1}{2}/8\pi^2                &  \mbox{higher-der. suppr.1} \\  
(N_{8})^{-1}\e^{-1/2 (\ptil^2/\om^2+1/z^2T^2)}\equiv W_8(\ptil,z),\ N_8=0.4177/8\pi^2 & \mbox{reciprocal suppr.1}\\
(N_{47})^{-1}\e^{-1/2 (\ptil^2/\om^2)(\ptil^2/\om^2+1/z^2T^2)}\equiv W_{47}(\ptil,z),\ N_{47}=0.1028/8\pi^2 & \mbox{higher-der. suppr.2}\\
(N_{56})^{-1}\e^{-1/2 (\ptil/z\om T)(\ptil/\om+1/zT)}\equiv W_{56}(\ptil,z),\ N_{56}=0.1779/8\pi^2 & \mbox{reciprocal suppr.2}\\
(N_{88})^{-1}\e^{-1/2 (\ptil^2/\om^2+1/z^2T^2)^2}\equiv W_{88}(\ptil,z),\ N_{88}=0.01567/8\pi^2 & \mbox{higher-der. reciprocal suppr.}\\
(N_{9})^{-1}\e^{-1/2 (\ptil/\om+1/zT)^2}\equiv W_9(\ptil,z),\ N_9=0.05320/8\pi^2 & \mbox{reciprocal suppr.3}\\
\end{array}
           \right.
\label{uncert1}
\eea
where 
$G_k^\mp(z,z) $ are defined in (\ref{HKA12}). 
The normalization constants $N_i$ are explained in App.E. 
\footnote
{
In the warped geometry we have, besides the cut-off parameter $\La$, 
two massive parameter $T$ and $\om$.   
We make all exponents in (\ref{uncert1}) dimensionless
by use of $T$ for $z$, and $\om$ for $\ptil$. 
}
In the above, 
we list some examples expected for the weight function $W(\ptil,z)$. 
$W_2$ and $W_3$ are regarded to correspond to the regularization taken by
Randall-Schwartz. 
How to specify the form of $W$ is the subject of the next section. 
We show the shape of the energy integrand $(-1/2)\ptil^3W(\ptil,z)F^-(\ptil,z)$ in 
Fig.\ref{W1L2mank5senT1}-\ref{W6L2mank5senT1} for various choices of $W$. 
We notice the valley-bottom line $\ptil\approx 0.75\La$, which appeared in the un-weighted 
case (Fig.\ref{p3FmL10000}-\ref{p3FmL40000}), is replaced by new lines:\ 
$\ptil^2+z^2\times \om^2T^2\approx \mbox{const}$(Fig.\ref{W1L2mank5senT1},$W_1$), $\ptil z\approx \mbox{const}$
(Fig.\ref{W3L2mank5senT1},$W_3$), 
$\ptil\approx \mbox{const}\times z$(Fig.\ref{W4L100k10T1},$W_4$), $\ptil\approx \mbox{const}\times\sqrt{z}$
(Fig.\ref{W6L2mank5senT1},$W_6$). They are all located {\it away from} the original $\La$-effected line
($\ptil\sim 0.75\La$).
\footnote{
For the graphical view, we take rather large values of $\om$'s. If we take 
a more smaller value for $\om$, the position of a valley-bottom line 
deviates more from that of the un-weighted case ($\ptil\sim 0.75\La$). 
} 

We can check the divergence (scaling) behavior of $E^{\mp~W}_{Cas}$ by 
{\it numerically} evaluating the $(\ptil,z)$-integral (\ref{uncert1}) for 
the rectangle region of Fig.\ref{zpINTregionW}. 
\footnote{
The data fitting is based on the numerical integration for the different 
cases of $(\La,\om,T)$. For example, the $W_1$ formula is based on the 
numerical values of $E_{Cas}$ for 
$T=0.01,\om=10^2,\La=10^3\times (1,2,4,8,16); 
 \om=10^3,\La=2\times 10^4, T=(1,1/2,1/4,1/8,1/16);
 T=1,\La=8\times 10^4, \om=10^2\times (8,16,32,64,128)$. 
See App.D for the numerical derivation. 
} 
\bea
-E^W_{Cas}=\mbox{\hspace{10cm}}\nn
\nn
\left\{
\begin{array}{cc}
\frac{\om^4}{T}\La\times 1.2\left\{  1+0.11~\ln\frac{\La}{\om}-0.10~ \ln\frac{\La}{T}  \right\} & \mbox{for}\q W_1 \\
\frac{\om^4}{T}\La\times 2.0\left\{  1+0.07~\ln\frac{\La}{\om}-0.10~ \ln\frac{\La}{T}  \right\}         & \mbox{for}\q W_{1b} \\
\frac{T^2}{\om^2}\La^4\times 0.062\left\{  1+0.03~\ln\frac{\La}{\om}-0.08~ \ln\frac{\La}{T}  \right\}   &\mbox{for}\q W_2\\
\frac{T^2}{\om^2}\La^4\times 0.14\left\{  1+0.01~\ln\frac{\La}{\om}-0.06~ \ln\frac{\La}{T}  \right\}    &\mbox{for}\q W_3 \\
\frac{\om^4}{T}\La\times 1.5\left\{  1+0.08~\ln\frac{\La}{\om}-0.10~ \ln\frac{\La}{T}  \right\}                 & \mbox{for}\q W_4\\
\frac{\om^4}{T}\La\times 1.5\left\{  1+0.07~\ln\frac{\La}{\om}-0.10~ \ln\frac{\La}{T}  \right\}          &\mbox{for}\q W_5 \\
\frac{\om^4}{T}\La\times 0.86\left\{  1+0.07~\ln\frac{\La}{\om}-0.07~ \ln\frac{\La}{T}  \right\}           & \mbox{for}\q W_6\\
\frac{\om^4}{T}\La\times 1.3\left\{  1+0.06~\ln\frac{\La}{\om}-0.08~ \ln\frac{\La}{T}  \right\}           & \mbox{for}\q W_{7} \\
\frac{\om^4}{T}\La\times 1.6\left\{  1+0.09~\ln\frac{\La}{\om}-0.10~ \ln\frac{\La}{T}  \right\}            & \mbox{for}\q W_8\\
\frac{\om^4}{T}\La\times 0.47\left\{  1+0.05~\ln\frac{\La}{\om}-0.07~ \ln\frac{\La}{T}  \right\}           & \mbox{for}\q W_{47}\\
\frac{\om^4}{T}\La\times 0.93\left\{  1+0.06~\ln\frac{\La}{\om}-0.07~ \ln\frac{\La}{T}  \right\}          & \mbox{for}\q W_{56}\\
\frac{\om^4}{T}\La\times 1.1\left\{  1+0.06~\ln\frac{\La}{\om}-0.07~ \ln\frac{\La}{T}  \right\}          & \mbox{for}\q W_{88}\\
\frac{\om^4}{T}\La\times 0.91\left\{  1+0.05~\ln\frac{\La}{\om}-0.07~ \ln\frac{\La}{T}  \right\}          & \mbox{for}\q W_{9}
\end{array}
           \right.
\label{uncert1bX}
\eea

The suppression behaviors of $W_2$ and $W_3$ improve, compared with (\ref{surfM1}) by 
Randall-Schwartz. The quintic divergence of (\ref{surfM1}) reduces to the quartic divergence
in the present approach of $W_2$ and $W_3$. 
The hyperbolic suppressions, however, are still insufficient for the renormalizability. 
After dividing by the normalization factor, $\La T^{-1}$, 
the cubic divergence remains. 
The desired cases are others. 
The Casimir energy for each case consists of three terms. 
The first terms give {\it finite} values after dividing by the overall normalization 
factor $\La T^{-1}$. The last two terms are proportional to $\log\La$ and show 
the {\it anomalous scaling}. Their contributions are order of $10^{-1}$ to the first leading terms. 
The second ones ($\ln\frac{\La}{\om}$) contribute positively while 
the third ones ($\ln\frac{\La}{T}$) negatively.   

They give, after normalizing the factor $\La/T$, {\it only} the {\it log-divergence}. 
\bea
E^W_{Cas}/\La T^{-1} =-\al \om^4\left( 1-4c\ln (\La/\om) -4c'\ln (\La/T) \right) 
\com
\label{uncert1c}
\eea
where $\al, c$ and $c'$ can be read from (\ref{uncert1bX}) depending on the choice of $W$. 
\footnote{
We should note that, for so many different forms of the suppression factor, 
the normalized $E^W_{Cas}$ takes this log-divergence behavior (\ref{uncert1c})
with similar coefficients ($\al,c,c'$). Exceptions are $W_2$ and $W_3$. 
}
This means the 5D Casimir energy is {\it finitely} obtained by the ordinary 
renormalization of the warp factor $\om$. (See the final section.)
In the above result of the warped case, the IR parameter $l$ in the flat result (\ref{uncert1cc}) 
is replaced by the inverse of the warp factor $\om$. 

At present, we cannot discriminate which weight is the right one. Here we list 
characteristic features 
(advantageous(Yes) or disadvantageous(No), independent (I) or dependent(D), 
singular(S) or regular(R)) for each weight from the following points.
\begin{description}
\item[point 1]
The behavior of $W$ for the limits:\ $T\ra\infty$ (4D limit)\ ;\ $T\ra 0$\ (5D limit)\ ;\ 
$\om\ra 0$\ (flat limit). This property is related to the continuation to the ordinary 
field-quantization.
\item[point 2]
How the path (bottom line of the valley) depends on the scales $T$ and $\om$. 
\item[point 3]
Regular(R) or singular(S) at $z=0$. This point is not important because the range of $z$ is 
$1/\om \leq z\leq 1/T$ or $-1/T\leq z\leq -1/\om$. 
\item[point 4]
Symmetric for $\ptil/\om\change Tz$.
\item[point 5]
Symmetric for $\ptil/\om \change 1/Tz$. (Reciprocal symmetry)
\item[point 6]
The value of $\al$.
\item[point 7]
The values of (-4$c$,-4$c'$). 
\item[point 8]
Under the Z$_2$-parity $z\leftrightarrow -z$, $W(\ptil,z)$ is even (E), odd (O) or none (N). 
\end{description}

\[
  \begin{array}{|c|c|c|c|c|c|c|c|c|}
\mbox{W type} & W_1& W_{1b} & W_2 & W_3 & W_4 & W_5 & W_6 & W_7 \\
\hline
\mbox{point 1}  &   &  &  &  &  &  &  &               \\
\mbox{$T\ra\infty$}  &  T^{-1}\times &  / & (\frac{\om}{T})^{-3+1/2}\times & 
                            (\frac{\om}{T})^{-3+1}\times & / & / & / & /              \\
                         &  \del(z) &  / &     \del(\sqrt{\ptil z}) & 
                                           \del(\ptil z) & / & / & / & /              \\
\mbox{$T\ra 0$}  &  / &  / & / & / & T\om\times    & T\sqrt{\om}\times    & \sqrt{T}\om\times 
& /        \\
                     &  / &  / & / & / & \del(\ptil/z) & \del(\sqrt{\ptil}/z) & \del(\ptil/\sqrt{z}) & /        \\
\mbox{$\om\ra 0$}  &  \om\times   &  \om\times   & (\frac{\om}{T})^{-3+1/2}\times 
& (\frac{\om}{T})^{-3+1}\times & T\om\times   & T\sqrt{\om}\times & \sqrt{T}\om\times &\om^2\times\\
                       &  \del(\ptil) &  \del(\ptil) & \del(\sqrt{\ptil z}) 
& \del(\ptil z) & \del(\frac{\ptil}{z}) & \del(\sqrt{\ptil}/z) & \del(\ptil/\sqrt{z}) & \del(\ptil^2)\\
\mbox{point 2}  &  \om,T &  \om & T/\om & T/\om & \om T & \om T^2 & \om^2 T & \om     \\
\mbox{point 3} &  (R) &  (R) & (R) & (R) & (S) & (S) & (S) & (R)                    \\
\mbox{point 4}  &  Y &  / & Y & Y & / & N & N & /          \\
\mbox{point 5}  &  / &  / & / & / & Y & N & N & /              \\
\mbox{point 6} &  1.2 &  2.0 & div. & div. & 1.5 & 1.5 & 0.86 & 1.3            \\
\mbox{point 7} &  &  &  &  &  &  &  &     \\
-4c & 0.11 & 0.07 & 0.03 & 0.01 & 0.08 & 0.07 & 0.07 & 0.06     \\
-4c' & -0.10 & -0.10 & -0.08 & -0.06 & -0.10 & -0.10 & -0.07 & -0.08     \\
\mbox{point 8}  & E  &  E & O & E & E & E & O & E      
   \end{array} 
\]

\[
  \begin{array}{|c|c|c|c|c|c|}
\mbox{W type} &   W_8 & W_{47} & W_{56} & W_{88} & W_9 \\
\hline
\mbox{point 1}       &  &  &  &  &       \\
\mbox{$T\ra\infty$}       & / & / & / & / & /      \\
                              & / & / & / & / & /      \\
\mbox{$T\ra 0$}           & T\times           & T\om\times            
                    & T\sqrt{\om}\times    & T^2\times           & T\times            \\
                              & \del(\frac{1}{z}) & \del(\frac{\ptil}{z}) 
                    & \del(\sqrt{\ptil}/z) & \del(\frac{1}{z^2}) & \del(\frac{1}{z})  \\
\mbox{$\om\ra 0$} & \om\times   & \om^2\times   & \om\sqrt{T}\times   & \om^2\times & \om\times \\
                      & \del(\ptil) & \del(\ptil^2) & \del(\ptil/\sqrt{z})& \del(\ptil^2)&\del(\ptil)\\
\mbox{point 2}                & \om,T & \om,T & \om,T & \om,T & \om,T       \\
\mbox{point 3}               & (S) & (S) & (S) & (S) & (S)       \\
\mbox{point 4}                & / & N & / & / & /     \\
\mbox{point 5}                & Y & N & Y & Y & Y      \\
\mbox{point 6}               & 1.6 & 0.47 & 0.93 & 1.1 & 0.91      \\
\mbox{point 7}               & & & & & \\
-4c               & 0.09 & 0.05 & 0.06 & 0.06 & 0.05 \\
-4c'               & -0.10 & -0.07 & -0.07 & -0.07 & -0.07 \\
\mbox{point 8}                 & E & E & N & E & N
   \end{array} 
\]
We notice $W_1$ and $W_8$ are specially important. 

So far as the legitimate reason of the introduction of $W(\ptil,y)$ is not clear, 
we should regard this procedure as a {\it regularization} to 
define the higher dimensional theories. 
We give a clear definition of $W(\ptil,y)$ and a legitimate explanation
in the next section. 
It should be done, in principle, in a consistent way with the bulk geometry and the gauge
principle.

\section{Meaning of Weight Function and Quantum Fluctuation of Coordinates and Momenta\label{weight}}

\begin{figure}
\caption{
The mesh of the dotted  lines shows the the ordinary integral 
of $\int d\ptil dz$. Two solid lines show two pathes $\ptil_1(z)$ and $\ptil_2(z)$. 
The path-integral $\int\Dcal\ptil(z)$ is the integral over all possible pathes. 
}
\includegraphics[height=8cm]{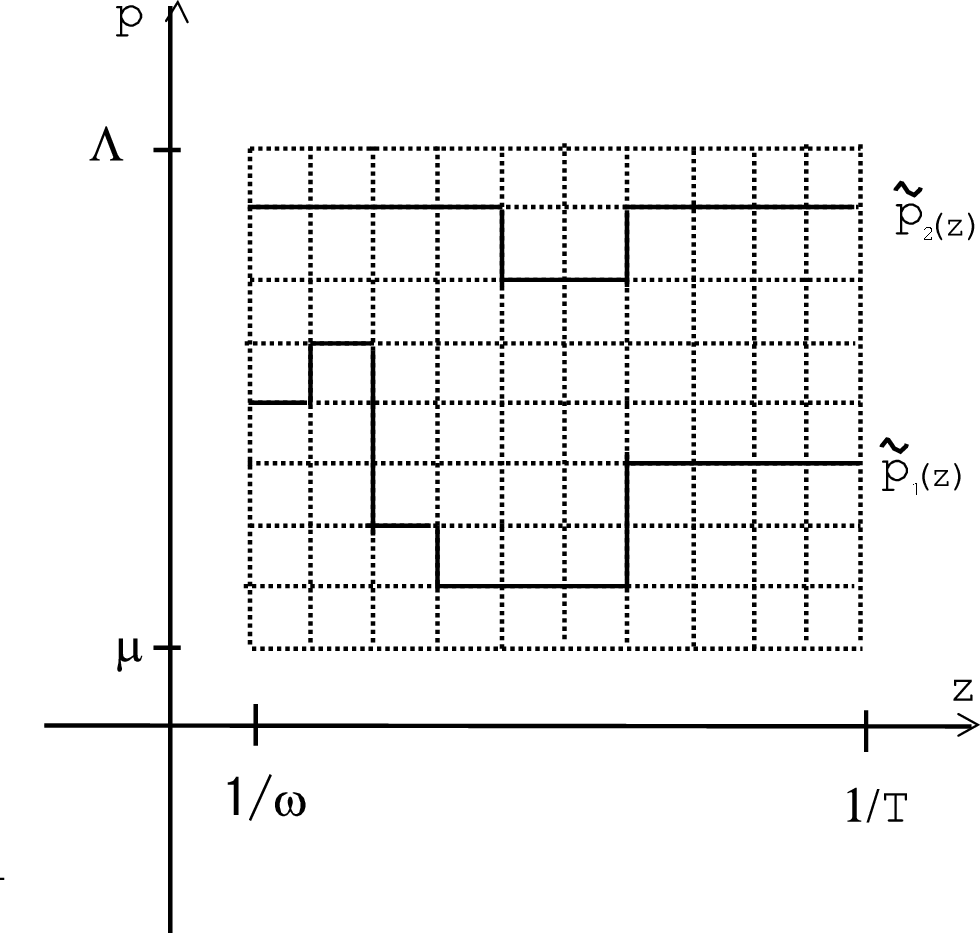}
\label{PathInt}
\end{figure}

In the previous work\cite{SI0801}, we have presented the following idea to 
define the weight function $W(\ptil,z)$. In the evaluation (\ref{uncert1}): 
\bea
-E^{W}_{Cas}(\om,T)=\intpE\int_{1/\om}^{1/T}dz~ W(\ptil,z)F^\mp (\ptil,z)\nn
=\frac{2\pi^2}{(2\pi)^4}\int d\ptil\int_{1/\om}^{1/T}dz~\ptil^3 W(\ptil,z)F^\mp (\ptil,z)\com
\label{weight1}
\eea
the $(\ptil,z)$-integral is over the rectangle region shown in Fig.\ref{zpINTregionW2} 
(with $\La\ra\infty$ and $\m\ra 0$). $F^\mp (\ptil,z)$ is explicitly given in (\ref{HKA13}). 
Following Feynman\cite{Fey72}, 
we can replace the integral by the summation over all possible pathes $\ptil(z)$ 
as schematically shown in Fig.\ref{PathInt}. 

\bea
-E^{W}_{Cas}(\om,T)=\int\Dcal\ptil(z)\int_{1/\om}^{1/T}dz~S[\ptil(z),z]\com\nn
S[\ptil(z),z]=\frac{2\pi^2}{(2\pi)^4}\ptil(z)^3 W(\ptil(z),z)F^\mp (\ptil(z),z)\pr
\label{weight1a}
\eea
Especially, in the figure, the mesh shows the independency of the integral-variables
$p^a$ and $z$. Two pathes $\ptil_1(z)$ and $\ptil_2(z)$ are shown as two solid lines. 
There exists the {\it dominant path} $\ptil_W(z)$ which is 
determined by the minimal principle
\footnote{
The valley-bottom line of the graph $S=S[\ptil,z]$ can be obtained by two steps.  
First we take the expression (\ref{weight1}). 
$-E^{W}_{Cas}(\om,T)=\int d\ptil  \int_{1/\om}^{1/T}dz~S[\ptil,z],\ 
S[\ptil,z]\propto\ptil^3 W(\ptil,z)F (\ptil,z)$. 
We do the variation, assuming the two independent coordinates $\ptil$ and $z$:\  
$
z\ra z+\Del z,\ \ptil\ra \ptil+\Del\ptil,\ S\ra S+\Del S,\ 
\Del S=\Del\ptil \pl_\ptil S+\Del z\pl_z S
$. 
Secondly we put the condition of path:\ $\ptil=\ptil(z)$. 
Then
$
\Del\ptil=\frac{d\ptil}{dz}\Del z,\ 
\Del S=(\frac{d\ptil}{dz}\pl_\ptil S+\pl_z S)\Del z
$
From the variation condition $\Del S=0$, we obtain (\ref{weight1b}). 
}
: 
$\del S=0$.
\bea
\mbox{Dominant Path }\ptil_W(z)\ :\ \q
\frac{d\ptil}{dz}=\frac{-\frac{\pl\ln(WF)}{\pl z}}{\frac{3}{\ptil}+\frac{\pl\ln (WF)}{\pl\ptil}}
\pr
\label{weight1b}
\eea
Hence it is fixed by $W(\ptil,z)$. Examples are the valley-bottom lines in Fig.\ref{W1L2mank5senT1}-\ref{W6L2mank5senT1}.  
On the other hand, there exists another independent path: the minimal surface  
curve $r_g(z)$. 
\bea
\mbox{Minimal Surface Curve }r_g(z)\ :\q
3+\frac{4}{z}r'r-\frac{r''r}{{r'}^2+1}=0\com\q
\frac{1}{\om}\leq z\leq \frac{1}{T}
\com\label{weight2}
\eea 
which is obtained by the {\it minimal area principle}: 
\bea
ds^2=\frac{1}{\om^2 z^2}(
\del_{ab}+\frac{x^ax^b}{(rr')^2} )dx^a dx^b\equiv g_{ab}(x)dx^adx^b 
\com\nn
\del A=0\com\q  
A=\int\sqrt{\det g_{ab}}~d^4x
=\int_{1/\om}^{1/T}\frac{1}{\om^4z^4}\sqrt{{r'}^2+1}~r^3 dz
\pr
\label{weight3}
\eea 
See App.A for detail. 
Hence $r_g(z)$ is fixed by the {\it induced geometry} $g_{ab}(x)$. 
Here we put the {\it requirement}\cite{SI0801}: 
\bea
\ptil_W(z)=\ptil_g(z)
\com\label{weight4}
\eea 
where $\ptil_g\equiv 1/r_g$. This means the following things. 
We {\it require} 
the dominant path coincides with the minimal surface line $\ptil_g(z)=1/r_g(z)$ which is 
defined independently of $W(\ptil,z)$. 
In other words, $W(\ptil,z)$ is defined here by 
the induced geometry $g_{ab}(x)$. 
In this way, we can connect the {\it integral-measure} over the 5D-space with the (bulk) geometry. 
We have confirmed the (approximate) coincidence by the 
numerical method.(See App.C) 

In order to most naturally accomplish 
the above requirement, we can go to a 
{\it new step}. Namely, 
we {\it propose} to {\it replace} the 5D space integral with the weight $W$, (\ref{weight1}), 
by the following {\it path-integral}. We 
{\it newly define} the Casimir energy in the higher-dimensional theory as follows.  
\bea
-\Ecal_{Cas}(\om,T,\La)\equiv \int_{1/\La}^{1/\m}d\rho\int_{\ptil(1/\om)=\ptil(1/T)=1/\rho}
\prod_{a,z}\Dcal p^a(z)F(\ptil,z)\nn
\times\exp\left[ 
-\frac{1}{2\al'}\int_{1/\om}^{1/T}\frac{1}{\om^4z^4}\frac{1}{\ptil^3}\sqrt{\frac{\ptil^{'2}}{\ptil^4}+1}~ dz
    \right]\nn
=\int_{1/\La}^{1/\m}d\rho\int_{r(1/\om)=r(1/T)=\rho}
\prod_{a,z}\Dcal x^a(z)F(\frac{1}{r},z)\nn
\times\exp\left[ 
-\frac{1}{2\al'}\int_{1/\om}^{1/T}\frac{1}{\om^4z^4}\sqrt{{r'}^2+1}~r^3 dz
    \right]\com
\label{weight5}
\eea 
where $\m=\La T/\om$ and the limit $\La T^{-1}\ra \infty$ is taken. 
The string (surface) tension parameter $1/2\al'$ is introduced.
\footnote{
$\al'$ is a free parameter of the theory. Two typical choices are considered:\ 
(a) $\al'=T^{-4}$ (soft surface);\ (b) $\al'=\om^{-4}$ (rigid surface). 
}
 (Note: Dimension of $\al'$ is [Length]$^4$. ) 
The square-bracket ($[ \cdots ]$)-parts of (\ref{weight5}) are \ 
$-\frac{1}{2\al'}$Area = $-\frac{1}{2\al'}\int\sqrt{\mbox{det}g_{ab}}d^4x$ 
(See (\ref{MiniSur4})) where $g_{ab}$ is the induced metric on the 4D surface. 
$F(\ptil,z)$ is defined in (\ref{uncert1}) or (\ref{HKA13}) and shows 
the {\it field-quantization} of the bulk scalar (EM) fields. 
In the above expression, we have followed 
the path-integral formulation of the {\it density matrix} (See Feynman's text\cite{Fey72}). 
The validity of the above definition is based on the following points: 
a)\ When the weight part (exp $[\cdots]$-part) is 1, the proposed quantity 
$\Ecal_{Cas}$ is equal to $E^W_{Cas}$, 
(\ref{weight1}), with $W=1$ 
;\ 
b)\ The leading path is given by $r_g(z) =1/p_g(z)$, (\ref{weight2});\ 
c)\ The proposed definition, (\ref{weight5}), clearly shows the 4D space-coordinates $x^a$ 
or the 4D momentum-coordinates $p^a$ are {\it quantized} (quantum-statistically, not field-theoretically) with 
the Euclidean time $z$ and the "{\it area} Hamiltonian" 
$A=\int\sqrt{\det g_{ab}}~d^4x$. Note that $F(\ptil,z)$ or $F(1/r,z)$ 
appears, in (\ref{weight5}), as the {\it energy density operator} in the quantum statistical system of
$\{ p^a(z)\}$ or $\{ x^a(z)\}$. 

In the view of the previous paragraph, the treatment of Sec.\ref{uncert} is an {\it effective} action 
approach using the (trial) weight function $W(\ptil,z)$. 
Note that the integral over $(p^\m,z)$-space, appearing in (\ref{HKA13}), 
is the summation over all degrees of freedom of the 5D space(-time) points using the "naive" measure 
$d^4pdz$. 
An important point is that we have the possibility to take another  
measure for the summation in the case of the higher dimensional QFT. 
We have adopted, in Sec.\ref{uncert}, the new measure $W(p^\m,z)d^4pdz$ in such a way that the Casimir energy 
{\it does not show physical divergences}. 
We expect the {\it direct} evaluation of (\ref{weight5}), numerically 
or analytically, leads to the similar result. 
\section{Discussion and Conclusion\label{conc}}

The log-divergence in (\ref{uncert1c}) is the familiar one 
in the ordinary QFT. It can be {\it renormalized} in the 
following way. 
\bea
E^W_{Cas}/\La T^{-1} =-\al\om^4\left( 1-4c\ln (\La/\om)-4c'\ln (\La/T) \right) 
=-\al (\om_r)^4\com\nn
\om_r=\om\sqrt[4]{1-4c\ln (\La/\om)-4c'\ln (\La/T) }\com
\label{conc1}
\eea
where $\om_r$ is the {\it renormalized} warp factor and $\om$ is the {\it bare} one. 
No local counterterms are necessary. 
Note that this renormalization relation is {\it exact} (not a perturbative result). 
In the familiar case of the 4D renormalizable theories, the coefficients $c$ and $c'$ depend on the coupling, but, 
in the present case, they are {\it pure numbers}. 
\footnote{
In the usual case, the log-terms (divergent terms) are separated 
and are canceled by the local counter-terms. It is difficult, in the present case, to take 
such a renormalization procedure 
because the theory is free and has no interaction terms (no couplings). 
The only choice, if we stick to the usual procedure, is the renormalization 
of the wave function and the mass parameter. It does not seem work well. 
Here we take a new approach. We regard the starting boundary parameter $\om$ is 
a {\it bare} quantity and the $\om_r$ defined in (\ref{conc1}) is 
a renormalized one. The boundary parameters flow {\it by themselves}. 
{\it No} local counter-terms are necessary. 
}
It reflects the {\it interaction between (EM) fields and the boundaries}. 
When $c$ and $c'$ are sufficiently small
\footnote{
In the list (\ref{uncert1bX}), all data show $|4c|\sim 10^{-1}$ and $|4c'|\sim 10^{-1}$ . 
}
we find the renormalization group function for the warp factor $\om$ as  
\bea
|c|\ll 1\com\q |c'|\ll 1\com\q 
\om_r=\om (1-c\ln (\La/\om)-c'\ln (\La/T) )\com\nn
\be (\mbox{$\be$-function})\equiv \frac{\pl}{\pl(\ln \La)}\ln \frac{\om_r}{\om}=-c-c'
\pr
\label{conc2}
\eea
We should notice that, in the flat geometry case, the IR parameter (extra-space size) $l$ 
is renormalized (see Sec.\ref{intro}). In the present warped case, however, the corresponding parameter $T$ 
is {\it not renormalized}, but the {\it warp parameter} $\om$ {\it is renormalized}. 
Depending on the sign of $c+c'$, the 5D bulk curvature $\om$ 
{\it flows} as follows. 
\footnote{
Both IR and UV boundary interactions ($c'$ and $c$ ) contribute to the scaling behavior of the system. 
This should be compared with the usual perturbative (w.r.t. the coupling) case where $\be$ is  calculated 
from one region, for example, UV-region. 
}
When $c+c'>0$, the bulk curvature $\om$ decreases (increases) as the 
the measurement energy scale $\La$ increases (decreases).
\footnote{
This is the case of "asymptotic free" in the usual renormalization of the 
gauge coupling of 4D YM. 
}
 When $c+c'<0$, the flow goes in the opposite way. 
When $c+c'=0$, $\om$ does not flow ($\be=0$) and is given by 
$\om_r=\om (1+c\ln(\om/T))$. 

The final result (\ref{conc1}) is the {\it new} type Casimir energy, $-\om^4$. 
$\om$ appears as a boundary parameter like $T$. 
The familiar one 
is $-T^4$ in the present context 
(See (\ref{uncert1cc}). Note $T$ is the IR parameter and is related to $l$ as  
$T=\om \e^{-\om l}$.). In ref.\cite{IM05NPB}, 
another type $T^2\om^2$ was predicted using a "quasi" Warped model (bulk-boundary theory). 

Through the Casimir energy calculation, in the higher dimension, we find a way to 
quantize the higher dimensional theories within the QFT framework. 
The quantization {\it with respect to the fields} (except the gravitational fields $G_{AB}(X)$) 
is done in the standard way. After this step, the expression has the summation 
{\it over the 5D space(-time) coordinates or momenta} 
$\int dz\prod_adp^a$. We have proposed that this summation should be replaced by 
the {\it path-integral} $\int \prod_{a,z}\Dcal p^a(z)$ with the {\it area} action (Hamiltonian) 
$A=\int\sqrt{\det g_{ab}}d^4x$ where $g_{ab}$ is the {\it induced} metric on the 4D surface. 
This procedure says the 4D momenta 
$p^a$ (or coordinates $x^a$) are {\it quantum statistical} operators and 
the extra-coordinate $z$ is the inverse temperature (Euclidean time). 
We recall the similar situation occurs in the standard string approach. 
The space-time coordinates obey some uncertainty principle\cite{Yoneya87}.  

Recently the dark energy ( as well as the dark matter ) in the universe is a hot subject. 
It is well-known that the dominant candidate is the cosmological term. 
We also know the proto-type higher-dimensional theory, that is, the 5D KK theory, 
has predicted so far the {\it divergent} cosmological constant\cite{AC83}. 
This unpleasant situation has been annoying us for a long time. 
If we apply the present result, the situation drastically improve. The cosmological
constant $\la$ appears as
\bea
R_\mn-\half g_\mn R-\la g_\mn =T_\mn^{matter}\com\nn
S=\int d^4x \sqrt{-g}\{\frac{1}{G_N}(R+\la) \}
+\int d^4x \sqrt{-g}\{\Lcal_{matter}\}\
\com\q
g=\mbox{det}~g_{\mn}\com
\label{conc3}
\eea
where $G_N$ is the Newton's gravitational constant, $R$ is the Riemann scalar 
curvature. 
We consider here the 3+1 dim Lorentzian space-time ($\mu,\nu=0,1,2,3$). 
The constant $\la$ observationally takes the value.
\bea
\frac{1}{G_N}\la_{obs}\sim \frac{1}{G_N{R_{cos}}^2}\sim m_\n^4\sim (10^{-3} eV)^4
\com\q
\la_{obs}\sim \frac{1}{R_{cos}^{~2}}\sim 4\times 10^{-66}(eV)^2\com
\label{conc4}
\eea
where $R_{cos}\sim 5\times 10^{32}\mbox{eV}^{-1}$ is the cosmological size (Hubble length), $m_\n$ is the neutrino mass.
\footnote{ 
The relation $m_\n\sim \sqrt{M_{pl}/R_{cos}}=\sqrt{1/R_{cos}\sqrt{G_N}}$, which appears  
in some extra dimension model\cite{SI0012,SI01Tohwa}, is used. The neutrino mass is, 
at least empirically, located 
at the {\it geometrical average} of two extreme ends of the mass scales in the universe.  
} 
On the other hand, we have theoretically so far
\bea
\frac{1}{G_N}\la_{th}\sim \frac{1}{{G_N}^2}={M_{pl}}^4\sim (10^{28} eV)^4
\pr
\label{conc5}
\eea
This is because the mass scale usually comes from the quantum 
gravity. (See ref.\cite{SI83NPB} for the derivation using the  
Coleman-Weinberg mechanism.) 
We have the famous huge discrepancy factor: 
\bea
\frac{\la_{th}}{\la_{obs}}\sim N_{DL}^{~2}\com\q N_{DL}\equiv M_{pl}R_{cos}\sim 6\times 10^{60}
\com
\label{conc5b}
\eea
where $N_{DL}$ is the Dirac's large number\cite{PD78}. 
If we use the present result (\ref{conc1}), we can obtain a natural choice of 
$T, \om$ and $\La$
as follows. By identifying 
$T^{-4}E_{Cas}=-\al_1\La T^{-1}\om^4/T^4$ with 
$\int d^4x\sqrt{-g}(1/G_N)\la_{ob}=R_{cos}^{~2}(1/G_N)$, we 
obtain the following relation. 
\bea
N_{DL}^{~2}=R_{cos}^{~2}\frac{1}{G_N}=-\al_1 \frac{\om^4\La}{T^5}\com\q \al_1\ :\ \mbox{some coefficient}
\pr
\label{conc6}
\eea
The warped (AdS$_5$) model predicts the cosmological constant {\it negative}, 
hence we have interest only in its absolute value.
\footnote{
This fact strongly suggests 
de Sitter (dS$_5$) version of the present work could solve this sign problem. 
Details are under way. 
} 
We take the following choice for $\La$ and $\om$. 
\bea
\La=M_{pl}\sim 10^{19}GeV\com\q
\om\sim
 \frac{1}{\sqrt[4]{G_N{R_{cos}}^2}}=\sqrt{\frac{M_{pl}}{R_{cos}}}\sim m_\n\sim 10^{-3}\mbox{eV}
\pr
\label{conc6b}
\eea
The choice for $\La$ is accepted in that the largest known energy scale is the Planck energy. 
The choice for $\om$ comes from the experimental bound for the Newton's gravitational force. 

As shown above, we have the standpoint that the cosmological constant is mainly made from 
the Casimir energy.  
We do not yet succeed in obtaining the value $\al_1$ negatively, but
succeed in obtaining  
the finiteness of the cosmological constant and its gross absolute value. 
The smallness of the value is naturally explained by the renormalization flow as follows. 
Because we already know the warp parameter $\om$ {\it flows} (\ref{conc2}), 
the $\la_{obs}\sim 1/R_{cos}^2$ expression (\ref{conc6b}), $\la_{obs}\propto \om^4$, says that the {\it smallness of the cosmological constant comes from 
the renormalization flow} for the non asymptotic-free case ($c+c'<0$ in (\ref{conc2})). 
\footnote{
A.M. Polyakov presented, in an early stage, 
the idea that the cosmological constant may be screened by the 
IR fluctuation of the metric\cite{Pol82}. 
It was clearly shown, using the 2 dim R$^2$-gravity, such thing really occurs\cite{SI95NPB}. 
}
\footnote{
We claim here the smallness of the cosmological constant is {\it dynamically} explained (without fine tuning). 
}

The IR parameter $T$, the normalization factor $\La/T$ in (\ref{uncert1c}) and the IR cutoff 
$\mu=\La\frac{T}{\om}$ are given by 
\bea
T=R_{cos}^{~-1}(N_{DL})^{1/5}\sim 10^{-20}eV\com\q
\frac{\La}{T}=(N_{DL})^{4/5}\sim 10^{50}\com\q
\mu=M_{pl}N_{DL}^{-3/10}\sim 1GeV\sim m_N
\com
\label{conc8}
\eea
where $m_N$ is the nucleon mass. 
\footnote{
Note that the present model predicts the nucleon mass scale (1 Gev) from the 3 data:\ 
Planck mass $M_{pl}$, the neutrino mass $m_\nu$, and the cosmological size $R_{cos}$ 
(or the cosmological constant $\la_{obs}\sim R_{cos}^{~-2}$).
}
The Fig.\ref{IRUVRegSurfW} strongly suggests that 
the degree of freedom of the universe (space-time) 
is given by 
\bea
\frac{\La^4}{\m^4}=\frac{\om^4}{T^4}=N_{DL}^{~6/5}\sim 10^{74}\sim (\frac{M_{pl}}{m_N})^4
\pr
\label{conc9}
\eea

In the recent exciting work by Ho\u{r}ava\cite{Hora0812,Hora0901}, the vastness of the string 
theory is stated as "the string theory represents a logical completion of quantum field theory, 
not a single theory". 
He tackles the renormalizability problem of the quantum gravity 
not from the string theory but from a "small" one, that is, a 3+1 dim local field theory
with spacially higher-derivative interactions. 
The idea comes from the success of Lifshitz theory\cite{Lifshitz41}, a spatially-higher-derivative scalar theory, in the condensed matter physics. 
The present approach has some points which can be compared with \Hora 's. 
Basically both are the gravity-matter local field theory inspired by the string, brane and 
membrane theories. \Hora 's one is basically 3+1 dimensional, while the present one 4+1 or 5 
dimensional. In both ones, the renormalization flow plays a key role in the renormalizability
(UV-completion). In the \Hora 's, the Lorentz symmetry is abandoned as the starting principle, 
but is regarded as the dynamically emergent one (in IR region of RG flow). 
At the cost 
of Lorentz symmetry, he introduces spatially higher-derivative terms in order to 
suppress divergences in UV-region. This anisotropy between space and time 
(non-relativistic aspect) does not appear in the present approach because 
we treat the 4D world isotropically. Instead we have anisotropy between the 4D world 
and the extra space. 
In the present case, the renormalizability 
is realized by the warped configuration (thickness) and the appropriate suppression 
by the weight function. We need {\it not} higher-derivative terms. 
The origin of the suppression is, at present, not established. 
Also in \Hora 's, an uncertain procedure "detailed balance condition" is introduced 
in order to reduce the number of independent coupling constants. 
It looks important to take a {\it new} standpoint or view, which has been overlooked 
so far, about some basic things, such as 
the starting symmetries, the quantum treatment of gravitational and matter fields, 
the regularization method  
and the meaning of the extra axis(es), to solve the 
{\it divergence problem} of gravity.


\section{App. A.\ \ Equation of Minimal Surface in AdS$_5$ Geometry and 
Classification of Surfaces\label{MiniSur}}

The present new idea is that the regularization surfaces are determined 
by the principle of the {\it minimal surface} in the bulk AdS$_5$ manifold. 
We {\it require} the ultraviolet and infrared
regularization surfaces obey the law of the higher dimensional (bulk) geometry. 
In this section, we examine the {\it minimal surface} for the warped and flat cases. 
We classify all paths (solutions). It is useful in drawing minimal surface curves 
in the text and in confirming the non-crossing
\footnote{
This requirement comes from the renormalization interpretation of the present regularization. 
See a few lines above (\ref{surf1}). 
}
 of curves. 

The AdS$_5$ geometry is described by 
\bea
ds^2=\frac{1}{\om^2 z^2}(
\del_{ab} dx^a dx^b+(dz)^2)\com\q
a,b=1,2,3,4\com\q 
\frac{1}{\om}\leq z\leq\frac{1}{T}
\com
\label{MiniSur1}
\eea 
where $\del_{ab}$ is the 4D Euclidean flat metric (not Minkowski) 
$(\del_{ab})=\mbox{diag}(1,1,1,1)$. 
$S^3$-sphere in the "3-brane" located on z of the extra coordinate is expressed as
\bea
(x^1)^2+(x^2)^2+(x^3)^2+(x^4)^2=r(z)^2
\com
\label{MiniSur2}
\eea 
where we allow the radius to change along $z$-axis. 
$r(z)$ is some function of $z$ which must be determined by 
the {\it minimal area principle} in the
AdS$_5$ geometry. On the 4D surface, in the bulk, defined by (\ref{MiniSur2}), 
which we call the {regularization surface} or {\it boundary surface}, 
the line element can be expressed as
\bea
ds^2=\frac{1}{\om^2 z^2}(
\del_{ab}+\frac{x^ax^b}{(rr')^2} )dx^a dx^b\equiv g_{ab}(x)dx^adx^b
\com\q r'\equiv \frac{dr}{dz} 
\pr
\label{MiniSur3}
\eea 
The surface area is given by
\bea
A=\int\sqrt{\det g_{ab}}~d^4x=\int\frac{1}{(\om^2z^2)^2}\sqrt{1+\frac{1}{{r'}^2}}d^4x\nn
=\int_{1/\om}^{1/T}\frac{1}{\om^4z^4}\sqrt{{r'}^2+1}~r^3 dz
\pr\nn
\left[ A=\int_{1/\om}^{1/T}\frac{1}{\om^4z^4}\frac{1}{\ptil^3}\sqrt{\frac{\ptil^{'2}}{\ptil^4}+1} dz\com\q
\ptil=\frac{1}{r}\pr\right]
\label{MiniSur4}
\eea 
Under the variation $r(z)\ra r(z)+\del r(z)$, $A$ changes as
\bea
\del A\propto \left[
\frac{1}{\om^4z^4}\frac{r'r^3}{  \sqrt{r^{'2}+1}  }\del r
              \right]_{1/\om}^{1/T}+
\int_{1/\om}^{1/T}\left\{
\frac{1}{\om^4z^4}\sqrt{{r'}^2+1}~3r^2 -(\frac{1}{\om^4z^4}\frac{r'r^3}{\sqrt{r^{'2}+1}})'
                   \right\}\del r dz
\pr
\label{MiniSur5}
\eea 
With the condition that the radii at the end-points are fixed:
\bea
\left.\del r\right|_{z=1/\om}=\left.\del r\right|_{z=1/T}=0
\com
\label{MiniSur6}
\eea 
we obtain the differential equation of the minimal surface.
\bea
3+\frac{4}{z}r'r-\frac{r''r}{{r'}^2+1}=0\com\q
\frac{1}{\om}\leq z\leq \frac{1}{T}
\pr\nn
\left[
3-\frac{4}{z}\frac{\ptil'}{\ptil^3}+\frac{\ptil\ptil''-2\ptil^{'2}}{\ptil^{'2}+\ptil^4}=0\com\q
\ptil=\frac{1}{r}\com\q
\frac{1}{\om}\leq z\leq \frac{1}{T}
\pr
\right]
\label{MiniSur7}
\eea 
In terms of the coordinate $y$ ($\om y=\ln (\om z)$ for $\om z\geq 1$, $y\geq 0$), 
(\ref{MiniSur7}) can be expressed as
\bea
3\e^{2\om y}+4\om r\rdot-\frac{r(\rddot-\om\rdot)}{1+\e^{-2\om y}\rdot^2}=0\com\q
\rdot\equiv \frac{dr}{dy}\com\q \rddot\equiv\frac{d^2r}{dy^2}\com\q 0\leq y\leq l
\pr
\label{MiniSur8}
\eea 
Let us examine the trajectory $r=r(z)$ or $r=r(y)$ which are the solution 
of (\ref{MiniSur7}) or (\ref{MiniSur8}) respectively.

(A) Flat limit\nl

In the $y$-expression (\ref{MiniSur8}), we can take the flat limit:\ $\om=0$. 
\bea
\mbox{flat limit}(\om=0)\nn
3-\frac{r\rddot}{1+\rdot^2}=0\com\q 0\leq y\leq l
\pr
\label{MiniSur9}
\eea 
In terms of $u\equiv 1/r^2$, the above one can be expressed as
\bea
u(y)\equiv\frac{1}{r(y)^2}=\frac{1}{x^ax^a}>0\com\nn
\mbox{flat limit}(\om=0)\q :\q{\ddot u}=-6u^2\leq 0\com\q 0\leq y\leq l
\pr
\label{MiniSur10}
\eea 
From this equation we know an important {\it inequality relation}:
\bea
{\dot u}|_{y=l}-{\dot u}|_{y=0}=-6\int_0^lu^2 dy<0
\pr
\label{MiniSur11}
\eea 
The inequality ${\ddot u}\leq 0$ in (\ref{MiniSur10}) implies 
$u(y)$ is convex upwards. 

\begin{figure}
\begin{center}
\begin{tabular}{cc}
\parbox{6cm}
{\caption{
Geodesic Curve (\ref{MiniSur9}) by Runge-Kutta. 
Type (ia).
$r(0)=4.472, \rdot(0)=-22.36$. 
         }
\label{5DFlatRvsY3}
}
&
\parbox{6cm}
{\caption{
Geodesic Curve (\ref{MiniSur9}) by Runge-Kutta. 
Type (ib$\al$). 
$r(0)=2.236, \rdot(0)=-5.590$. 
         }
\label{5DFlatRvsY2}
}
\\
\includegraphics[height=4cm]{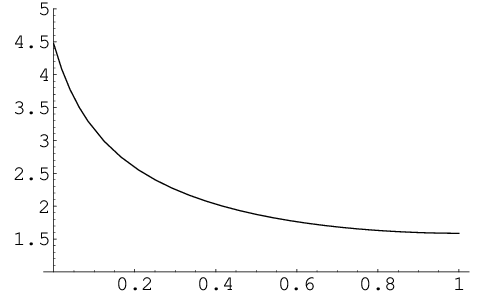}
& 
\includegraphics[height=4cm]{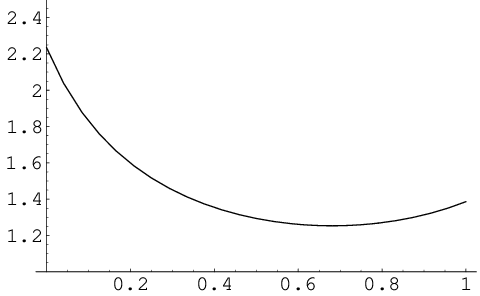}
\\
\parbox{6cm}
{\caption{
Geodesic Curve (\ref{MiniSur9}) by Runge-Kutta. 
Type (ib$\be$). 
$r(0)=1.4142, \rdot(0)=-1.4142$. 
        }
\label{5DFlatRvsY1}
}
&
\parbox{6cm}
{\caption{
Geodesic Curve (\ref{MiniSur9}) by Runge-Kutta. 
Type (ii).
$r(1.0)=10.0, \rdot(1.0)=350.0$ 
        }
\label{5DFlatRvsY7}
}
\\
\includegraphics[height=4cm]{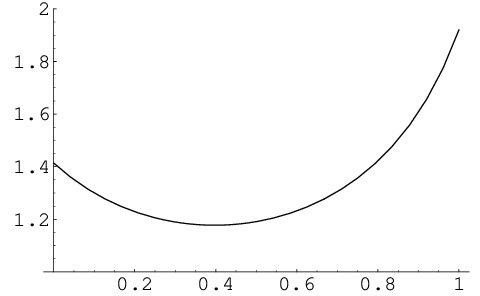}
&
\includegraphics[height=4cm]{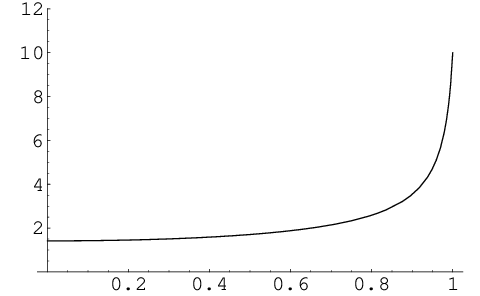}
\end{tabular}
\end{center}
\end{figure}

Making use of the above relation, we can classify all solutions
as follows. \nl
(i)$\udot(y=0)>0$\nl
\hspace*{20pt} (ia)$\udot(l)>0$\nl
\hspace*{40pt}    In this case $\udot(y)>0 \mbox{\ for\ } 0\leq y\leq l$. 
$u(y)$ is simply increasing (r(y) is simply decreasing),\nl
\hspace*{40pt}    
                  Fig.\ref{5DFlatRvsY3}\nl
\hspace*{20pt} (ib)$\udot(l)<0$\nl
\hspace*{40pt}    (ib$\al$)\ $u(0)<u(l)$\ ,    
                  Fig.\ref{5DFlatRvsY2}\nl 
\hspace*{40pt}    (ib$\be$)\ $u(0)>u(l)$\ 
                   , Fig.\ref{5DFlatRvsY1}\nl 
(ii)$\udot(y=0)<0$\nl
\hspace*{20pt} $u(y)$ is simply decreasing (r(y) is simply increasing), Sample 7, Fig.\ref{5DFlatRvsY7}\nl


Although numerical solutions are displayed in 
Fig.\ref{5DFlatRvsY3}-\ref{5DFlatRvsY7}  
the flat limit case is exactly solved and was explained in Appendix A of ref.\cite{SI0801}. 
We have confirmed the high-precision equality between the numerical curves
and the analytical ones. 

(B) Warped Case\nl
In terms of $u\equiv 1/r^2$, eq.(\ref{MiniSur7}) can be rewritten as
\bea
-\frac{1}{z}\frac{u'}{u^3}
+\frac{6u^2+u''}{{u'}^2+4u^3}=0\com\q '=\frac{d}{dz}
\pr
\label{MiniSur12}
\eea 
From this equation, we obtain the following {\it inequality relations}.
\bea
\frac{d}{dz}~\ln({u'}^2+4u^3)
=2\frac{1}{z}\frac{{u'}^2}{u^3}\ >\ 0\com\q
[\ln({u'}^2+4u^3)]_{1/\om}^{1/T}
=2\int_{1/\om}^{1/T}\frac{1}{z}\frac{{u'}^2}{u^3}dz\ >\ 0\com\nn
u'(6u^2+u'')=\frac{1}{z}\frac{{u'}^2({u'}^2+4u^3)}{u^3}\ >\ 0
\pr
\label{MiniSur13}
\eea 
Note that the second equation implies 
$({u'}^2+4u^3)|_{1/T}>({u'}^2+4u^3)|_{1/\om}$. 

B1) z-flat limit\nl
Before the classification of all solutions, we note here, in the case that the 
sphere radius slowly changes (flows) in the following way,
\bea
\left| \frac{dr}{dz}\right|\ll \frac{z}{r}
\q\q\mbox{equivalently}\q\q \left|\frac{d(r^2)}{d(z^2)}\right|\ll 1\com
\label{MiniSur14}
\eea 
the equation (\ref{MiniSur7}) reduces to the {\it $z$-flat limit}:\ 
\bea
3-\frac{r''r}{{r'}^2+1}=0\com\q '=\frac{d}{dz}\com\q
\frac{1}{\om}\leq z\leq \frac{1}{T}
\com
\label{MiniSur15}
\eea 
which is the same as (\ref{MiniSur9}) except the variable $z$.\footnote
{  
It is interesting that, in this limit, the parameter $1/\om$ appears 
as the UV-cutoff and $1/T$ appears as the IR-cutoff of z-integral. 
Compare with the flat geometry case summarized in Sec.\ref{intro}. 
} 
The classification
of the z-flat solutions goes as in the previous case.\nl
(i)$u'(z=1/\om)>0$\nl
\hspace*{20pt} (ia)$u'(z=1/T)>0$\nl
\hspace*{40pt}    Simply Increasing (r(z) is simply decreasing).\nl
\hspace*{20pt} (ib)$u'(z=1/T)<0$\nl
\hspace*{40pt}    (ib$\al$)\ $u(z=1/\om)<u(z=1/T)$\nl 
\hspace*{40pt}    (ib$\be$)\ $u(z=1/\om)>u(z=1/T)$\nl 
(ii)$u'(z=1/\om)<0$\nl
\hspace*{20pt} Simply Decreasing (r(z) is simply increasing)\nl

B2) General Case\nl
Let us consider the general warped case. The first inequality relation of (\ref{MiniSur13})
implies ${u'}^2+4u^3$ is simply increasing for $1/\om < z < 1/T$.

(i)$({u'}^2+4u^3)|_{1/T}>1$\nl
\hspace*{20pt} (ia)$({u'}^2+4u^3)|_{1/\om}>1$\nl
\hspace*{40pt}    ${u'}^2+4u^3>1$\ for $1/\om < z < 1/T$.\nl
\hspace*{20pt} (ib)$({u'}^2+4u^3)|_{1/\om}< 1$\nl
\hspace*{40pt}    $\al < {u'}^2+4u^3 < \be$ for $1/\om < z < 1/T$, where $\al$ and $\be$
are the constants defined by $0<\al\equiv ({u'}^2+4u^3)|_{1/\om}     <1,\ 
\be\equiv ({u'}^2+4u^3)|_{1/T} >1$\nl  
(ii)$({u'}^2+4u^3)|_{1/T}<1$\nl
\hspace*{20pt} ${u'}^2+4u^3<1$\ for $1/\om < z < 1/T$.\nl
Note the relations: ${u'}^2+4u^3=4(1+{r'}^2)/r^6=zu^3(6u^2+u'')/u'=
-z(3+3{r'}^2-rr'')/r'r^7$\nl

Taking into account the last inequality relation of (\ref{MiniSur13}), the above
three cases satisfy the following relations between $u$, $u'$ and $u''$.

\begin{figure}
\caption{
Numerical Solution by Runge-Kutta. (\ref{MiniSur12}), $1/\om=10^{-4}\leq z\leq 1.0=1/T$.  
 $u(1)=0.9^{-2}, u'(1)=0.0$ . 
Graph Type (ia). Vertical axis is $u=1/r^2$. 
}
\includegraphics[height=8cm]{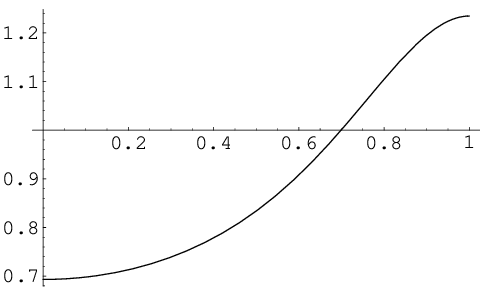}
\label{uVz09m2du0}
\end{figure}

\begin{figure}
\caption{
Numerical Solution by Runge-Kutta. (\ref{MiniSur12}), $1/\om=10^{-4}\leq z\leq 1.0=1/T$.  
 $u(1)=0.9^{-2}, u'(1)=0.0$ . 
Graph Type (ia). Vertical axis is $r=1/\sqrt{u}$
}
\includegraphics[height=8cm]{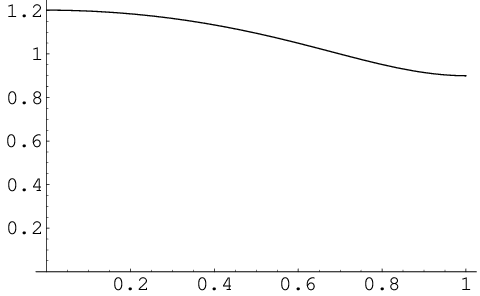}
\label{rVz09dr0}
\end{figure}

(ia)\ When $u'>0$ is valid, $u''>-6u^2\ ,\ {u'}^2>1-4u^3$\nl
\q\q  When $u'<0$ is valid, $u''<-6u^2\ ,\ {u'}^2>1-4u^3$\nl
See Fig.\ref{uVz09m2du0}-\ref{rVz09dr0}

\begin{figure}
\caption{
Numerical Solution by Runge-Kutta. (\ref{MiniSur12}), $1/\om=10^{-4}\leq z\leq 1.0=1/T$.  
 $u(1)=1.0, u'(1)=0.5$ . 
Graph Type (ib). Vertical axis is $u=1/r^2$. 
}
\includegraphics[height=8cm]{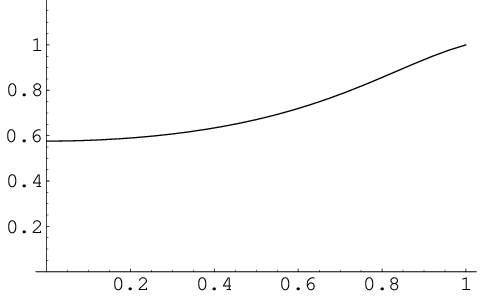}
\label{uVz1du05}
\end{figure}

\begin{figure}
\caption{
Numerical Solution by Runge-Kutta. (\ref{MiniSur12}), $1/\om=10^{-4}\leq z\leq 1.0=1/T$.  
 $u(1)=1.0, u'(1)=0.5$ . 
Graph Type (ib). Vertical axis is $r=1/\sqrt{u}$
}
\includegraphics[height=8cm]{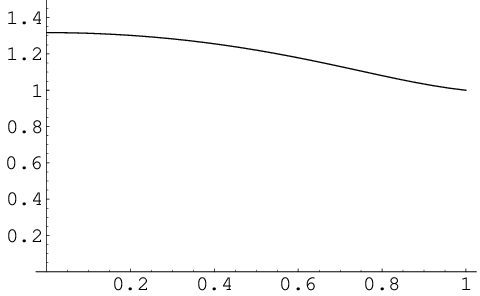}
\label{rVzu1du05}
\end{figure}

(ib)\ When $u'>0$ is valid, $u''>-6u^2\ ,\ \be-4u^3>{u'}^2>\al-4u^3$\nl
\q\q  When $u'<0$ is valid, $u''<-6u^2\ ,\ \be-4u^3>{u'}^2>\al-4u^3$\nl
See Fig.\ref{uVz1du05}-\ref{rVzu1du05}.

\begin{figure}
\caption{
Numerical Solution by Runge-Kutta. (\ref{MiniSur12}), $1/\om=10^{-4}\leq z\leq 1.0=1/T$.  
 $u(1)=0.5, u'(1)=0.5$ . 
Graph Type (ii).Vertical axis is $u=1/r^2$.  
}
\includegraphics[height=8cm]{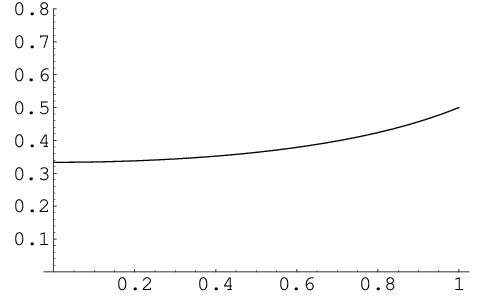}
\label{uVz0505}
\end{figure}

\begin{figure}
\caption{
Numerical Solution by Runge-Kutta. (\ref{MiniSur12}), $1/\om=10^{-4}\leq z\leq 1.0=1/T$.  
 $u(1)=0.5, u'(1)=0.5$ . 
Graph Type (ii). Vertical axis is $r=1/\sqrt{u}$
}
\includegraphics[height=8cm]{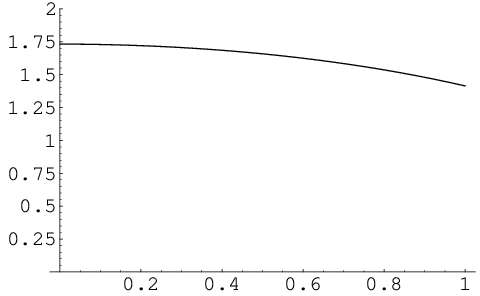}
\label{rVz0505}
\end{figure}

(ii)\ When $u'>0$ is valid, $u''>-6u^2\ ,\ {u'}^2<1-4u^3$\nl
\q\q  When $u'<0$ is valid, $u''<-6u^2\ ,\ {u'}^2<1-4u^3$\nl
See Fig.\ref{uVz0505}-\ref{rVz0505}




\section{App. B.\ \ Casimir Energy of 4D Electromagnetism\label{4dEM}}

We review the ordinary Casimir energy of the 4D electromagnetism in the way 
comparable to the 5D analysis in the text.
\footnote{
See, for example, the text\cite{IZ80}. 
}
 The content is described in a way suggestive to corresponding quantities appearing in the text. 
The Lagrangian is given by
\bea
\Lcal^{inv}_{EM}=-\fourth F_\mn F^\mn
=\half A_\n\pl_\m\pl^\m A^\n +\half (\pl_\m A^\m)^2+\mbox{tot. deri.}\com\nn
F_\mn=\pl_\m A_\n-\pl_\n A_\m\com\q -\infty < x^\m <\infty
\com
\label{4dEM1}
\eea 
where the space-time is flat (Minkowski):\ $(\eta_\mn)=\mbox{diag}(-1,1,1,1)$ and $\m,\n=0,1,2,3.$ 
This theory has the U(1) local gauge symmetry.
\bea
A_\m\ra A_\m+\pl_\m\La(x)
\com
\label{4dEM1b}
\eea 
where $\La(x)$ is the local gauge parameter. 
We  take Lorentz gauge. 
\bea
\pl_\m A^\m=0
\pr
\label{4dEM2}
\eea 
Then the gauge-fixed Lagrangian is given by
\bea
\Lcal_{EM}=\Lcal^{inv}_{EM}-\half(\pl_\m A^\m)^2-\mbox{tot. deri.}=  \half A_\m\pl^2 A^\m
\com
\label{4dEM3}
\eea 
where $\pl^2=\pl_\n \pl^\n=-(\pl/\pl t)^2+(\pl/\pl x)^2+(\pl/\pl y)^2+(\pl/\pl z)^2,\ (x^\m)=(t,x,y,z)$. 
Now we take the periodic boundary condition of the electromagnetic field $A^\m(t,x,y,z)$ as follows: 
for the x and y coordinates it is {\it periodic} with the periodicity $2L$, and 
for the z coordinate with $2l$. The former is for the IR-regularization of the two (x,y)-planes, 
while the latter is the separation-length of the two planes.  We consider the case $L\gg l$. 
\bea 
\mbox{Periodic B.C.:}\q x\ra x+2L\com\q y\ra y+2L\com\q z\ra z+2l\com\nn
L\gg l
\pr
\label{4dEM4}
\eea 
Then $A^\m$ is expanded as
\bea 
A^\m(t,x,y,z)=\sum_{m_xm_yn\in \bfZ}  \left( 
a^\m_{m_xm_yn}(t)
\exp\{i\frac{\pi}{L}(m_x x+m_y y)+i\frac{\pi}{l}nz\}
                                    +\mbox{c.c.}\right)\nn
=\sum_{m_x,m_y,n\in \bfZ}  \left( 
a^\m_{m_xm_yn}(t)
\exp\{i\omvec_{m_xm_yn}\cdot\rvec\}+\mbox{c.c.}\right)\com\nn
\omvec_{m_xm_yn}=(\frac{\pi}{L}m_x,\frac{\pi}{L}m_y,\frac{\pi}{l}n),\ 
\rvec=(x,y,z)
\com
\label{4dEM5}
\eea 
where $\bfZ$ is the set of all integers, and "c.c." means "complex conjugate". 
The gauge-fixing condition (\ref{4dEM2}) says
\bea
-\frac{da^0_{m_xm_yn}}{dt}+i\omvec_{m_xm_yn}\cdot\avec_{m_xm_yn}=0
\com
\label{4dEM6}
\eea 
where $(a^\m)=(a^0,\avec)$. We take the polarization vector $\avec$ 
perpendicular to the wave-number (3D momentum) vector $\omvec$. 
\bea
\omvec_{m_xm_yn}\cdot\avec_{m_xm_yn}=0
\pr
\label{4dEM7}
\eea 
Then from the gauge condition (\ref{4dEM6}), 
\bea
a^0_{m_xm_yn}=\mbox{const.(independent of t)}
\pr
\label{4dEM8}
\eea 
Hence $a^0$ is not a dynamical variable. 
Finally we obtain the action of the system.
\bea
S=\int_{-\infty}^\infty dt~ L_{EM}\com\nn
L_{EM}=\int_{-L}^{L}dx\int_{-L}^{L}dy\int_{-l}^{l}dz\Lcal_{EM}-\mbox{tot.div.}-\mbox{div.const.}\nn
=\half
\sum_{m_x,m_y,n\in \bfZ}\{\frac{d\avec_{m_xm_yn}}{dt}\cdot\frac{d\avec^*_{m_xm_yn}}{dt}
-{\omtil_{m_xm_yn}}^2~\avec_{m_xm_yn}\cdot\avec^*_{m_xm_yn}\}  
\com
\label{4dEM9}
\eea 
where ${\omtil_{m_xm_yn}}^2=(\omvec_{m_xm_yn})^2=(m_x\frac{\pi}{L})^2+(m_y\frac{\pi}{L})^2+(n\frac{\pi}{l})^2$. 
\ $\avec^*$ is the complex conjugate of $\avec$. 
In the above action, two out of three components of $\avec_{m_xm_yn}$ are independent 
due to the relation (\ref{4dEM7}).  

The system (\ref{4dEM9}) is the set of harmonic oscillators with different frequencies
$\om_{m_xm_yn}$ corresponding to the degree of freedom of the 3D continuous space. Let us consider 
the {\it quantum mechanical} system of the {\it harmonic oscillator}.

\bea
L_{HO}(\bdot,b)=\half \bdot^2-\half\omtil^2 b^2\com\q \bdot\equiv \frac{db}{dt}\com\q -\infty < t < \infty
\pr
\label{4dEM10}
\eea 
In order to to examine the quantum statistical property at the temperature
$T=1/\be$, we use the well-known correspondence:\ 
Quantum statistical mechanics in D-dim space versus 
Euclidean quantum field theory in (D+1)-dimension spacetime, $0\leq\tau\leq\be$. 
(See the book\cite{Zee03}. D=0 is the present case. )
\bea
i\int dt L_{HO}(\frac{db}{dt},b)=-\int d\tau\Ltil_{HO}(\frac{db}{d\tau},b)\com\q t=-i\tau\com\nn
\Ltil_{HO}(b',b)=\half {b'}^2+\half\omtil^2b^2\com\q b'=\frac{db}{d\tau}\com\nn
b(\tau)=b(\tau+\be):\ \mbox{periodic with {\it periodicity}}\ \be 
\pr
\label{4dEM11}
\eea 
The energy of the harmonic oscillator, $E_{HO}$, is given by
\footnote{
In terms of the heat-kernel or the density matrix (See the text by Feynman\cite{Fey72})
\bea
<x|\e^{-\be H}|x'>=\rho(x,x';\be)=\int_{b(0)=x,b(\be)=x'}\Dcal b(\tau)
\exp \left[-\int_0^\be d\tau\Ltil_{HO}(b',b;\omtil) \right]
\com
\label{4dEM12b}
\eea 
(Note that, in text, the periodicity (\ref{4dEM11}) requires $b(0)=b(\be)=x$. ) 
we can express
\bea 
E_{HO}=\Tr <x|H\e^{-\be H}|x'>=\int_{-\infty}^\infty dx
<x|H\e^{-\be H}|x>\com\nn
W(\be,\omtil)=\Tr \e^{-\be H}=\int_{-\infty}^\infty dx <x|\e^{-\be H}|x>
\pr
\label{4dEM12c}
\eea 
          }
\bea
E_{HO}=\frac{1}{W(\be,\omtil)}\int\Dcal b~\Ltil_{HO}(b',b)|_\be\exp \left[ -\int_0^\be d\tau\Ltil_{HO}(b',b;\omtil) \right]
=-\frac{\pl}{\pl\be}\ln W(\be,\omtil)\com\nn
W(\be,\omtil)=\int\Dcal b~\exp \left[ -\int_0^\be d\tau\Ltil_{HO}(b',b;\omtil) \right]
\com
\label{4dEM12}
\eea 
where the path-integral is done over all paths which satisfy the {\it periodic} condition $b(0)=b(\be)$. 
Using the periodic property (\ref{4dEM11}), $b(\tau)$ is expressed as 
\bea
b(\tau)=\sum_{n\geq 1}c_n\sin~\tnpb\tau +\sum_{n\geq 0}\ctil_n\cos~\tnpb\tau
\pr
\label{4dEM13}
\eea 
The first part is odd for the Z$_2$-parity: $\tau\ra -\tau$ and 
the second one is even. 
Then $W$ can be clearly defined and is evaluated as
\bea
W(\be,\omtil)=\int\prod_{n\geq 1}dc_n\prod_{n\geq 0}d\ctil_n
\exp\left[
-\half\sum_{n\geq 1}\half\{(\tnpb)^2+\omtil^2 \}{c_n}^2
-\half\sum_{n\geq 0}\half\{(\tnpb)^2+\omtil^2 \}{\ctil_n}^2
    \right]\nn
=\exp\left[ -\half\sum_{n\geq 1}\ln \{(\tnpb)^2+\omtil^2 \}-\half\sum_{n\geq 0}\ln \{(\tnpb)^2+\omtil^2 \}\right]
\pr
\label{4dEM14}
\eea 
We normalize W at $\omtil=0$ (free motion). 
\bea
\What(\be,\omtil)=\frac{W(\be,\omtil)}{W(\be,0)}
\pr
\label{4dEM15}
\eea 
Hence, in order to evaluate $\What$, it is sufficient to consider the following quantity. 
\bea
\half\sum_{n\in \bfZ}\ln \frac{(\tnpb)^2+\omtil^2}{(\tnpb)^2}
=\half\int_{-\infty}^\infty dz\ln\frac{(\frac{2\pi}{\be}z)^2+\omtil^2}{(\frac{2\pi}{\be}z)^2}
+\int_{-\infty}^\infty dz\frac{\ln\frac{z^2+(\frac{\be\omtil}{2\pi})^2}{z^2}}{\e^{-2\pi iz}-1}\nn
=\frac{\be}{2}\omtil+2\pi\int_0^{\be\omtil/2\pi}dx\frac{1}{\e^{2\pi x}-1}
\pr
\label{4dEM16}
\eea 
We reach the familiar formula of the energy spectrum of the radiation.
\bea
E_{HO}(\be,\omtil)=-\frac{\pl}{\pl\be}\ln\What(\be,\omtil)
=\frac{\omtil}{2}+\frac{\omtil}{\e^{\be\omtil}-1}\equiv E_0(\omtil)+E_1(\be,\omtil)
\pr
\label{4dEM17}
\eea 
The first term $E_0$ is the {\it zero-point oscillation energy} and does {\it not} 
depend on $\be$, 
while the second one
$E_1$ does depend on $\be$. 
We realize the summation 
over the "Kaluza-Klein modes" along the $\tau$-direction corresponds to 
the familiar way of the statistical-procedure over the canonical ensemble 
$\{\exp (-E_n/T),E_n=(\half+n)\omtil\ |~n=0,1,2\cdots \}$. This simply means 
the equivalence of the statistical system in the equilibrium (at a temperature $T=1/\be$) and the Euclidean field theory 
with the {\it periodic} Euclidean-time (periodicity $\be$). 

Going back to the energy evaluation of the 4D EM, we obtain, using the results (\ref{4dEM17}), 
\bea
E_{4dEM}=E_{Cas}+E_\be\com\nn
E_{Cas}=2\sum_{m_x,m_y,n\in\bfZ}E_0(\omtil_{m_xm_yn})=\sum_{m_x,m_y,n\in\bfZ}\omtil_{m_xm_yn}\com\nn
E_{\be}=2\sum_{m_x,m_y,n\in\bfZ}E_1(\omtil_{m_xm_yn})=2\sum_{m_x,m_y,n\in\bfZ}\frac{\omtil_{m_xm_yn}}{\e^{\be\omtil_{m_xm_yn}}-1}
\pr
\label{4dEM18}
\eea 
The factor 2, in front of the middle equations above, reflects the degree of freedom 
of the polarization vector $\avec$ (\ref{4dEM7}). 
$E_{Cas}$ is the sum of zero-point energy over all frequency modes (vacuum energy of the 4D EM). 
It is {\it Casimir energy}. It is that part of the vacuum energy which is {\it independent
of the coupling} and is {\it dependent on the boundaries}.   
$E_\be$ gives us Stefan-Boltzmann's law. 
\bea
E_{\be}=2\sum_{m_x,m_y,n\in\bfZ}\frac{\omtil_{m_xm_yn}}{\e^{\be\omtil_{m_xm_yn}}-1}
=2\int_0^\infty\frac{dk4\pi k^2}{(\frac{\pi}{L})^2\frac{\pi}{l}}\frac{\omtil(k)}{\e^{\be\omtil(k)}-1}
=(2L)^2(2l)\int_0^\infty dk \Pcal(\be,k)    \nn
=\frac{3! \zeta(4)}{\pi^2}(2L)^2\cdot 2l\times\frac{1}{\be^4}\com\q 
\Pcal(\be,k)=\frac{1}{\pi^2}\frac{k^3}{\e^{\be k}-1}\com\q 
\zeta(4)=\frac{\pi^4}{90}
\com
\label{4dEM19}
\eea 
where $\omtil(k)=k$ and a formula $\int_0^\infty x^s/(\e^x-1)dx=s!\zeta(s+1)$ is used. 
$\Pcal(\be,k)$ is the Planck's radiation formula. The behavior of $\Pcal(\be,k)$ is 
graphically shown in Fig.\ref{PlanckDistB}. (We see similar graphs in the 
5D case of the text. The extra axis corresponds to the $\be$-axis. ) 
The peak curve of the graph is {\it hyperbolic}, $\be k$=const., in the $(\be,k)$-plane 
(Wien's displacement law). 
We note $E_\be\propto \be^{-4}=T^4$, hence it vanishes for $T=0$. And $(2L)^2(2l)$ is the volume
of the region bounded by the two planes.

$E_{Cas}$ does not vanish for $T=0$. 
It is, however, formally {\it divergent}.  We need a proper {\it regularization} 
for the {\it summation over the infinite degree of freedom} due to the continuity 
of the space-time.  It corresponds to the {\it renormalization} procedure in the 
local field theories. 
From the explanation so far, $E_{Cas}$ is given by 
\bea
E^\La_{Cas}=\sum_{m_x,m_y,n\in\bfZ}\omtil_{m_xm_yn}~g(\frac{\omtil_{m_xm_yn}}{\La}) \nn
=\sum_{m_x,m_y,n\in\bfZ}\sqrt{
(m_x\frac{\pi}{L})^2+(m_y\frac{\pi}{L})^2+(n\frac{\pi}{l})^2
                              }~g(\frac{\omtil_{m_xm_yn}}{\La})
\com
\label{4dEM20}
\eea 
where we introduce the {\it cut-off} function: 
$g(\om/\La)=1 \mbox{\ for\ } 0\leq\om\leq\La,\q =0 \mbox{\ for\ } \om>\La$. $\La$ is the 
cut-off parameter for the absolute value of the 3D (x,y,z) momentum. 
We will take the limit $\La\ra\infty$ at an appropriate stage. 
We first fix the reference point, $L\ra\infty, L\gg l\ra\infty$, from which we "measure" the energy.
\bea
E^{\La 0}_{Cas}
=\int_{-\infty}^\infty\int_{-\infty}^\infty\frac{dk_xdk_y}{(\pi/L)^2} 
\int_{-\infty}^\infty\frac{dk_z}{\pi/l} 
\sqrt{k_x^2+k_y^2+k_z^2}~g(\frac{k}{\La}) \nn
=\int\int\int_{k\leq \La}
\frac{dk_xdk_ydk_z}{(\pi/L)^2\pi/l} 
\sqrt{k_x^2+k_y^2+k_z^2}
\com
\label{4dEM21}
\eea 
This quantity diverges quartically. 
\footnote{
The last expression of (\ref{4dEM21}) shows the introduction of the cut-off 
function $g(\om/\La)$ is equivalent to the usual one ($k\leq \La$) taken in the text.  
}
Hence the energy density (the energy per unit area of xy-plane) $u$ is given by
\bea
u=\frac{E^\La_{Cas}-E^{\La 0}_{Cas}}{(2L)^2}
=\int_{-\infty}^\infty \int_{-\infty}^\infty \frac{dk_xdk_y}{(2\pi)^2}
      \left[
\sum_{n\in\bfZ} \sqrt{k_x^2+k_y^2+(n\frac{\pi}{l})^2} 
~g\left(\frac{1}{\La}\sqrt{k_x^2+k_y^2+(n\frac{\pi}{l})^2} \right)
       \right.\nn
       \left. 
-\int_{-\infty}^\infty\frac{dk_z}{\pi/l} \sqrt{k_x^2+k_y^2+k_z^2}
~g\left(\frac{1}{\La}\sqrt{k_x^2+k_y^2+k_z^2}\right)
      \right]    \nn
=\frac{1}{\pi}\int_0^\infty kdk\left[
\frac{k}{2}g\left(\frac{k}{\La}\right)
+\sum_{n=1}^{\infty}\sqrt{k^2+(n\frac{\pi}{l})^2} 
~g\left(\frac{1}{\La}\sqrt{k^2+(n\frac{\pi}{l})^2}\right)
                               \right.   \nn
                               \left.
-\int_0^\infty dn\sqrt{k^2+(n\frac{\pi}{l})^2} 
~g\left(\frac{1}{\La}\sqrt{k^2+(n\frac{\pi}{l})^2}\right)
                       \right]      \nn
\equiv \frac{1}{\pi}\left[
\frac{1}{2}X(0)+\sum_{n=1}^{\infty}X(n)
           -\int_0^\infty dnX(n)
                       \right] \com\nn
X(n)=\int_0^\infty kdk\sqrt{k^2+(n\frac{\pi}{l})^2} 
~g\left(\frac{1}{\La}\sqrt{k^2+(n\frac{\pi}{l})^2}\right)                   
      \pr
\label{4dEM22}
\eea 
Using the Euler-MacLaurin formula 
\bea
\half X(0)+X(1)+X(2)+\cdots -\int_0^\infty dn X(n)
=-\frac{1}{2!}B_2X'(0)-\frac{1}{4!}B_4X'''(0)+\cdots\com\nn
B_2=\frac{1}{6}\com\q B_4=-\frac{1}{30}
\com
\label{4dEM23}
\eea 
where $B_n$ is the Bernoulli number, 
we finally obtain the {\it finite} result.
\bea
u=\frac{\pi^2}{(2l)^3}\frac{B_4}{4!}=-\frac{\pi^2}{720}\frac{1}{(2l)^3}
\com
\label{4dEM24}
\eea 
which does not depend on $\La$. Especially there remains no log$\La$ divergences. 
\footnote{
This means the interaction between (4D) free fields and the boundary is so simple 
that there is no anomalous scaling behavior. 5D free fields, however, turn out to
have the anomalous scaling behavior as is shown in the text. 
}
\footnote{
Note that the {\it positive definite} expression (\ref{4dEM20}) is, 
after the change of the energy origin (\ref{4dEM22}), assigned  
to a {\it negative} value. The experimentally-observed {\it attractiveness} of the Casimir force 
tells the importance of this regularization procedure. 
}
This point is contrasting with the ordinary renormalization of {\it interacting} theories 
such as 4D QED and 4D YM. 
Hence we need {\it not} the renormalization of the wave-function and the parameter $l$. 
As is shown in the text, the {\it renormalization of the boundary parameter(s)} is necessary 
in the 5D case.


\section{App. C.\ \ Numerical Confirmation of the Relation between Weight Function 
and Minimal Surface Curve \label{WFaGC}}

\begin{figure}
\caption{Contour of 
$(-1/2)\ptil^3W_4(\ptil,z)F^-(\ptil,z)$(linear suppression, Fig.\ref{W4L100k10T1}). 
$\La=100,\ \om=10,\ T=1$\ . 
$1.0001/\om\leq z\leq 0.9999/T ,\ \m=\La T/\om\leq \ptil\leq 25$. 
}
\includegraphics[height=8cm]{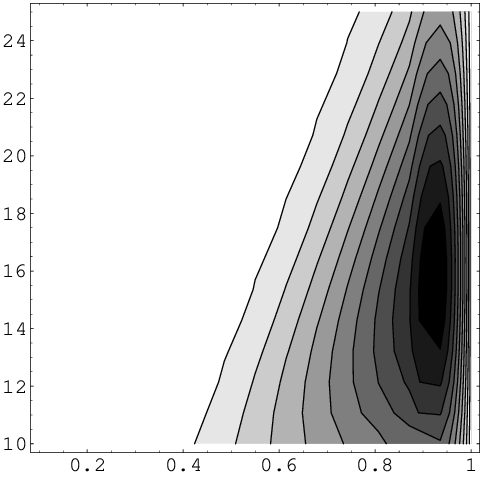}
\label{ContW4L100k10T1}
\end{figure}

\begin{figure}
\caption{
Minimal Area Curve $\ptil(z)$, (\ref{MiniSur7}). $\ptil(1.0)=30.0$, $\ptil'(1.0)=10.0$. 
Horizontal axis:\ 0.0001$\leq$ z $\leq$ 1.0\ ;\ 
Vertical axis:  $0.0 \leq \ptil\leq 30$. 
}
\includegraphics[height=8cm]{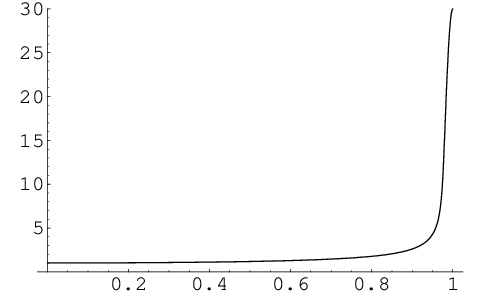}
\label{Geo30P1a10DP1}
\end{figure}

In this appendix, we numerically confirm the proposal in Sec.\ref{weight}. 
In order to define the weight function $W(\ptil,z)$, 
we presented the requirement (\ref{weight4}) , that is, 
the valley-bottom line of the $(\ptil,z)$-integral of (\ref{weight1}) should be 
equal to the minimal surface line (\ref{weight2}). 
For the case of the linear suppression (the weight $W_4$, Fig.\ref{W4L100k10T1}), 
its valley-bottom line is read from the contour graph of Fig.\ref{ContW4L100k10T1}. 
In Fig.\ref{Geo30P1a10DP1}, the line is numerically reproduced as the minimal 
surface line. For most of other suppression forms, we confirm 
their valley-bottom lines can be reproduced as the minimal surface lines 
by taking the boundary conditions appropriately. 


\section{App. D:\ Numerical Evaluation of Scaling Laws: $E_{Cas}$ (\ref{UIreg5}), $E_{Cas}^{RS}$ (\ref{surfM1}), 
and $E_{Cas}^W$ (\ref{uncert1bX})\label{appC}}
In the text, (regularized) Casimir energy is numerically calculated in three ways:\ 
1) Original version (Rectangle-region integral), 2) Restricted-region integral ( Randall-Sundrum type ), 
3) Weighted version. The final expressions show the {\it scaling} behaviors about the boundary 
(extra-space) parameters $T$ , $\om$ and the 4D momentum cut-off $\La$. The results are crucial 
for the present conclusion. Hence we explain here how the numerical results are obtained.

\begin{figure}
\caption{
Casimir Energy  $E_{Cas}^{\La,-}$ of (\ref{AppD1}) for various $\La$.  
$T=1,\ \om=10^4$. 
Horizontal axis: $\ln \La$ ($\La=10^4\times (1,2,4,8,16)$),\ Vertical Axis: 
$-\ln (|2^3\pi^2\times(1/2)E_{Cas}^{\La,-}|)$.  
}
\includegraphics[height=8cm]{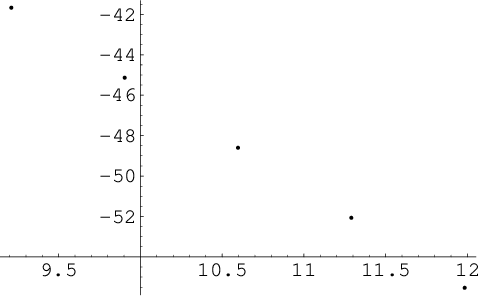}
\label{T1k10000Ln}
\end{figure}

\begin{figure}
\caption{
Casimir Energy  $E_{Cas}^{\La,-}$ of (\ref{AppD1}) for various $T$.  
$\om=10^4,\ \La=2\times 10^4$. 
Horizontal axis: $\ln T$ ($T=(1,2,4,8,16)$),\ Vertical Axis: 
$\ln (|2^3\pi^2\times(1/2)E_{Cas}^{\La,-}|)$.  
}
\includegraphics[height=8cm]{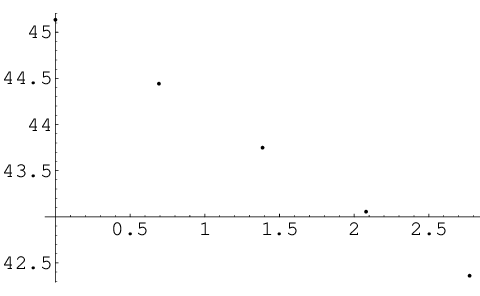}
\label{k1manL2manLn}
\end{figure}

First, let us take the un-weighted case with the rectangle integral-region (original form) of 
Casimir energy (\ref{UIreg2b}). 
\bea
-E_{Cas}^{\La,\mp}(\om,T)=\frac{2\pi^2}{(2\pi)^4}\int_{\m}^{\La}d\ptil\int_{1/\om}^{1/T}dz~\ptil^3 F^\mp (\ptil,z)\com\nn
F^\mp (\ptil,z)
=\frac{2}{(\om z)^3}\int_\ptil^\La\ktil~ G^\mp_k(z,z)d\ktil
\pr
\label{AppD1}
\eea
where $G^\mp_k$ is explicitly given in (\ref{HKA12}). The integral region 
is graphically shown, in Fig.\ref{zpINTregionW}, as the rectangle ($[1/\om,1/T]\times[\m=\La T/\om, \La]$). 
The graphs of the integrand of (\ref{AppD1}), $(-1/2)\ptil^3F(\ptil,z)$, are shown for 
$(T,\om,\La)=(1,10^4,10^4) \mbox[Fig.\ref{p3FmL10000}], (1,10^4,2\times 10^4) \mbox[Fig.\ref{p3FmL20000}],  (1,10^4,4\times 10^4) \mbox[Fig.\ref{p3FmL40000}], $
in the text. From the behaviors we can expect $E_{Cas}^\La(\om,T)$, (\ref{AppD1}), leadingly behaves as $\La^5/T$, because 
the depth of the valley, shown in Fig.\ref{p3FmL10000}-\ref{p3FmL40000}, proportional to $\La^4$ and their behaviors 
are monotonous (except near the boundaries $z=1/\om$ and $1/T$) along the extra axis. 
It is confirmed by directly evaluating (\ref{AppD1}) 
numerically (the numerical integral in \cite{Mathema}). We plot the numerical results 
in Fig.\ref{T1k10000Ln} for various $\La$'s with fixed $(T,\om)$, and 
in Fig.\ref{k1manL2manLn} for various $T$'s with fixed $(\om, \La)$.  
Furthermore we have confirmed sufficient $\om$-indepedence for the 
region $T=1, \om=10^3\times (1,2,4,8,16), \La=10^4\times (2,4)$. 
Let us fit $E_{Cas}^\La(\om,T)$ from the above numerical results. 
First we can regard it as the function of one massive parameter $\La$, 
and two massless parameters $\La/T$ and $\La/\om$. From the linear 
dependences in Fig.\ref{T1k10000Ln} and Fig.\ref{k1manL2manLn} we may put
\bea
E_{Cas}^\La(\om,T)=\frac{\La^5}{T}(a_1+a_2\ln\frac{\La}{T}+a_3\ln\frac{\La}{\om})
\pr
\label{AppD2}
\eea
From the $\om$-independence, we take $a_3=0$. 
From the numerical results, the best fit is given by 
(Manipulating Numerical Data in \cite{Mathema})
\bea
E^{\La,-}_{Cas}(\om,T)=\frac{1}{8\pi^2}\times \left[ 
-0.02500\frac{\La^5}{T}(1-4.685\times 10^{-9}\ln\frac{\La}{T}) \right]
\pr
\label{AppD3}
\eea
From the precision of the numerical integral, we may safely regard the 
log term in (\ref{AppD3}) vanishing ($a_2=0$). Finally we obtain (\ref{UIreg5}).

\begin{figure}
\caption{
Casimir Energy  $E_{Cas}^{-RS}$ of (\ref{AppD4}) for various $\La$ with fixed $(\om=10^3,T=1)$.  
Horizontal axis: $\ln \La$ ($\La=(1,2,4,8,16)\times 10^4$),\ Vertical Axis: $\ln (|2^3\pi^2\times \frac{1}{2}E_{Cas}^{-RS}|)$.  
The results are placed on a straight line with the slope 5 for different $\La$'s. 
}
\includegraphics[height=8cm]{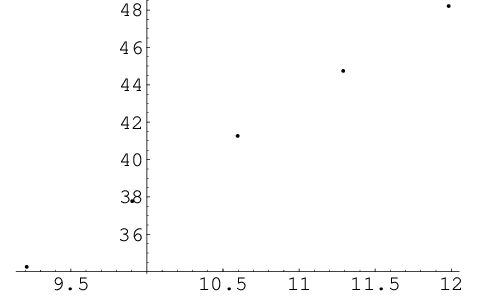}
\label{om1000T1xLnLa}
\end{figure}

\begin{figure}
\caption{
Casimir Energy  $E_{Cas}^{-RS}$ of (\ref{AppD4}) for various $\om$ with fixed $(T=1,\La=2\times 10^4)$. 
Horizontal axis: $\ln \om$ ($\om=(1,2,4,8,16)\times 10^2$),\ Vertical Axis: $\ln (|2^3\pi^2\times \frac{1}{2}E_{Cas}^{-RS}|)$.  
The results are placed on a straight line with the slope -1 for different $\om$'s. 
}
\includegraphics[height=8cm]{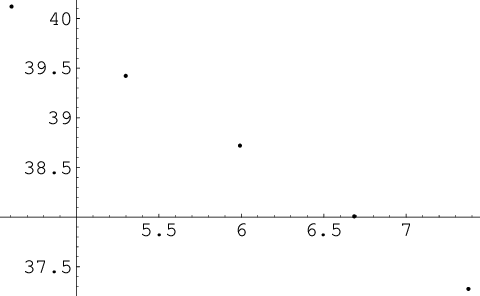}
\label{L2manT1xLnom}
\end{figure}

For the restricted region case, (\ref{surfM1}), we do the numerical integral of 
the following expression.
\bea
-E^{-RS}_{Cas}(\om,T)=
\frac{2\pi^2}{(2\pi)^4}\int_{\m}^{\La}dq\int_{1/\om}^{\La /\om q}dz~q^3 F^- (q,z)
=\frac{2\pi^2}{(2\pi)^4}\int_{1/\om}^{1/T}du\int_{\m}^{\La /\om u}d\ptil~\ptil^3 F^- (\ptil,u)
\com
\label{AppD4}
\eea
where $\m=\La T/\om$. 
We plot the results, in Fig.\ref{om1000T1xLnLa}, for various $\La$ with fixed $(\om,T)$ and 
in Fig.\ref{L2manT1xLnom}, for vaious $\om$ with fixed $(T,\La)$. 
Furthermore 
we have confirmed T-independence for $T=(1,2,3)$ with $(\om=10^4,\La=2\times 10^4)$ and 
for $T=(1,2,3,4,8,16)$ with $(\om=10^3,\La=2\times 10^4)$. 
From the straight line behaviors and the T-independence, we can safely fit the curve as 
$E_{Cas}^{-RS}=(\La^5/\om)(b_1+b_2\ln(\La/T)+b_3\ln(\La/\om)), b_2=0$. The best fit is given by 
\bea
2^3\pi^2E_{Cas}^{-RS}(\om,T)=\frac{\La^5}{\om}\left(
(-1.59,-1.56)\times 10^{-2} + (-1.41,-1.97)\times 10^{-4}~\ln(\La/\om)
                                               \right)
\pr
\label{AppD5}
\eea
The first component of the above coefficients comes from Fig.\ref{om1000T1xLnLa} data, 
the second one from Fig.\ref{L2manT1xLnom} data. The "width" of the coefficient-values tells us 
the first-term coefficient has the significant digit number 2, while the second-term one has 
the number 1. 
In the text, we take the average values (\ref{surfM1}).

\begin{figure}
\caption{
Casimir Energy  $E_{Cas}^{-W_1}$ of (\ref{AppD6}) for various $\La$ with fixed $(T=1,\om=10^3)$(RightDown) 
and $(T=0.01,\om=10^2)$(LeftUp).  
Horizontal axis: $\ln\La$ ($\La=10^3\times(1,2,4,8,16)$ and $10^4\times(1,2,4,8,16)$),\ Vertical Axis: 
$-\ln|2^3\pi^2(\frac{N_1}{2})E_{Cas}^{-~W_1}|$.  
}
\includegraphics[height=8cm]{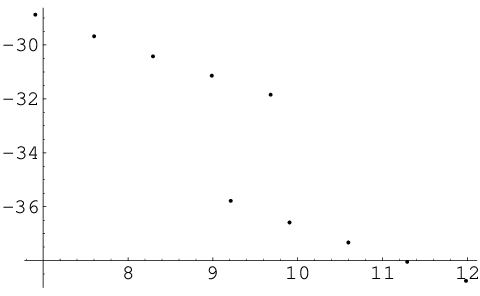}
\label{T001k100T1k1000}
\end{figure}

\begin{figure}
\caption{
Casimir Energy  $E_{Cas}^{-W_1}$ of (\ref{AppD6}) for various $T$ with fixed $(\La=2\times10^4,\om=10^3)$. 
Horizontal axis: $\ln~T$ ($T=(1,1/2,1/4,1/8,1/16)$),\ Vertical Axis: 
$-\ln|2^3\pi^2(\frac{N_1}{2})E_{Cas}^{-~W_1}|$.  
}
\includegraphics[height=8cm]{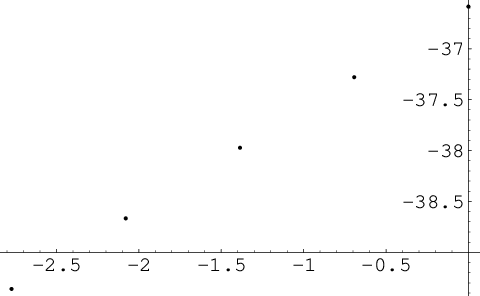}
\label{L20000k1000Ln}
\end{figure}

\begin{figure}
\caption{
Casimir Energy  $E_{Cas}^{-W_1}$ of (\ref{AppD6}) for various $\om$ with fixed $(\La=8\times10^4,T=1)$. 
Horizontal axis: $\ln~\om$ ($\om=10^2\times(4,8,16,32,64,128)$),\ Vertical Axis: 
$-\ln|2^3\pi^2(\frac{N_1}{2})E_{Cas}^{-~W_1}|$.  
}
\includegraphics[height=8cm]{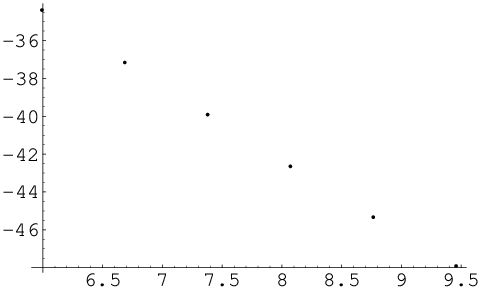}
\label{L80000T1Ln}
\end{figure}

Finally we explain the weighted case (\ref{uncert1}) taking the elliptic type, $W_1$, as an example. 
\bea
-E^{-~W_1}_{Cas}(\om,T)\equiv\int_{\ptil\leq\La}\frac{d^4p_E}{(2\pi)^4}\int_{1/\om}^{1/T}dz~ W_1(\ptil,z)F^- (\ptil,z)\com\nn
(N_1)^{-1}\e^{-(1/2) \ptil^2/\om^2-(1/2) z^2 T^2}\equiv W_1(\ptil,z),\ N_1=1.711/8\pi^2 
\com
\label{AppD6}
\eea
where the UV cut-off $\La$ is introduced to see the scaling behavior.  
In Fig.\ref{T001k100T1k1000}, we show 
the numerical results of $E^{-W_1}_{Cas}(\om,T)$ for different $\La$'s with fixed $(T,\om)$. 
The two lines both are straight ones with the slope -1. For the T-dependence (with fixed $(\om,\La)$) and 
$\om$-dependence (with fixed $(T,\La)$) we show them in Fig.\ref{L20000k1000Ln} and in Fig.\ref{L80000T1Ln}, respectively. 
They are straight lines with slopes +1 and -4, respectively. 
From the straight-lines behavior of Fig.\ref{T001k100T1k1000}-\ref{L80000T1Ln}, we can safely fit the curve as 
$E_{Cas}^{-W1}=(\om^4\La/T)(c_1+c_2\ln(\La/\om)+c_3\ln(\La/T))$. 
The best fit is given by 
\bea
(2^3\pi^2N_1/2)\times E_{Cas}^{-W_1}(\om,T)=\frac{\om^4}{T}\La \left( -1.04-0.11~\ln\frac{\La}{\om}+0.099~\ln\frac{\La}{T}
                                               \right)
\pr
\label{AppD7}
\eea
Taking into account the present precision, we take $c_1=1.04\times (2/1.711)=1.22,\ c_2=0.11\times (2/1.711)=0.13,\ c_3=-0.099\times (2/1.711)=-0.12$ in the text. 
As for other types of $W$'s, the best fit scaling behaviors are listed in (\ref{uncert1bX}) of the text.

\section{App. E.\ \ Normalization Constants of Weight Functions (\ref{uncert1}) \label{NorConst}}

In Sec.\ref{uncert}, we introduce the weight function $W$ to evaluate the Casimir energy. 
For the comparison between the Casimir energy values obtained by different $W$'s,  
the normalization constants are important. 
The normalization constants $N_i$'s (\ref{uncert1}) are defined by the following condition: 
\bea
\int_{\m=\La T/\om <\ptil<\La}\frac{d^4p}{(2\pi)^4}\int_{1/\om}^{1/T}dz~ W_i(\ptil,z)=
\frac{1}{8\pi^2}\frac{\om^4}{T}\int_{\La T/\om^2}^{\La/\om}dx\int_{T/\om}^1dw~ x^3 W_i(\ptil=\om x,z=w/T)\nn
\equiv\frac{\om^4}{T}\com\q\q
\La\gg\om\gg T\com
\label{NorConst1}
\eea
For the ends of the integral-regions, we practically may take $\frac{\La}{\om}=\infty$ and $\frac{T}{\om}=0$ 
except for $W_2$ and $W_3$. As for 
the starting end of $x$-integral, we have two choices depending on 
what range of the value $\om$ is considered (\ (A)\ $\om\sim\sqrt{\La T}$,\ (B)\ $\om\gg\sqrt{\La T}$\ 
(C)\ $\om\ll\sqrt{\La T}$\ ) in the numerical data-taking. 
They are explicitly given by

(A)\ $\om\sim\sqrt{\La T}$\ (geometrically averaged point)\nl
In this case, we take $\frac{\La T}{\om^2}=1, \frac{T}{w}=0, \frac{\La}{\om}=\infty$ in (\ref{NorConst1}). 
\bea
8\pi^2N_1=\frac{3}{\sqrt{\e}}\int_0^1dw~ \e^{-w^2/2}=1.557\com\q
8\pi^2N_{1b}=\frac{3}{\sqrt{\e}}=\int_1^\infty dx~ x^3 \e^{-x^2/2}=1.820\com\nn
8\pi^2N_{2}=\int_1^\infty dx \int_{T/\om}^1~ dw x^3 \e^{-xw}=2(\frac{\om}{T})^3\com\nn
8\pi^2N_{3}=\int_1^\infty dx \int_{T/\om}^1~ dw x^3 \e^{-x^2w^2/2}=\frac{2}{3}(\frac{\om}{T})^3\com\nn
8\pi^2N_{4}=\int_1^\infty dx \int_0^1 dw~ x^3 \e^{-x^2/2w^2}=0.3222\com\q
8\pi^2N_{5}=\int_1^\infty dx \int_0^1 dw~ x^3 \e^{-x/w^2}=0.6342\com\nn
8\pi^2N_{6}=\int_1^\infty dx \int_0^1 dw~ x^3 \e^{-x^2/2w}=0.5521\com\q
8\pi^2N_{7}=\int_1^\infty dx~ x^3 \e^{-x^4/2}=\frac{1}{2\sqrt{\e}}=0.3033\com\nn
8\pi^2N_8=\frac{3}{\sqrt{\e}}\int_0^1dw~ \e^{-1/2w^2}=0.3800\com\q
8\pi^2N_{47}=\int_1^\infty dx \int_0^1 dw~ x^3 \e^{-x^2(x^2+1/w^2)/2}=0.03893\com\nn
8\pi^2N_{56}=\int_1^\infty dx \int_0^1 dw~ x^3 \e^{-(x/w)(x+1/w)/2}=0.1346\com\nn
8\pi^2N_{88}=\int_1^\infty dx \int_0^1 dw~ x^3 \e^{-(1/2)(x^2+1/w^2)^2}=0.005006\com\nn
8\pi^2N_{9}=\int_1^\infty dx \int_0^1 dw~ x^3 \e^{-(1/2)(x+1/w)^2}=0.03921\com
\label{NorConst2}
\eea

(B)\ $\om\gg\sqrt{\La T}$\nl
In this case, we take $\frac{\La T}{\om^2}=0, \frac{T}{w}=0, \frac{\La}{\om}=\infty$ in (\ref{NorConst1}). 
\bea
8\pi^2N_1=2\times\int_0^1dw~ \e^{-w^2/2}=1.711\com\q
8\pi^2N_{1b}=\int_0^\infty dx~ x^3 \e^{-x^2/2}=2\com\nn
8\pi^2N_{2}=\int_0^\infty dx \int_{T/\om}^1~ dw x^3 \e^{-xw}=2(\frac{\om}{T})^3\com\nn
8\pi^2N_{3}=\int_0^\infty dx \int_{T/\om}^1~ dw x^3 \e^{-x^2w^2/2}=\frac{2}{3}(\frac{\om}{T})^3\com\nn
8\pi^2N_{4}=\int_0^\infty dx \int_0^1 dw~ x^3 \e^{-x^2/2w^2}=\frac{2}{5}\com\q
8\pi^2N_{5}=\int_0^\infty dx \int_0^1 dw~ x^3 \e^{-x/w^2}=\frac{2}{3}\com\nn
8\pi^2N_{6}=\int_0^\infty dx \int_0^1 dw~ x^3 \e^{-x^2/2w}=\frac{2}{3}\com\q
8\pi^2N_{7}=\int_0^\infty dx~ x^3 \e^{-x^4/2}=\frac{1}{2}\com\nn
8\pi^2N_8=2\times\int_0^1dw~ \e^{-1/2w^2}=0.4177\com\nn
8\pi^2N_{47}=\int_0^\infty dx \int_0^1 dw~ x^3 \e^{-x^2(x^2+1/w^2)/2}=0.1028\com\nn
8\pi^2N_{56}=\int_0^\infty dx \int_0^1 dw~ x^3 \e^{-(x/w)(x+1/w)/2}=0.1779\com\nn
8\pi^2N_{88}=\int_0^\infty dx \int_0^1 dw~ x^3 \e^{-(1/2)(x^2+1/w^2)^2}=0.01567\com\nn
8\pi^2N_{9}=\int_0^\infty dx \int_0^1 dw~ x^3 \e^{-(1/2)(x+1/w)^2}=0.05320\com
\label{uncert1ab}
\eea
In the text, we take the case (B).

(C)\ $\om\ll\sqrt{\La T}$\nl
In this case, we may take $\frac{\La T}{\om^2}=0, \frac{T}{w}=0, \frac{\La}{\om}=\infty$ in (\ref{NorConst1}). 
This is the same as the case (B). In Sec.\ref{conc}, we apply the results to the cosmological constant or 
the dark energy. We take there $\La=10^{28}$ eV, $\om=10^{-3}$ eV and $T=10^{-20}$ eV.

\newpage
\section{Acknowledgment}
 Parts of the content of this work have been already presented at
the international conference on "Progress of String Theory and Quantum Field Theory" (07.12.7-10, Osaka City Univ., Japan), 
63rd Meeting of Japan Physical Society (08.3.22-26,Kinki Univ.,Osaka,Japan),  
Summer Institute 2008 (08.8.10-17, Chi-Tou, Taiwan), the international conference
on "Particle Physics, Astrophysics and Quantum Field Theory"(08.11.27-29, Nanyang 
Executive Centre, Singapore), 1st Mediterranean Conference on Classical and Quantum Gravity 
(09.9.14-18,Greece) and the international workshop on "Strong Coupling Gauge Theories in LHC Era" 
(09.12.8-11, Nagoya, Japan). 
The author thanks 
T. Appelquist (Yale Univ.), S.J. Brodsky (SLAC), 
K. Fujikawa (Nihon Univ.), T. Inagaki (Hiroshima Univ.), 
K. Kanaya (Tsukuba Univ.), T. Kugo (Kyoto Univ.), S. Moriyama (Nagoya Univ.), 
N. Sakai (Tokyo Women's Univ.), M. Tanabashi (Nagoya Univ.) and H. Terao (Nara Women's Univ.) 
for useful comments on the occasions.

\end{document}